\documentclass[11pt]{article}
\usepackage{jheppub}
\usepackage{mathtools,amssymb,amsthm}
\usepackage[utf8]{inputenc}
\usepackage{enumitem}
\usepackage{amsmath}
\usepackage{amssymb}
\usepackage{physics}
\usepackage[all]{xy}
\usepackage{xcolor}
\usepackage{simpler-wick}
\usepackage{tikz}
\usepackage{cancel}
\usetikzlibrary{decorations.pathmorphing}
\usepackage{ragged2e}
\usetikzlibrary{decorations.pathmorphing,calc}

\usepackage{babel}
\begin{document}

\global\long\def\ad{{\rm ad}}%

\global\long\def\ads{{\rm AdS}}%

\global\long\def\dd{{\rm d}}%

\global\long\def\del{\mathcal{\delta}}%

\global\long\def\Del{\Delta}%

\global\long\def\da{\dagger}%

\global\long\def\id{{\rm id}}%

\global\long\def\me{\mathcal{E}}%

\global\long\def\ml{\mathcal{L}}%

\global\long\def\mm{\mathcal{\mathcal{M}}}%

\global\long\def\mc{\mathcal{C}}%

\global\long\def\mf{\mathcal{\mathcal{F}}}%

\global\long\def\la{\langle}%

\global\long\def\ra{\rangle}%

\global\long\def\lr{\leftrightarrow}%

\global\long\def\lf{\leftrightarrow}%

\global\long\def\ma{\mathcal{A}}%

\global\long\def\mb{\mathcal{B}}%

\global\long\def\md{\mathcal{D}}%

\global\long\def\mbr{\mathbb{R}}%

\global\long\def\mbz{\mathbb{Z}}%

\global\long\def\mh{\mathcal{\mathcal{H}}}%

\global\long\def\mi{\mathcal{\mathcal{I}}}%

\global\long\def\ms{\mathcal{\mathcal{\mathcal{S}}}}%

\global\long\def\mg{\mathcal{\mathcal{G}}}%

\global\long\def\mfa{\mathcal{\mathfrak{a}}}%

\global\long\def\mfb{\mathcal{\mathfrak{b}}}%

\global\long\def\mfb{\mathcal{\mathfrak{b}}}%

\global\long\def\mfg{\mathcal{\mathfrak{g}}}%

\global\long\def\mj{\mathcal{\mathcal{J}}}%

\global\long\def\mk{\mathcal{K}}%

\global\long\def\mmp{\mathcal{\mathcal{P}}}%

\global\long\def\mn{\mathcal{\mathcal{\mathcal{N}}}}%

\global\long\def\mq{\mathcal{\mathcal{Q}}}%

\global\long\def\mo{\mathcal{O}}%

\global\long\def\qq{\mathcal{\mathcal{\mathcal{\quad}}}}%

\global\long\def\ww{\wedge}%

\global\long\def\ka{\kappa}%

\global\long\def\nn{\nabla}%

\global\long\def\nb{\overline{\nabla}}%

\global\long\def\we{\wedge}%

\global\long\def\pp{\prime}%

\global\long\def\sq{\square}%

\global\long\def\vp{\varphi}%

\global\long\def\te{\theta}%

\global\long\def\tr{{\rm Tr}}%

\global\long\def\nt{\hat{n}_{{\rm tot}}}%

\global\long\def\sh{{\rm {\rm sh}}}%

\global\long\def\ch{{\rm ch}}%

\global\long\def\Si{{\rm {\rm \Sigma}}}%

\global\long\def\si{{\rm {\rm \sigma}}}%

\global\long\def\sch{{\rm {\rm Sch}}}%

\global\long\def\vol{{\rm {\rm {\rm Vol}}}}%

\global\long\def\reg{{\rm {\rm reg}}}%
\newenvironment{malign}{\begin{equation}\begin{aligned}}{\end{aligned}\end{equation}\noindent}
\newcommand{\CirclePoints}[1]{%
  \def\R{#1}
  \coordinate (P)  at ({\R*cos(175)},{\R*sin(175)}); 
  \coordinate (BL) at ({\R*cos(225)},{\R*sin(225)}); 
  \coordinate (B)  at ({\R*cos(270)},{\R*sin(270)}); 
  \coordinate (RU) at ({\R*cos(25)},{\R*sin(25)});   
  \coordinate (RM) at ({\R*cos(0)},{\R*sin(0)});     
  \coordinate (RL) at ({\R*cos(-25)},{\R*sin(-25)}); 
}
\tikzset{
  circ/.style={line width=0.9pt},
  chord/.style={black, line width=0.7pt},
  dashedcon/.style={black, dashed, line width=0.7pt}, 
  blueprop/.style={blue, line width=0.9pt, -{Stealth[length=3.2pt,width=4.2pt]}},
  dot/.style={circle, fill=black, inner sep=1.6pt},
  lab/.style={font=\small},
}

\newcommand{\be}{\begin{equation}}
\newcommand{\ee}{\end{equation}}
\newcommand{\cH}{\mathcal{H}}

\newcommand{\prpsi}[2][2]{\Psi_{#1}^{(#2)}}

\def\JX#1{{\color{purple}{[#1]}}}
\def\jx#1{{\color{purple}{[#1]}}}

\newcommand\HR[1]{{\color{blue}{ [HR: #1]}}}

\title{Emergent States and Algebras from the Double-Scaling limit of Pure States in SYK}

\author[a]{Harshit Rajgadia}
\author[b]{and Jiuci Xu}

\affiliation[a]{Department of Theoretical Physics, Tata Institute for Fundamental Research, 1 Homi Bhabha Road, Mumbai, Maharashtra 400005, India}
\affiliation[b]{Department of Physics, University of California, Santa Barbara, CA 93106, USA}
\emailAdd{harshit.rajgadia@tifr.res.in}
\emailAdd{Jiuci\_Xu@ucsb.edu}

\justify \abstract{
Recent work has emphasized a subtlety of large-$N$ limits in AdS/CFT: a sequence of pure states in the microscopic theory need not remain pure with respect to the emergent algebra of observables.  We study this phenomenon for Kourkoulou--Maldacena (KM) states in the double-scaling limit of the SYK model, and show that their ensemble-averaged algebraic description depends crucially on which observables are retained in the limit. For fermionic operators of size $p\sim N^{1/2}$, generic operators converge to the usual chord operators of double-scaled SYK. The resulting von Neumann algebra is the standard Type II$_1$ factor, and the KM pure states at infinite temperature converge to the tracial state, so generic probes lose access to microscopic purity.  We then identify a class of operators adapted to the KM state that also survive the double-scaling limit. Since the KM state may be viewed as a projection inside the tracial state, these operators become dressed chord creation and annihilation operators, with the dressing encoding the distance to the KM state through the bulk.  Once included, the limiting algebra becomes Type I$_\infty$ and the sequence of KM states converges to a pure state. This provides a solvable example in which adding a sufficiently state-adapted operator to the emergent algebra restores access to the purity of the underlying state.

We derive exact modified chord-diagram rules for correlators of the dressed operators, including general uncrossed $2n$-point functions and crossed four-point functions. In the semiclassical limit these correlators exhibit a novel factorization pattern, while in the Schwarzian limit they reduce to correlators of boundary primaries dressed to Schwarzian modes. We also study a deformation of the chord Hamiltonian, the emergence of bound states above a critical coupling, and its relation to JT gravity coupled to an EOW brane of general tension. We identify an emergent $U(1)$ global symmetry in the Type I$_\infty$ algebra as well as its violation by $1/N$ corrections from double-trace wormholes.  Finally, we discuss analogies with boundary algebras proposed for black hole interiors and closed universes, and suggest lessons from our construction for both.
}

\enlargethispage{3\baselineskip}

\maketitle

\section{Introduction}
Understanding how semi-classical gravity emerges from large-$N$ quantum systems remains one of the central problems in holography. In the usual AdS/CFT setting~\cite{Maldacena:1997re,Witten:1998qj,Gubser_1998}, one considers a sequence of theories and states for which suitably normalized correlators of simple operators converge as $N\to\infty$. The limiting algebra of observables, together with its state, is then interpreted as the semi-classical bulk description. Recent work has sharpened the extent to which this procedure constrains the possible emergent physics, especially in questions involving purity, mixedness, and the algebraic structure seen by low-complexity probes~\cite{Antonini:2023hdh,Antonini:2024mci,Antonini:2025ioh,Gesteau:2024rpt,Engelhardt:2025vsp,Kudler-Flam:2025cki,Liu:2025close}.

A notable recent result in this direction is Gesteau's no-go theorem~\cite{Gesteau:2025obm}, which shows that for a sequence of pure states $\Psi_N$ with energy of order $O(N^0)$ in a holographic CFT, the usual large-$N$ limit cannot turn the state into a mixed state, provided the single-trace correlators $\langle \Psi_N | S_N | \Psi_N\rangle$ admit a well-defined large-$N$ limit. Thus, within the traditional notion of large-$N$ convergence, purity is preserved in the emergent algebraic description under the stated assumptions. On the other hand, recent work~\cite{Liu:2025close,Kudler-Flam:2025cki} showed that this conclusion can be evaded if one considers a suitably averaged large-$N$ limit instead. In that framework, a sequence of pure states may converge, as seen by a restricted algebra of observables, to a mixed state.

The purpose of this paper is to study this issue in a different physical regime and in a more refined algebraic setting. We consider the double-scaling limit of SYK,\footnote{Recent studies of this model include~\cite{Berkooz:2018jqr,Berkooz:2018qkz,Berkooz:2022mfk,Berkooz:2024lgq,Berkooz:2025ydg,Lin:2022rbf,Lin:2022nss,Rabinovici:2023yex,Goel:2023svz,Narovlansky:2023lfz,Okuyama:2023bch,Blommaert:2023opb,Blommaert:2023wad,Milekhin:2023bjv,Xu:2024von,Okuyama:2024yya,Almheiri:2024ayc,Almheiri:2024xtw,Milekhin:2024vbb,Aguilar-Gutierrez:2024oea,Narovlansky:2025tpb,Okuyama:2025hsd,Sergio:2025Complexity-Dictionary,Aguilar-Gutierrez:2025hty,Xu:2025Complexity,Cao:2025pir,Blommaert:2025avl,Aguilar-Gutierrez:2026ogo}.} and study the algebra generated by operators which are linear combinations of products of Majorana fermions of size
\[
p \sim N^{1/2}
\]
acting on Kourkoulou--Maldacena (KM) states~\cite{kourkoulou2017purestatessykmodel}. Unlike the $O(N^0)$ states considered above, KM states have energy and entropy of order $O(N)$, as is true of typical finite-temperature states in SYK. Correspondingly, a generic operator in the ensemble\footnote{By a \emph{generic operator} in SYK, we mean a finite linear combination of products of Majorana fermions with random Gaussian couplings, where each term has the same length $p\sim O(N^{1/2})$.} has support predominantly on states whose energy is also of order $N$. From the bulk point of view, the large-$N$ limit of such states is therefore expected to lie above the black-hole threshold, rather than in the regime of thermal AdS relevant to low-energy closed-universe constructions.\footnote{For recent developments of holographic descriptions of closed universes, see~\cite{Usatyuk:2024isz,Engelhardt:2025azi,Harlow:2025pvj,Abdalla:2025gzn,Blommaert:2025bgd,Balasubramanian:2025akx,Nomura:2025whc,Belin:2025ako,Harlow:2026hky,Zhao:2026mpl,Nomura:2026igt}.}

Our main result is that the same family of finite-$N$ pure states admits two inequivalent limiting descriptions, depending on which observables are retained in the double-scaled limit. If one considers only generic operators, then after ensemble averaging the resulting algebra converges to the familiar Type II$_1$ factor with the standard chord-diagram description. With respect to this algebra, the sequence of KM pure states converges to the tracial state. Thus generic probes lose access to the microscopic purity of the state, and the averaged large-$N$ limit produces an effectively mixed description.

This is not the full story, however. We identify a distinguished class of operators of the same size $O(N^{1/2})$ that can probe correlations of the KM state hidden from generic operators. We show that these operators also survive the double-scaling limit, and refer to them as \textit{dressed} operators. Once they are included, the averaged limit remains convergent, but the algebraic type changes drastically: the limiting von Neumann algebra is no longer Type II$_1$, but Type I$_\infty$. Correspondingly, the limiting state is no longer tracial and is instead pure. In this sense, purity and mixedness are not intrinsic properties of the sequence of states alone; rather, they depend crucially on the choice of limiting observable algebra. The KM states therefore provide a concrete example in which the same microscopic family of pure states gives rise to two different emergent algebraic descriptions: a Type II$_1$ description for generic probes, and a Type I$_\infty$ description once one includes the dressed operators.

A dressed operator is a non-Hermitian operator $M$ defined so as to annihilate the KM state at finite $N$. In the double-scaling limit, $M$ and its Hermitian conjugate $M^\dagger$ become creation and annihilation operators for a distinguished type of chord, dressed by the chord-number operator:
\begin{equation}
M^\dagger \;\to\; b^\dagger q^{\frac{\Delta}{2}\hat n},
\qquad
M \;\to\; q^{\frac{\Delta}{2}\hat n} b .
\end{equation}
This dressing is one of the central structural features of the limit. It shows that the operator capable of probing the purity of the KM state is not an ordinary chord operator, but one whose action depends on the surrounding chord configuration. The novelty is therefore not merely the appearance of state-dependent correlators, but the emergence of a state-adapted operator algebra in the double-scaled limit.

From this point of view, the generic operators are analogous to the universal single-trace operators in AdS/CFT. They survive the limit in a universal way\footnote{We mean that they survive the large-$N$ limit in any state admitting a semi-classical bulk description in that limit. For a general discussion of the algebraic formulation of the large-$N$ limit in AdS/CFT, see section VI of the recent lecture~\cite{Liu:2025krl}.} and generate an emergent algebra naturally interpreted as probing only the exterior region. By contrast, the dressed operators also survive the limit, but their GNS representation is adapted to the KM state. Once they are included, the emergent algebra is enlarged, and in the bulk interpretation this corresponds to extending the accessible region beyond the horizon. The extended portion of the entanglement wedge has been referred to as ``no man's island'' in recent work~\cite{Cao:2025pir}, where attempts were made to probe the island using the one-sided double-scaled algebra. The construction developed here provides a microscopic realization of the algebra considered there. The information distinguishing the pure KM state from the thermal state is encoded not in an additional label attached to the emergent state, but in the very structure of the emergent algebra of observables generated by the surviving operators. In this sense, the double-scaling limit furnishes a concrete realization of a state-dependent emergent operator algebra, paralleling the idea that different classes of boundary operator algebras in AdS/CFT can reconstruct different bulk regions.

To establish these results, we first analyze the combinatorics governing correlation functions with fermionic insertions in KM states. We show that in the double-scaling limit these correlators can be mapped to expectation values in the tracial state, supplemented by modified chord-diagram rules. Generic operators reduce to the standard chord operators and are insensitive to the microscopic details of the KM state, whereas the adapted operators reduce to dressed chord operators that remain sensitive to the projector defining the KM state.

We then study the correlation functions of the dressed operators. We derive analytic expressions for general uncrossed $2n$-point functions and crossed four-point functions, and use these exact formulas to analyze two distinct semi-classical limits. In the finite-temperature regime, the extra state-dependent contractions modify the usual factorization pattern of matter-chord correlators. In the Schwarzian regime, the two-point function of dressed operators reduces to the three-point function of boundary primaries dressed by Schwarzian modes in JT gravity coupled to matter~\cite{Kolchmeyer:2023gwa}.

Since the dressed chord operators belong to the double-scaled algebra, one can also deform the original chord Hamiltonian by a Hermitian combination of them and study the resulting spectral problem. Above a critical coupling, a finite number of bound states separates from the continuum, closely paralleling the appearance of bound states in traversable-wormhole constructions~\cite{Maldacena:2017axo,Maldacena-Qi:2018lmt,Maldacena:2018gjk,Maldacena:2019ufo,Maldacena:2020sxe}. We analyze the resulting spectrum, derive exact expressions for observables associated with the wormhole-length operator, and study the Schwarzian interpretation in terms of JT gravity coupled to an End-of-the-World brane, including the regime of negative brane tension.

Finally, we identify an emergent global $U(1)$ symmetry of the Type I$_\infty$ double-scaled algebra, whose conserved charge is the number of $M$-chords. This symmetry exists only in the strict double-scaling limit. We show how it is violated at finite $N$ by computing the leading correction to overlaps between states of different charge sectors, and identify this violation with the contribution of double-trace wormholes.

\paragraph{Organization of the paper.}
In section~\ref{sec:review} we review thermal Kourkoulou--Maldacena states, their bulk interpretation in the conventional large-$N$ limit, and the emergence of standard chord operators in double-scaled SYK.

In section~\ref{sec:KM} we study the double-scaling limit of KM states and the combinatorics of fermionic correlators in such states. We show that correlators involving both generic operators and the adapted operators $M^\dagger, M$ admit a chord-diagram description, with the latter reducing to dressed chord creation and annihilation operators.

In section~\ref{sec:vn_algebras} we study the von Neumann algebras generated by these operators acting on the KM state. We show that generic operators lead to the standard Type II$_1$ chord algebra, whereas including $M,M^\dagger$ enlarges the limiting algebra to a Type I$_\infty$ factor, with the KM state remaining pure with respect to this enlarged algebra.

In section~\ref{sec:correlators} we evaluate uncrossed $2n$-point functions and crossed four-point functions of dressed operators. In section~\ref{sec:classical} we use the exact formulas to analyze both the finite-temperature semi-classical limit and the Schwarzian limit near the spectral edge.

In section~\ref{sec:branes} we deform the Hamiltonian by a Hermitian combination of dressed chord operators and study the resulting spectral phases, with particular emphasis on the regime above the threshold for bound-state formation and its interpretation in JT gravity coupled to an EOW brane.

Finally, in section~\ref{sec:symmetry} we discuss the emergent global $U(1)$ symmetry of the double-scaled theory and show how its leading finite-$N$ violation arises from a double-trace wormhole contribution.

Various technical details are collected in the appendices. Appendix~\ref{app:useful} reviews analytic results in the zero- and one-particle sectors of the chord Hilbert space used in the correlator calculations. Appendix~\ref{app:spectral} studies the spectrum of the deformed chord Hamiltonian. Appendices~\ref{app:derivation-S} and~\ref{app:distance} contain details of the derivation of the length-operator observables in the continuous and discrete sectors.

\textbf{Note added:} Shortly after this paper appeared, Ref.~\cite{Sontag:2026iiu} constructed a baby-universe phase in the double-scaled regime of coupled SYK models. In that setup, a Hamiltonian chord can traverse the wormhole by contracting with the fermions in a heavy operator that supports the wormhole. Since the Kourkoulou--Maldacena projection may be viewed as the insertion of an infinitely heavy matter operator, it would be interesting to clarify the connection between the two setups by considering a ``lighter'' insertion that projects onto a subspace of spins. Applying the operator-algebraic analysis developed in the present paper to the emergence of baby universes remains an interesting open direction.

\section{Review of pure states in SYK and the Chord Diagram Expansion} \label{sec:review}

\subsection{Pure States in SYK and their Holographic Dual}

The SYK model \cite{SachdevYe1993, Kitaev2015SimpleModel, MaldacenaStanford2016,PolchinskiRosenhaus2016} is a theory of $2N$ Majorana fermions $\{\psi_i\,:\, 1 \leq i \leq 2 N\}$, which interact via the following Hamiltonian\begin{malign}\label{eq:SYK_Hamiltonian}
    H = \sum_{I} J_{I} \prpsi[I]{p} \,.
\end{malign}Here $I$ runs over all distinct sets of $p$ Majorana fermions and we define $\prpsi[I]{s}$ as a product of all Majorana fermions in $I$:
\begin{malign}
    \prpsi[I]{s} &= i^{s/2} \prod_{k = 1}^s \psi_{i_k} \quad \text{for} \quad  I = \{i_1 \dots i_s\}  \,.
\end{malign}The Majorana fermions satisfy the anti-commutation relation
\begin{malign}
    \{\psi_i , \psi_j\} = 2 \delta_{ij} \quad \forall \; 1 \leq i, j \leq 2N\,.
\end{malign}The couplings $J_I$ are sampled from a Gaussian distribution with the following two-point function.\footnote{Throughout this paper, we will use overline to denote the ensemble average.}
\begin{malign}\label{eq:SYK_ensemble}
 \overline{ J_I J_K} = \mathcal{J}^2 N \delta_{IK}  \binom{2N}{p}^{-1}\,.
\end{malign}At large $N$, the SYK model is solvable in the saddle point approximation. At low temperatures $\beta^{-1}$ such that $1 \ll \beta \mathcal J  \ll N$, the effective dynamics is described by the Schwarzian action, which also describes the dynamics of Nearly-AdS$_2$ geometries \cite{Maldacena:2016upp, KitaevSuh2018}. In particular, the thermal state of the SYK model at low temperatures serves as the holographic dual of black holes in Nearly-AdS$_2$ spacetime. 

In \cite{kourkoulou2017purestatessykmodel}, Kourkoulou and Maldacena constructed pure states in the SYK model, which look approximately thermal at large $N$. Their construction goes as follows: Consider the following set of spin operators
\begin{equation}
    Z_i = - i \psi_{2i - 1} \psi_{2i}, \quad  1 \leq  i \leq N \,.
\end{equation}
Since $Z_i^2 =1$ and $\Tr(Z_i ) = 0$,  $Z_i$ have eigenvalues $\pm 1$.  Let $\vec s$ be an $N$-dimensional vector whose components can be either $1$ or $-1$. We define a spin state $\ket{\vec s\,}$ as the simultaneous eigenstate of $Z_i$ that satisfies
\begin{equation}\label{eq:spin_state}
    Z_i \ket{\vec s\, } = s_i \ket{\vec s \,} \, \quad \forall \; 1 \leq i  \leq N.
\end{equation}The average energy of any spin state is zero whereas the energy of a thermal state in SYK at small temperatures is $-\alpha N$ at large $N$ for some $\alpha > 0$ \footnote{Note that $\bra{\vec{s}\,} H^2 \ket{\vec{s}\,} = O(N)$, but the ground state energy scales linearly with $N$. Thus, in the large $N$ limit, the spin state has significant overlap only with high energy eigenstates of the SYK Hamiltonian.}${}^,$\footnote{ \cite{Gur-Ari:2018okm} numerically found that the variance of the ground state energy in the $p=4$ SYK model at large $N$ scales as $N^{-3.43}$, thus, providing the evidence for the convergence of the ground state energy of the SYK model. }. 

We can reduce the average energy of the spin state by a Euclidean evolution with the SYK Hamiltonian. The resulting pure state is called the Kourkoulou-Maldacena (KM) state.
\begin{malign}\label{eq:KM_state}
    \ket{\beta}_s = \frac{1}{\sqrt{N_\beta}} e^{-\frac{\beta}{2} H}\ket{\vec{s}\,}, \qquad N_\beta = \bra{\vec{s}\,} e^{-\beta H} \ket{\vec{s}\,}\,.
\end{malign}At large $N$, the SYK model has an emergent permutation symmetry which acts by permuting the Majorana fermions as follows:
\begin{malign}
    \pi_{\sigma}^{-1} \psi_i \pi_{\sigma} = \psi_{\sigma(i)}  \quad \forall \; 1 \leq  i \leq 2N 
\end{malign}for any $\sigma$ that belongs to the permutation group of $2N$ elements. The permutation symmetry is a result of the fact that the correlation functions of fermions in a single instance of the SYK ensemble approach their corresponding ensemble average in large $N$ limit.  Thus, the $n$-point functions of single fermion operators which are related by permutation of fermions are also equal to each other at $O(N^0)$ in the $1/N$ expansion. In particular, the $n$-point function of fermions in a thermal state is non-zero at $O(N^0)$ if and only if it is permutation-invariant.   

An element of the permutation group that exchanges $\psi_{2k-1}$ and $\psi_{2k}$ acts on a spin state by flipping its $k$-th spin. The invariance of correlation function under the permutation group allows us to relate correlation functions in the spin state to the thermal correlation functions. For instance, correlation function of a permutation invariant combination of Majorana fermions in the KM state agree with the corresponding thermal correlation function at the same temperature up to $1/N$ corrections: 
\begin{malign}
N_\beta = \bra{\vec{s}\,} e^{-\beta H} \ket{\vec{s}\,} \approx \frac{1}{2^N} \sum_{\vec{s}\,'} \bra{\vec{s}\,' } e^{-\beta H} \ket{\vec{s}\,'} &= Z(\beta), \quad Z(\beta) =  \Tr(e^{-\beta H})\,, \\ 
\frac{1}{(2N)^k} \sum_{i_1 \dots i_k} {}_s\bra{\beta}  \mathcal{T} \prod_{l = 1}^k   \psi_{i_l}(\tau_{2 l-1}) \psi_{i_l}(\tau_{2 l}) \ket{\beta}_{s} &\approx \frac{1}{(2N)^k} \sum_{i_1 \dots i_k}  \frac{\Tr \left( e^{-\beta H} \mathcal{T} \prod_{l=1}^k   \psi_{i_l}(\tau_{2l-1}) \psi_{i_l}(\tau_{2l}) \right) }{ Z(\beta) },\\
\end{malign}where $\mathcal T$ is the time ordering operator under forward time evolution. 
The above correlation functions cannot distinguish the KM state $\ket{\beta}_s$ from the thermal state at temperature $\beta^{-1}$. Combining this observation with the fact that the SYK model has a holographic dual at small temperatures, the authors of~\cite{kourkoulou2017purestatessykmodel} concluded that $\ket{\beta}_s$ is the holographic dual of a single-sided black hole at inverse temperature $\beta$ in the Nearly-AdS$_2$ geometry. Correlation functions of the permutation invariant operators can only probe the geometry of the black hole exterior. 

However, not all correlation functions are thermal. Indeed, spin operators $Z_i$ have non-zero expectation values in the KM state while their thermal expectation values vanish. Given a spin state $\ket{\vec{s}\,}$, KM defined the following operator adapted to $\ket{\vec{s}\,}$ \begin{malign}\label{eq:length_operator}
    \hat S = \frac{1}{N}\sum_{i} s_i \hat Z_i\,.
\end{malign}The one-point function of $\hat S$ in the KM state $\ket{\beta}_s$ is related to the square of the fermion two-point function in the thermal state:
\begin{malign}
 {}_s \hspace{-0.5 mm} \bra{\beta} \hat S(\tau) \ket{\beta}_s  &= \frac{1}{N} \sum_{i} {}_s \hspace{-0.5 mm}\bra{\beta}(-i s_i \psi_{2i-1} (\tau)\psi_{2i}(\tau)) \ket{\beta\,}_{s} \\ &= \frac{1}{N \sqrt{N_\beta} }\sum_{i} {}_s \hspace{-0.5 mm}\bra{\beta\,} (-i s_i \psi_{2i-1}(\tau) \psi_{2i}(\tau)) e^{-\frac{\beta}{2} H}  \ket{\vec{s}\,}  \\&= \frac{1}{N \sqrt{N_\beta} }\sum_{i} {}_s \hspace{-0.5 mm}\bra{\beta\,} (-i s_i \psi_{2i-1}(\tau) \psi_{2i}(\tau)) e^{-\frac{\beta}{2} H} (-i s_i \psi_{2i-1} \psi_{2i}) \ket{\vec{s}\,} \\&=  -\frac{1}{N N_\beta} \sum_{i} {}_s \hspace{-0.5 mm}\bra{\beta\,} \psi_{2i-1}(\tau) \psi_{2i}(\tau) e^{-\frac{\beta}{2} H} \psi_{2i-1} \psi_{2i} \ket{\vec{s}\,} \\&\approx  \frac{1}{N} \sum_i \frac{\tr(e^{-\beta H}\psi_{2i-1}(\tau + \beta/2) \psi_{2i-1}(0) \psi_{2i}(\tau + \beta/2) \psi_{2i}(0) )}{Z(\beta)} \\ &\approx   \frac{1}{Z(\beta)}\tr\left(e^{-\beta H} \left( \sum_{i}\frac{\psi_{i}(\tau + \beta/2) \psi_i(0) }{2N}\right)^2\right) \\ &\approx \left(\frac{1}{Z(\beta)}\tr\left(e^{-\beta H} \sum_{i}\frac{\psi_{i}(\tau + \beta/2) \psi_i(0)}{2N }\right) \right)^2
 \end{malign}\begin{figure}
     \centering
     \includegraphics[width=0.4\linewidth]{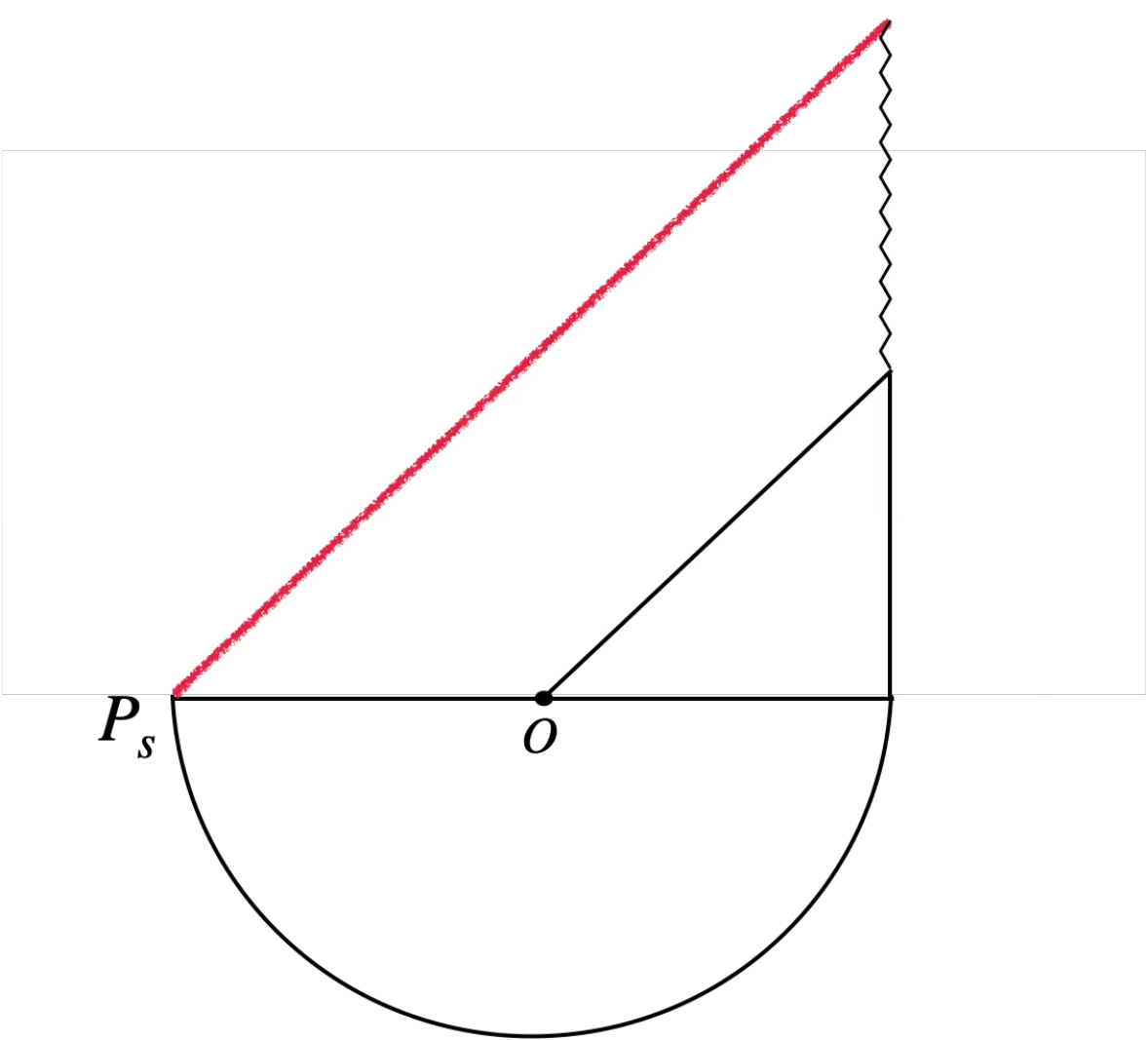}
     \caption{We illustrate the bulk dual of the Kourkoulou-Maldacena state. The boundary state is prepared by a Euclidean path integral along a semicircle followed by an insertion of the projection operator $P_s$. The bulk dual is a black hole in $AdS_2$ with a bifurcate horizon at $O$. The bulk  contains a shock (shown by the red line) that emanates from the location of the projection operator and propagates along the null direction in the black hole interior. The boundary is shown by a straight line on the right.}
     \label{fig:KM_bulk_dual}
 \end{figure}In the second step, we have used the definition of the KM state (see equation \eqref{eq:KM_state}). In the third step, we inserted the product $-is_i \psi_{2i-1} \psi_{2i} $ in the correlation function for free because the spin state $\ket{\vec{s}\,}$ is invariant under this action. In the fifth step, we used the permutation symmetry to replace the expectation value in the spin state by the normalized trace. To write the sixth step, we have again used permutation symmetry to generate the double sum over Majorana fermions.  In the final step, we used the large $N$ factorization of the correlation function.

 
 The above analysis crucially relied on the fact that $\ket{\vec{s}\,}$ is an eigenstate of the spin operators $Z_i$. This makes it clear that $\hat S$ can probe correlations in the KM state which are otherwise hidden to the permutation invariant observables.  This is in contrast with the thermal state in which the expectation value of $\hat S$ vanishes, once again, as a result of the permutation symmetry. 
 

 We sketch the bulk dual of the KM state \(\ket{\beta}_s\) in figure \ref{fig:KM_bulk_dual}. It is obtained by slicing the gravitational path integral for \(N_\beta\) along the time-reflection-symmetric slice.  On the boundary, the path integral defining \(N_\beta\) consists of Euclidean evolution around a circle of length \(\beta\), followed by the insertion of the projector \(P_s = \ket{\vec{s}\,}\!\bra{\vec{s}\,}\). The dominant bulk saddle is the Euclidean AdS\(_2\) disk with boundary length \(\beta\), with the projector represented by a marked point on the left boundary. After continuation to Lorentzian signature, the geometry is a black hole with inverse temperature \(\beta\). The insertion of the projection operator on the left boundary produces a null shock that propagates through the black-hole interior towards the boundary.

Boundary operators whose correlators in the KM state appear thermal are dual to operators supported in the black hole exterior. This suggests that the algebra generated by such operators does not provide complete access to the KM state. The full operator algebra also includes operators in the black-hole interior, which can be accessed by bringing the interior into causal contact with the boundary. Concretely, this is achieved by adding a small deformation to the SYK Hamiltonian, proportional to \(\hat S\), during the Lorentzian evolution of the KM state (see section 6 of \cite{kourkoulou2017purestatessykmodel}). 

The upshot is that the KM state appears mixed when one restricts attention to observables accessible from the exterior, such as those arising in the large-$N$ limit of single-fermion insertions at the boundary. Once the observable algebra is enlarged to include operators adapted to the KM state, such as $\hat S$, one instead recovers the full algebra with respect to which the KM state is pure. In section~\ref{sec:vn_algebras}, we make this statement precise. In particular, we characterize the limiting state by analyzing the Murray--von Neumann type of the corresponding double-scaled algebras of fermionic operators acting on the KM states in DSSYK.

\subsection{The Double Scaled SYK}\label{sec:review_dssyk}

The double-scaled limit of the SYK model, known as the double-scaled SYK (DSSYK), corresponds to taking the limit  $p \rightarrow \infty$, $ N \rightarrow \infty$ and $\mathcal J \rightarrow 0$ in equations \eqref{eq:SYK_Hamiltonian} and \eqref{eq:SYK_ensemble} such that
\begin{equation} \label{eq:ds-limit-def}
    \lambda =  \lim_{p , N \rightarrow \infty}\frac{ p^2}{N} , \quad {\mathcal{J}'}^2 = N \mathcal J^2\quad \text{remain finite.}
\end{equation} We will set $\mathcal{J}^\prime =1$ for convenience.  A useful parameter that controls interactions in the double-scaled limit is $e^{-\lambda}$. For the rest of this paper, we will denote this parameter by $q$. 
\begin{malign}
    q = e^{-\lambda} \,.
\end{malign}The DSSYK model also contains matter operators defined as
\begin{malign}
    O = \sum_{\bar K} J'_{\bar K}  \Psi_{\bar K}^{(l)} \,.
\end{malign}Here, $\bar K$ runs over all distinct sets of $l$ fermions and $J'_{\bar {K}}$ are   sampled from a Gaussian distribution that satisfies\begin{malign}
    \overline{ J'_{\bar K}  J'_{\bar K'}} = \binom{2N}{l}^{-1} \delta_{\bar K \bar K'} \,.
\end{malign}
In the double-scaling limit, $l, p\rightarrow \infty$ such that the ratio
\begin{malign}
   \lim_{l,p \rightarrow \infty} l/p  = \alpha \quad \text{is finite.}
\end{malign}A simple quantity of interest in DSSYK is the thermal partition function:
\begin{malign}
    Z(\beta) = \tr(e^{-\beta H}) ,
\end{malign}where $\tr(\cdot)$ is the normalized trace over the Hilbert space such that $\tr(\mathbb{I}) = 1$\footnote{We can view the normalized trace of an operator as its expectation value in the maximally mixed state.}. The major simplification in the double-scaled limit comes from the fact that correlation functions in this limit can be computed by taking the ensemble average over the couplings of $H$.  In particular, the partition function is exactly equal to its ensemble average in the double scaling limit as a consequence of self-averaging
\begin{malign}
Z(\beta) = \overline{Z(\beta)}\,.
\end{malign}
We can compute $Z(\beta)$ by expanding it in a power series in $\beta$\begin{malign}
   Z(\beta) = \sum_{n} \frac{(-\beta)^n}{n!}\, \overline{\tr(H^n)}\,.  
\end{malign}
\begin{figure}
    \centering
    \includegraphics[width=0.25\linewidth]{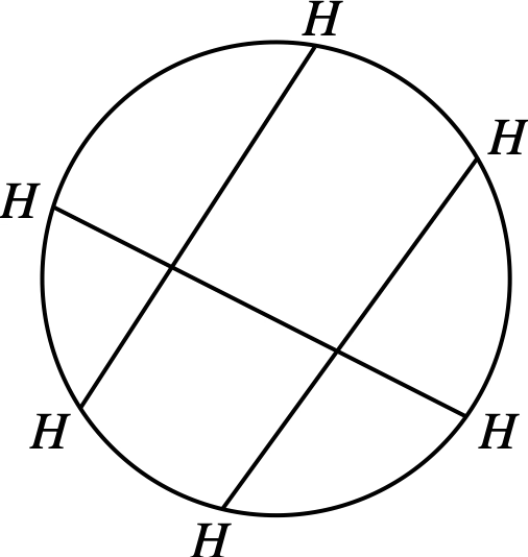}
    \caption{A chord diagram representation of $\tr(\protect\wick{\c1 H \c3 H \c2 H \c1 H \c2 H \c3 H})$. Since the diagram contains two intersections of the $H$-chords, its value is $q^2$.}
    \label{fig: example_chord_diagram}
\end{figure}
\cite{Berkooz:2018jqr} found a representation of $\overline{\tr(H^n)}$ as a sum of chord diagrams. $\overline{\tr(H^n)}$ involves an ensemble average over the product of couplings of $H$:
\begin{malign}
    \overline{\tr(H^n)} = \sum_{I_1 \dots I_n} \overline{\prod_{k = 1}^n J_{I_k}} \,  \tr\left(\prod_{l = 1}^n \Psi_{I_l}\right)\,.
\end{malign}Since the couplings are sampled independently from a Gaussian ensemble, the average over products of couplings is given by a sum over products of Wick contractions of the couplings. Diagrammatically, we represent $\tr(H^n)$ by a circle with $n$ insertions of $H$. Wick contractions are represented by ``$H$-chords" connecting pairs of $H$ on the circle and the resulting diagram is called a chord diagram  (see figure \ref{fig: example_chord_diagram} for an example of a chord diagram).  If a chord diagram contains $n_{HH}$ pairwise intersections of $H$-chords\footnote{We always assume that no two chords intersect each other more than once.}, then the value of a chord diagram is given by $q^{n_{HH}}$. The value of $\overline{\tr (H^n)}$ is the sum over all chord diagrams with $n$ insertions of $H$ along the circle.

A general lesson of the analysis of \cite{Berkooz:2018jqr} is that in the double-scaled limit, the weight of intersection between two chords in a chord diagram does not depend on the presence of other chords. Therefore, to find the weight of intersection between two $H$ chords, it is sufficient to consider a chord diagram with a single chord intersection. Such a chord diagram would correspond to the following Wick contraction in the four-point function of $H$.
\begin{malign}
    \tr(\wick{\c1 H \c2 H \c1 H \c2 H}) &= \binom{2N}{p}^{-2} \sum_{I, K} \tr \left(\Psi_I \Psi_K \Psi_I \Psi_K\right) \\ &= \binom{2N}{p}^{-1}  \sum_K \left( \binom{2N}{p}^{-1} \sum_{I} (-1)^{|I \cap K|}\right) \\ 
\end{malign}We can evaluate the sum by organizing it in a series of the number of common fermions in $\Psi_I$ and $\Psi_K$:
\begin{malign}
    \binom{2N}{p}^{-1} \sum_I (-1)^{|I \cap K|} &= \binom{2N}{p}^{-1} \sum_{r =0}^p (-1)^r \sum_I \delta_{|I \cap K|, r} \\  &= \binom{2N}{p}^{-1}  \sum_{r \leq p} (-1)^r \binom{2N - p}{p-r} \binom{p}{r}
\end{malign}In the double-scaled limit, the sum approaches $q$. Therefore,
\begin{malign}
    \tr(\wick{\c1 H \c2 H \c1 H \c2 H}) = q\,.
\end{malign}

Correlation functions of $O$ with the Hamiltonian have a similar chord diagram expansion where Wick contractions of $O$ have chord representation in terms of $O$-chords. In addition to intersections between $H$-chords, a general chord diagram also contains intersections between $O$-chords and intersections between an $H$-chord and an $O$-chord. The weight of such intersections is determined by the following Wick contractions in the four-point functions of $H$ and $O$:
\begin{malign}
    \tr(\wick{\c1{O} \c2{O} \c1{O} \c2{O}}) = q^{\alpha^2}, \quad 
      \tr(\wick{\c1{O} \c2{H} \c1{O} \c2{H}}) = q^{ \alpha} \,.
\end{malign}
The value of a chord diagram containing $H$ and $O$ chord are given by
\begin{malign}
    q^{(n_{H H} + \alpha \,n_{H O}+ \alpha^2\, n_{OO})}\,,
\end{malign}where $n_{A B}$ is the number of chord intersections between $A$ and $B$ chords. 

\begin{figure}
    \centering
    \includegraphics[width=0.25\linewidth]{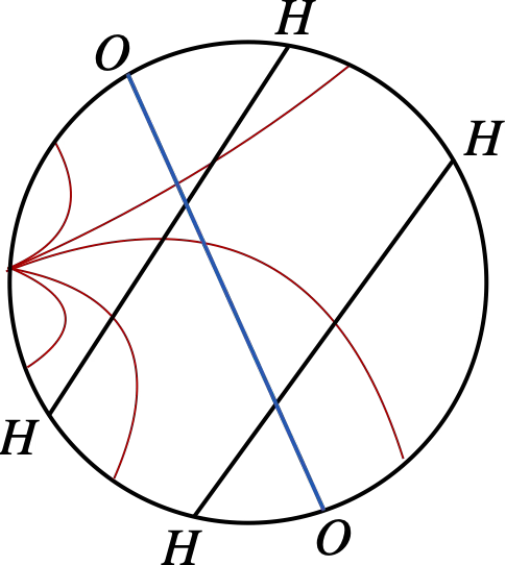}
    \caption{States in the doubled Hilbert space can be generated by cutting chord diagrams along the red slices. The state on a slice is fixed by the chords crossing it. In this figure, the state on the red slices evolves in the anti-clockwise direction as $\ket{0} \rightarrow a^\dagger \ket{0} \rightarrow  a^\dagger \tilde a^\dagger a^\dagger {} \ket{0} \rightarrow a^\dagger \tilde a^\dagger \ket{0} \rightarrow \ket{0}.$}\label{fig:chord_hilbert_space}
\end{figure}
There is a systematic way to sum over all terms in the chord diagram expansion of a given correlation function of $H$ and $O$. The basic idea is to rewrite the normalized trace of an operator in DSSYK as its expectation value in a maximally entangled state $\ket{0}$ between two copies of the Hilbert space of DSSYK. \begin{malign}
    \tr(f(O,H)) = \bra{0} f(O,H) \ket{0}\,.
\end{malign}We call $\ket{0}$ the zero chord state. More general states in the doubled Hilbert space are generated by insertions of $H$ and $O$ along the circle and by slicing the circle at the two points as shown in figure \ref{fig:chord_hilbert_space}. As we move along the circle, every insertion of $H$ (resp. $O$) can either create or annihilate an $H$-chord (resp. $O$-chord). Thus,  the action of $H$ (similarly, $O$) is naturally mapped to a sum of chord creation and annihilation operators: 
\begin{malign}
   H = a+ a^\dagger, \quad  O = \tilde a +\tilde a^\dagger\,.
\end{malign}The chord creation and annihilation operators are defined to annihilate $\bra{0}$ and $\ket{0}$ respectively.
\begin{malign}
    a \ket{0} = \tilde a \ket{0} = 0, \quad \bra{0}a^\dagger = \bra{0} \tilde a^\dagger = 0\,.
\end{malign}When a chord is annihilated, it may intersect other chords. The $q$-deformed commutation rules generate a sum over all possible ways in which chords can be annihilated while also taking into account the weights of intersection among different chords. \begin{malign}\label{eq:chord_commuation_rule}
    [a, a^\dagger]_{q} \equiv a a^\dagger - q a^\dagger a = 1, \quad [\tilde a, \tilde a^\dagger]_{q^{\alpha^2}} = 1, \quad [a,\tilde a^\dagger]_{q^{ \alpha}} = [\tilde a, a^\dagger]_{q^{ \alpha}} = 0 \,.
\end{malign}Representation of $H$ and $O$ as sums of creation and annihilation operators and the $q$-deformed commutation relations generate the sum over all chord diagrams in the correlation function. To summarize, correlation functions of the Hamiltonian and the matter operators in the maximally mixed state in DSSYK are given by the following formula:
\begin{malign}
    \Tr(f(O,H)) = \bra{0} f(\tilde a+ \tilde a^\dagger, a + a^\dagger)\ket{0} \,.
\end{malign}
where $f(O,H)$ is some polynomial of $O$ and $H$. 

Action with arbitrary polynomials of $H$ and $O$ on the zero chord state $\ket{0}$ defines the Hilbert space relevant to the dynamics of DSSYK. From the fact that $H$ and $O$ can be expressed as linear combinations of chord creation and annihilation operators, it is clear that the resulting Hilbert space is
\begin{align}
    \cH = \text{span} \left \{ \prod_{i =0}^k (a^\dagger)^{m_i} (\tilde a^\dagger)^{n_i} \ket{0}: m_i, n_i, k \in \mathbb{Z}_{\geq 0} \right\} \,.
\end{align}
Inner product on $\cH$ naturally follows from the condition that $H$ and $O$ act as Hermitian operators on $\cH$ in the double-scaling limit.  Thus, according to this inner product, $a$ and $\tilde a $ are related to $a^\dagger$ and $\tilde a^\dagger$ respectively by a Hermitian conjugation.

\section{Kourkoulou-Maldacena State and its Double-Scaling Limit}\label{sec:KM}
In this section, we study the double-scaled limit of KM states. Without loss of generality, we will work with the spin state $\ket{\omega}$ for which
\be
{Z}_i |\omega\ra = -|\omega\ra, \quad \forall i\in\{1,2,\dots ,N\}.
\ee
Excitations above $\ket{\omega}$ are generated by flipping its spins. A natural way to implement a spin flip is to introduce the following creation and annihilation operators.
\begin{malign}\label{eq:creation_operators}
   \quad c_i^\dagger = \frac{\psi_{2i -1} + i \psi_{2i}}{2}, \quad   c_i = \frac{\psi_{2i -1} - i \psi_{2i}}{2}, \quad 1 \leq i \leq N  \,.
\end{malign}In addition to the standard anti-commutation relation $\{c_i, c_j^\dagger\} = \delta_{ij}$, $c_i$ and  $c_i^\dagger$ satisfy the following commutation relation with $Z_i$:
\begin{malign}
    [Z_i, c^\dagger_j] = 2 \delta_{ij} c^\dagger_j, \quad [Z_i, c_j] = -2 \delta_{ij} c_j \,. 
\end{malign}By construction $c_i$'s annihilate $\ket{\omega}$: \begin{malign}c_i \ket{\omega} = 0\,.\end{malign}We can measure the number of spin excitations above $\ket{\omega}$ with an operator $\hat Q$ defined as
\begin{malign} 
   \hat Q = \sum_{i = 1}^N c_i^\dagger c_i = \sum_{i = 1}^N \frac{1 + Z_i}{2}  \,. 
\end{malign}Note that $\hat Q$ is related to $\hat S$ defined in equation \eqref{eq:length_operator} as:
\begin{malign}
    \hat S = 1 - 2 \frac{\hat Q}{N}.
\end{malign}Since we are interested in the double-scaled limit of the pure states and excitation above them created by operators such as $H$ and $O$, we need to normalize $\hat Q$ to ensure that it has a well-defined limit. To fix the normalization, consider the expectation value of $\hat Q$ in the state $H \ket{\omega}$ in the double-scaled limit
\begin{malign}  
 \overline{   \bra{\omega} H   \hat Q H \ket{\omega} } &= \sum_{I,K} \overline{J_I J_K  }\bra{\omega} \Psi_I \hat Q \Psi_K  \ket{\omega} \\  &= \binom{2N}{p}^{-1}  \sum_{I} \bra{\omega} \Psi_I \, \hat Q  \, \Psi_I \ket{\omega} \,.
\end{malign}The above quantity computes the average number of spin flips when $\ket{\omega}$ is acted upon by a product of $p$ Majorana fermions chosen at random. In the double-scaled limit, the probability that $\Psi_I$, when chosen at random contains two fermions acting on the same spin is suppressed by $1/\sqrt{N}$. Thus, a typical $\Psi_I$ flips $p$ spins of $\ket{\omega}$ and we get the following result\begin{malign}
   \overline{\bra{\omega} H  \hat Q H \ket{\omega} }=  p \,. \end{malign}We must rescale $\hat Q$ with $1/p$ to get a well-defined operator in the double-scaling limit~\eqref{eq:ds-limit-def}. We will call this operator the ``chord number" operator and denote it by $\hat n_{\text{tot}}$. 
\begin{malign}
     \hat n_{\text{tot}} = \frac{1}{p} \hat Q \,. 
\end{malign}We can also construct the following ``dressed" matter operators which excite $p'$ spins above the state $\ket \omega$:
\begin{malign}\label{eq:dressed_operators}
   M =  \sum_I M_I C_I, \quad 
    M^\dagger = \sum_I M_I^* C_I^\dagger\,.
\end{malign}Here, $I$ runs over all distinct sets of $p'$ spins and $C_I, C_I^\dagger$ are products of annihilation and creation operators defined as follows:\begin{malign} 
    C_I = i^{p'/2} \prod_{k = 1}^{p'} c_{i_k}, \quad C_I^\dagger = i^{p'/2} \prod_{k = 1}^{p'} c_{i_k}^\dagger, \quad I = \{i_1, \dots, i_{p'}\}.
\end{malign}$M_I$ are complex random numbers drawn from Gaussian distribution with the following two-point function
\begin{malign}
    \overline{M_I M_J^*} = \delta_{I J} \binom{N}{p'}^{-1}\,.
\end{malign}We refer to $M$ and $M^\dagger$ as dressed matter operators, since their definition depends explicitly on the spin configuration of $\ket{\omega}$. In the discussion below, we will give a geometric interpretation of this dressing within the bulk Hilbert space that emerges in the double-scaling limit.

By construction, the dressed operators $M$ and $M^\dagger$ satisfy the following commutation relation with $Q$:
\begin{malign}\label{eq:charge_relation}
    [\hat Q, M] = - p' M, \quad [\hat Q, M^\dagger] = p' M^\dagger\,. 
\end{malign}$M$ and $M^\dagger$ also survive the limit \eqref{eq:ds-limit-def} if $p'$ is taken to be constant multiple of $p$, with the ratio \begin{malign}
    \lim_{p \rightarrow \infty} \frac{p'}{p} = \Delta \quad \text{is fixed.}
\end{malign}  Then the double-scaled limit of equations \eqref{eq:charge_relation} results in the following equations:
\begin{malign}  
   [\hat n_{\text{tot}}, M] &= -\Delta M, \quad [ \hat n_{\text{tot}}, M^\dagger] &=  \Delta M^\dagger\,.
\end{malign}Since $M$ is defined as a linear combination of products of annihilation operators with Gaussian random couplings, the ensemble average of any correlation function of $M, M^\dagger$ and $H$ in $\ket{\omega}$ is non-vanishing only if it contains an equal number of $M$ and $M^\dagger$. Since the correlation functions of double-scaled operators approach their corresponding ensemble average in the double-scaling limit, any non-zero correlation function in the double-scaling limit must contain an equal number of $M$ and $M^\dagger$.


Said equivalently, $M$ and $M^\dagger$ carry a $U(1)$ charge that is conserved in the double-scaled limit. In section \ref{sec:symmetry}, we will study the violation of this symmetry at leading order in $1/N$.

\subsection{Chord Rules for the Kourkoulou-Maldacena State}
\begin{figure}
    \centering
\includegraphics[width=0.25\linewidth]{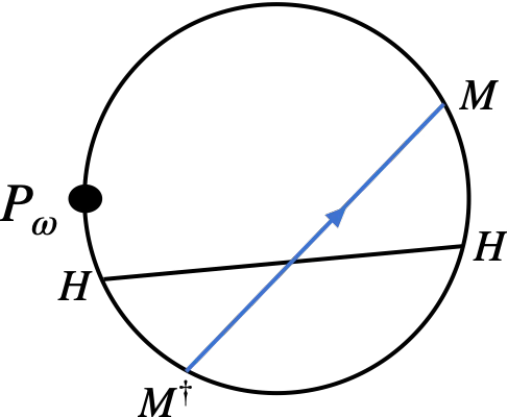}
    \caption{Chord diagram for the correlation function $\tr(P_\omega \protect \wick{\c1 M \c2 H \c1 M^\dagger \c2 H})$.}\label{fig:matter_chord_example}
\end{figure}

In this section, we derive the chord rules for computing correlation functions of the Hamiltonian and the dressed matter operators in the state $\ket{\omega}$. To obtain a chord-diagram representation of a general correlator of the form $\bra{\omega} f(M,M^\dagger,H)\ket{\omega}$, where $f$ is any polynomial in these three operators, we first observe that such correlators can always be rewritten as correlation functions in the tracial state with an insertion of the projection operator $P_\omega = \ket{\omega}\hspace{-1mm}\bra{\omega}$:
\begin{malign}
    \bra{\omega} f(M,M^\dagger, H) \ket{\omega} = \text{tr} (  P_\omega \, f(M,M^\dagger, H) ).
\end{malign}where tr$(\cdot)$ is the \textit{un-normalized} trace so that tr$(P_\omega) = 1$. The trace is represented diagrammatically by a circle with a marked point corresponding to an insertion of $P_\omega$ on the leftmost point on the circle. This is followed by insertions of $H$ and $M, M^\dagger$ in the clockwise direction in the same order as they appear in the expansion of $f(M,M^\dagger, H)$ (see figure \ref{fig:matter_chord_example} for an example). As a result of the ensemble average over couplings of $M, M^\dagger$ and $H$, we get a sum over all possible ways of Wick contracting $M$ with $M^\dagger$ and all possible Wick contracting pairs of $H$. Wick contractions of $H$ are represented by $H$-chords connecting pairs of $H$ and Wick contraction of $M$ and $M^\dagger$ has a representation as an $M$-chord connecting insertions of $M$
with $M^\dagger$. To distinguish $M$ from $M^\dagger$, we label the $M$-chord with an arrow that points towards $M$. There are no chords connecting $M$ or $M^\dagger$ with $H$ as they are sampled from different ensembles but they do have a non-trivial commutation relation in the double-scaling limit that we analyze below. 

\begin{figure}
    \centering
    \includegraphics[width=\linewidth]{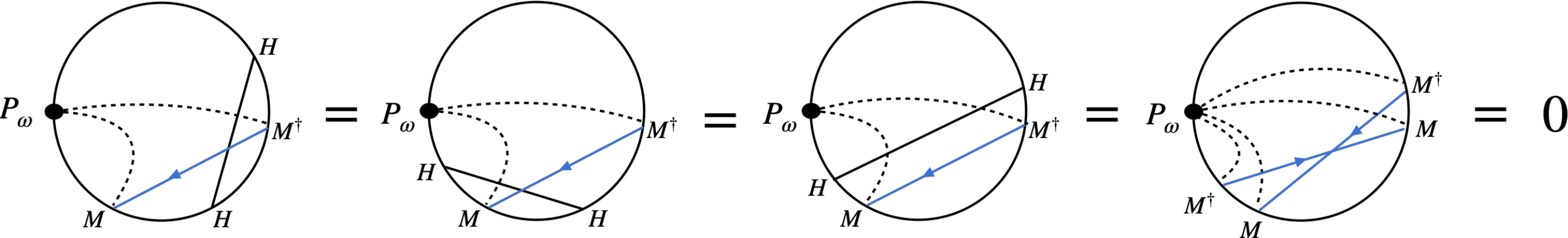}
    \caption{The value of any chord diagram in which an $M$-chord connects $M^\dagger$ and $M$ such that the sequence $M^\dagger - M - P_\omega$ are arranged in a clockwise direction is zero.}
    \label{fig:chord_rules_zero}
\end{figure}

However, this is not yet an adequate representation of the Wick contractions. Correlation functions of the dressed matter operators remain sensitive to the location of the projection operator $P_\omega$ on the circle. Accordingly, for each insertion of $M$ (and similarly of $M^\dagger$), we draw a dashed chord connecting the insertion to $P_\omega$. These dashed chords play an essential role in the evaluation rules for the chord diagrams. For notational convenience, we refer to them as $D$-chords, and represent them by dashed lines as in Fig.~\ref{fig:chord_rules_zero}.

Chord rules in the double-scaled limit are fixed entirely by the four-point functions. Below, we enumerate and evaluate all four-point functions of the Hamiltonian and dressed matter operators. Note that the Wick-contraction symbol in the correlators below indicates that we are considering a particular contraction pattern after ensemble averaging.  For this reason, we do not place an additional overline on the correlators: the quantities of interest are already defined in the averaged theory.

\begin{enumerate}
    \item  Let us begin with the four-point function of $H$:   \begin{malign} \bra{\omega} \wick{\c1{H} \c2{H} \c1{H} \c2{H}} \ket{\omega} &= \binom{2N}{p}^{-2} \sum_{I,J} \bra{\omega} \Psi_I \Psi_J \Psi_I \Psi_J \ket{\omega} \\ &=\binom{2N}{p}^{-2} \sum_{I,J} \tr\left( \Psi_I \Psi_J \Psi_I \Psi_J \right)  \\ &= q .\end{malign}In the second line, we have used the fact that the product $\Psi_I \Psi_J \Psi_I \Psi_J$ is proportional to the identity operator. Hence, its expectation value in $\ket{\omega}$ is the same as its normalized trace. 
    
    The crossing rules for the $H$-chords in $\ket{\omega}$ and the maximally mixed state are identical. Therefore, all correlation functions of $H$ in $\ket{\omega}$ agree with those in the infinite temperature state. In other words, correlation functions of $H$ in $\ket{\omega}$ cannot distinguish it from a maximally mixed state. 
    \item \label{it:zero_1} \begin{malign} \bra{\omega} \wick{\c1{M}^\dagger \c2{H} \c1{M} \c2{H}} \ket{\omega} =  \binom{N}{p'}^{-1} \binom{2N}{p}^{-1} \sum_{I,J} \bra{\omega}  C^\dagger_J \Psi_I  C_J \Psi_I \ket{\omega} = 0 \end{malign} 
    This is because $\bra{\omega} C_J^\dagger = 0$. Similarly, $\bra{\omega} \wick{ \c2{H} \c1{M}^\dagger \c2{H} \c1{M} } \ket{\omega} = 0$
    \begin{figure}
        \centering
        \includegraphics[width=\linewidth]{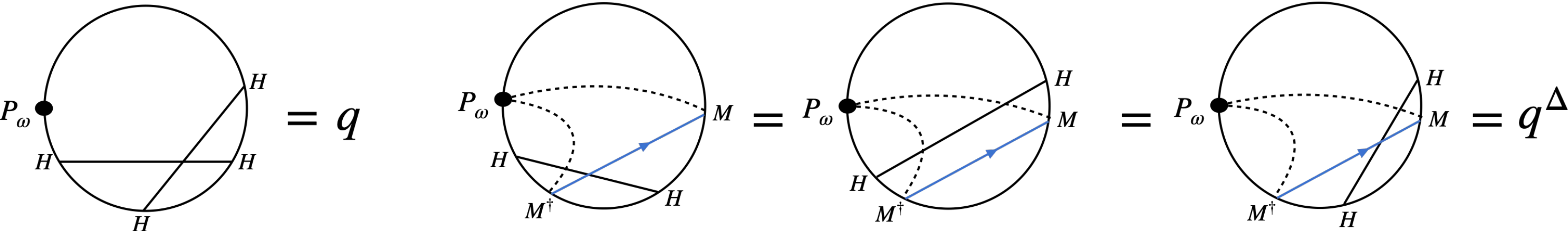}
        \caption{The left chord diagram fixes the weight of intersection between two $H$-chords to  $q$. The three diagrams on the right illustrate the three different ways in which $M$ and $H$ chords may have a non-trivial interaction. It follows from the diagrams that the intersection between $M$ and $H$ chords is weighted by $q^{\Delta/2}$. Similarly, intersection between $H$ and $D$ chords has weight $q^{\Delta/2}$. }
        \label{fig:chord_rules_MH}
    \end{figure}
    \item \label{it:zero_2} \begin{malign} \bra{\omega} \wick{ \c2{H} \c1{M}^\dagger \c1{M} \c2{H}} \ket{\omega} =  \binom{N}{p'}^{-1} \binom{2N}{p}^{-1} \sum_{I,J} \bra{\omega} \Psi_I C_J^\dagger C_J \Psi_I \ket{\omega}   \end{malign}The non-zero terms in the sum are those for which $\Psi_I$ flips all the spins of $\ket{\omega}$ which are annihilated by $C_J$. For a given $C_J$ the number of such terms is 
    \begin{malign}
        2^{p'}  \times \binom{2N-2p'}{p-p'} 
    \end{malign}
    In the double scaled limit, the overall contribution of these terms is suppressed as $N^{-p'}$.  Hence, the value of this two-point function approaches zero in the double-scaled limit. 
    \item \label{it:zero_3} Consider the following four-point function of matter operators
\begin{malign}
    \bra{\omega} \wick{\c2{M} \c1{M}^\dagger  \c1{M} \c2{M}^\dagger}\ket{\omega} = \binom{N}{p'}^{-2} \sum_{I,J} \bra{\omega} C_I C_J^\dagger C_J C_I^\dagger \ket{\omega}
\end{malign}The terms in the sum over $I,J$ have a non-zero contribution only when $I=J$. The sum over such contributions is suppressed in the double-scaled limit. Therefore, 
\begin{malign}
    \bra{\omega} \wick{\c2{M} \c1{M}^\dagger  \c1{M} \c2{M}^\dagger}\ket{\omega} =  0
\end{malign}
in the strict double-scaling limit.
    \item \begin{malign}\label{eq:rule_5} \bra{\omega} \wick{ \c2{H} \c1{M} \c2{H}\c1{M}^\dagger } \ket{\omega} =  \binom{N}{p'}^{-1} \binom{2N}{p}^{-1} \sum_{I,J} \bra{\omega}  \Psi_I C_J \Psi_I C_J^\dagger\ket{\omega} \end{malign} Notice that $\bra{\omega} \Psi_I C_J \Psi_I C_J^\dagger \ket{\omega}$ can be non-zero if and only if $\Psi_I C_J \Psi_I \propto C_J$. Since $\Psi_I^2 = 1$, $\Psi_I$ must either commute or anti-commute with $C_J$.  This is possible if and only if it either commutes or anti-commutes with every annihilation operator in $C_J$. Consider an annihilation operator $c_j$ that belongs to $C_J$. $c_j$ is a linear combination of $\psi_{2j-1}$ and $\psi_{2j}$ as defined in equation \eqref{eq:creation_operators}.  $\Psi_I$ anti-commutes with $c_j$ if it contains the product $\psi_{2j-1} \psi_{2j}$ and it commutes if it contains none of the two fermions. Thus, $\Psi_I$ commutes or anti-commutes with $C_J$ if and only if it has either two common fermions or no common fermion with every annihilation operator in $C_J$.
The number of such $\Psi_I$ which anti-commute with exactly $r$ annihilation operators in $C_J$ is  $\binom{p'}{r} \binom{2 N - 2p'}{p - 2 r}$. Their contribution in equation \eqref{eq:rule_5} is weighted by a factor of $\binom{2N}{p}^{-1}$. In the double-scaled limit, their overall contribution scales as 
\begin{align}\label{eq:suppression}
   \binom{p'}{r} \binom{2 N - 2p'}{p-2 r} \binom{2N}{p}^{-1} \sim {p'}^{r}\frac{p^{2r}}{N^{2r}} \sim \frac{1}{N^{r/2}} \,.
\end{align}
For $r > 0$, this is suppressed in the double-scaled limit and it can be ignored. Thus, the non-zero contribution to the sum is from  those $\Psi_I$ operators which commute with all annihilation operators in  $C_J$. The number of such operators is $\binom{2N - 2p'}{p}$. In the double-scaled limit, their contribution in the double-scaling limit is finite:
\begin{malign}
    \binom{2 N}{p}^{-1} \sum_I \bra{\omega} \Psi_I C_J \Psi_I C_J^\dagger \ket{\omega} \rightarrow  \binom{2N - 2p'}{p}\binom{2 N}{p}^{-1}  \rightarrow q^\Delta \,.
\end{malign}
and the correlation function in \eqref{eq:rule_5} in the double-scaling limit becomes:
\begin{malign}
    \bra{\omega} \wick { \c1 H  \c2 M \c1 H \c2 M^\dagger }  \ket{\omega} \to \binom{N}{p^\prime}^{-1} \sum_{J} q^{\Delta}  = q^\Delta\,.
\end{malign}Similarly, $\bra{\omega} \wick{\c1{M} \c2{H} \c1{M}^\dagger \c2{H}} \ket{\omega} = q^{\Delta}$.
\item \begin{malign}
   \bra{\omega} \wick{\c2{H} \c1{M} \c1{M}^\dagger \c2{H}}  \ket{\omega} &= \binom{N}{p'}^{-1}  \binom{2N}{p}^{-1} \sum_{I, J}  \bra{\omega} \Psi_I C_J  C_J^\dagger \Psi_I \ket{\omega} 
   \end{malign}For given $I$ and $J$, each operator $\Psi_I C_J C_J^\dagger \Psi_I$ is a projector. When $\Psi_I$ has one common Majorana fermion with some annihilation operator in $C_J$, $\Psi_I C_J C_J^\dagger \Psi_I$ projects onto an orthogonal subspace to $\ket{\omega}$. Therefore, such a projector annihilates $\ket{\omega}$.  Once again, the correlation function is non-zero if and only if $\Psi_I$ shares either zero or two common fermions with every annihilation operator in $C_J$. However, as argued near equation \eqref{eq:suppression}, the contribution of terms with two common fermions with some annihilation operator in $C_J$ is suppressed in the limit $N \rightarrow \infty$ and they can be ignored. Thus,
\begin{malign}
    \binom{2N}{p}^{-1} \sum_{I} \bra{\omega} \Psi_I C_J C_J^\dagger \Psi_I \ket{\omega} \rightarrow q^\Delta
\end{malign}and the correlation function in the strictly double-scaling limit becomes \begin{malign} \bra{\omega} \wick{\c2{H} \c1{M} \c1{M}^\dagger \c2{H}}  \ket{\omega} = q^\Delta\,. \end{malign}Notably, in the standard chord-diagram representation of this Wick contraction, the $H$-chord does not intersect the $M$-chord. The effect of their interaction is instead encoded by introducing intersections between the $H$-chord and the dashed lines connecting $M$ and $M^\dagger$ to $P_\omega$, which we denote as $D$-chords. This is the essential modification of the chord rules for the operators $M$ and $M^\dagger$: the nontrivial penalty factor $q^\Delta$ arises because the pair $M M^\dagger$ is sensitive to the internal correlations of the KM state through its relation to the projection operator (see the third chord diagram in figure \ref{fig:chord_rules_MH}). In this precise sense the operators are \emph{dressed}. We will later explain the geometric meaning of this dressing in the emergent bulk Hilbert-space description.
 \begin{figure}
     \centering
     \includegraphics[width=0.6\linewidth]{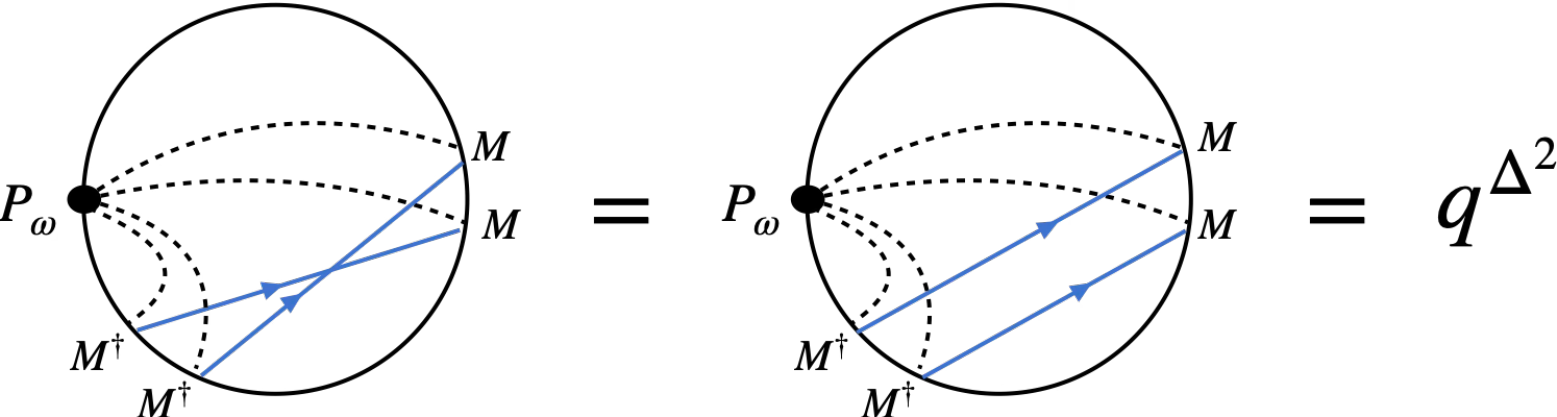}
     \caption{Matter chords can intersect non-trivially in two distinct ways. However, their contribution to the chord diagram is the same. From the two diagrams, it follows that the weight of intersection between a dashed chord and an $M$ chord is $q^{\Delta^2/2}$ while the weight of intersection between two $M$ chords is $q^{0} = 1$. }
     \label{fig:chord_rules_MM}
 \end{figure}

Chord rules for intersection between the matter chord and the Hamiltonian chord are illustrated in figure \ref{fig:chord_rules_MH}.

\item 
\begin{malign} \bra{\omega} \wick{\c1 M \c2 M \c1 M^\dagger \c2 M^\dagger} \ket{\omega} = \sum_{I,J} \binom{N}{p'}^{-2} \bra{\omega} C_I C_J C_I^\dagger C_J^\dagger \ket{\omega}\end{malign}Whenever $I \cap J = \emptyset$, $C_I C_J$ is a non-trivial product of $2p'$ annihilation operators. Every product of $2p'$ annihilation operators appears $\binom{2p'}{p'}$ times in the sum. Thus, the above sum is given by \begin{malign}
\binom{N}{p'}^{-2} \sum_{I,J}\bra{\omega} C_I C_J C_J^\dagger C_I^\dagger \ket{\omega} &= \binom{N}{p'}^{-2}  \binom{2p'}{p'} \sum_{K, |K| = 2p'} \bra{\omega} C_K C_K^\dagger \ket{\omega} \\&=  \binom{N}{p'}^{-2}  \binom{2p'}{p'} \binom{N}{2p'}
\end{malign}In the double-scaled limit, it approaches $q^{\Delta^2}$. Thus, we have \begin{malign}
    \bra{\omega} \wick{\c1 M \c2 M \c1 M^\dagger \c2 M^\dagger} \ket{\omega}   = q^{\Delta^2}\,.
\end{malign}
\item Similarly, we can evaluate $\bra{\omega} \wick{\c2 M \c1 M \c1 M^\dagger \c2 M^\dagger} \ket{\omega}$ and we get the same value\begin{malign}
    \bra{\omega} \wick{\c2 M \c1 M \c1 M^\dagger \c2 M^\dagger} \ket{\omega} = q^{\Delta^2}\,.
\end{malign}
\end{enumerate}
We illustrate the chord rules for intersection between matter operators in figure \ref{fig:chord_rules_MM}.

Chord rules \ref{it:zero_1}, \ref{it:zero_2} and  \ref{it:zero_3} imply that any correlation function in $\ket{\omega}$ where $M$ is contracted with an insertion of $M^\dagger$ to its left vanishes in the double-scaled limit i.e.
\begin{malign}
    \bra{\omega} O_1 \wick{\c1 M^\dagger O_2 \c1 M} O_3 \ket{\omega} = 0
\end{malign}where $O_i$ are some operators made of $H$ and $M, M^\dagger$. In the chord diagram expansion of such correlation functions, each chord diagram has an $M$-chord that connects $M^\dagger$ and $M$ such that the sequence $M^\dagger-M-P_\omega$ is ordered in the clockwise order along the circle (See figure \ref{fig:chord_rules_zero}). Now consider the state
\begin{malign}
 \ket{\chi} =   \prod_{i = 1}^n M^{p_i} H^{q_i} M^{\dagger\, r_i}  H^{s_i} \ket{\omega}  \quad \text{for} \quad \sum_i (p_i - r_i)> 0 ,
\end{malign}and none of the exponents $\{p_i,q_i,r_i,s_i\}$ above scale with $N$.  Inner product of $\ket \chi$ with any  state of the form $f(M,M^\dagger, H) \ket{\omega}$ vanishes.  This is because every term in the chord diagram expansion of the inner product contains at least one $M$-chord which connects $M$ with an insertion of $M^\dagger$ to its left. Thus, $\ket\chi \rightarrow 0$ in the double-scaled limit. 

In summary, any state constructed by acting with a product of $H$ and $M, M^\dagger$ on $\ket{\omega}$ such that there are more insertions of $M$ than $M^\dagger$ vanishes in the double-scaled limit. Roughly this result suggests that $M^\dagger$ creates a matter excitation on $\ket{\omega}$ while $M$ annihilates it. 

\begin{figure}
    \centering
    \includegraphics[width=0.99\linewidth]{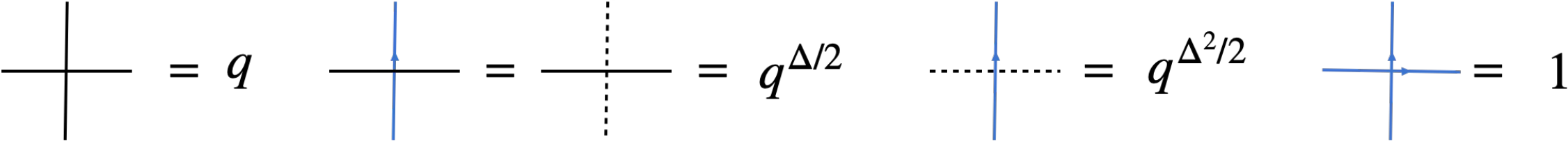}
    \caption{A summary of chord intersection rules.}
    \label{fig:chord_rules}
\end{figure}

Four-point functions of operators in the KM state fix all the weights of intersection between chords that may appear in a general correlation function of the Hamiltonian and the matter operator. We provide a summary of these weights in figure \ref{fig:chord_rules}. The value of a general chord diagram is given by the formula\begin{malign}
    q^{n_{HH} + \frac{\Delta}{2} (n_{HM} + n_{HD} )+ \frac{\Delta^2}{2} n_{DM} }
\end{malign}where $n_{AB}$ is the number of intersections between chords of type $A$ and $B$. 

In addition to the weights, we need the extra rule that the value of any chord diagram where an $M$-chord connects $M^\dagger$ and $M$ such that the sequence $M^\dagger-M-P_\omega$ appears in a clockwise direction on the circle is zero.
\subsection{Summing Over the Chord Diagrams with Dressed Operators} \label{sec:sum-cds}
\begin{figure}
    \centering
    \includegraphics[width=0.25\linewidth]{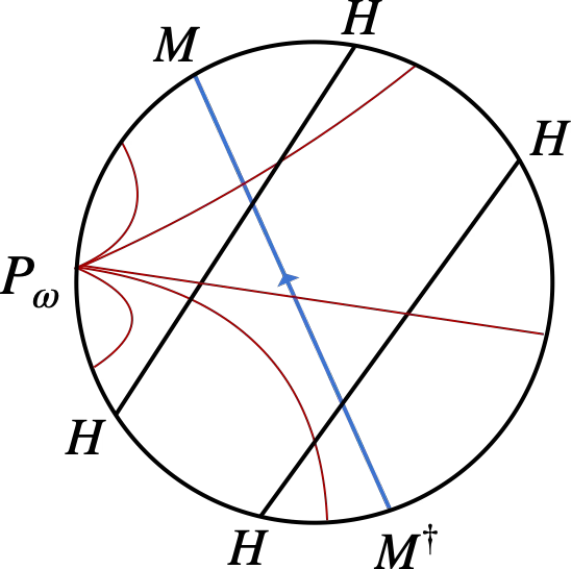}
    \caption{States in the Hilbert space of excitations about $\ket{\omega}$ are prepared by cutting chord diagrams along the red slices. The state on a red slice is fixed by the chords crossing it. The state on the red slice evolves as $\ket{0} \rightarrow (a^\dagger)^2 \ket{0}\rightarrow  a^\dagger b^\dagger a^\dagger \ket{0}  \rightarrow a^\dagger b^\dagger \ket{0} \rightarrow \ket{0}.$      }
\label{fig:KM_state_bulk_slice}
\end{figure}
Now that we have the rules to evaluate chord diagrams, we want to evaluate the sum over all terms in the chord diagram expansion of a correlation function in the KM state. Once again the approach is to map evolution along the boundary circle to the evolution of a bulk slice as shown in figure \ref{fig:KM_state_bulk_slice}. State on a bulk slice is fixed by the chords passing through that slice. Since the state dual to $\ket{\omega}$ contains no chord, it is naturally mapped to the zero chord state $\ket{0}$\footnote{Unlike the zero chord state defined in section \ref{sec:review_dssyk}, $\ket{0}$ is defined on a single copy of the Hilbert space of DSSYK.}. \begin{malign}
    \ket{\omega} \rightarrow \ket{0}. 
\end{malign}The next step is to map the Hamiltonian and the matter operators. Consider the action of $H$: Every insertion of $H$ in the correlation function either creates or annihilates an $H$-chord. Therefore, we map $H$ to the sum of a chord creation and a chord annihilation operator:
\begin{malign}
    H \rightarrow a + a^\dagger.
\end{malign}Moreover, chords can only emerge from $\ket{\omega}$ and end at  $\bra{\omega}$. This is encoded by the following equations:
\begin{malign}
    a \ket{0} = \bra{0} a^\dagger = 0.
\end{malign}The chord commutation rules generate a sum over all possible ways of annihilating chords while also taking into account the weight of intersection between chords. Since the weight of intersection between $H$-chords in the KM state is the same as in the maximally mixed state, the commutation relations between $a$ and $a^\dagger$ are the same as equation \eqref{eq:chord_commuation_rule}. \begin{malign}
    [a, a^\dagger]_q = 1.
\end{malign}Now let us consider the map of $M$ and $M^\dagger$. Every insertion of $M^\dagger$ creates an $M$-chord while an insertion of $M$ annihilates it.  Moreover, there are dashed chords connecting $M$ and $M^\dagger$ with $P_\omega$. This suggests the following map for $M$ and $M^\dagger$ \begin{malign}
    M^\dagger \rightarrow b^\dagger \, d,  \quad M \rightarrow d\,  b.
\end{malign}where $b^\dagger$ and $b$
 create and annihilate an $M$-chord respectively, while $d$ creates a dashed chord. Notice that the ordering of $b$ and $d$ in the definition of $M$ is chosen to ensure that an $M$-chord does not intersect a dashed chord emerging from the same insertion of $M$. The ordering of $b^\dagger$ and $d$ in the definition of $M^\dagger$ is chosen similarly. The fact that $M$  annihilates $\ket{\omega}$ is captured by the following relation:
 \begin{malign}
     b \ket{0} = \bra{0} b^\dagger = 0.
 \end{malign}To find the commutation relation between $b$ and $b^\dagger$, it is convenient to isolate the $M$-chords from the dashed chords. Since the weight of intersection between $M$-chords is 1, $b$ and $b^\dagger$ follow the standard commutation relation:
\begin{malign}
    [b, b^\dagger] = 1.
\end{malign}Weight of intersection between $H$-chords and $M$-chords are encoded in the $q$-deformed commutation relations between the creation and annihilation operators of different chord species:\begin{malign}
   \quad b a^\dagger - q^{\Delta/2} a^\dagger b \equiv [b,a^\dagger]_{q^{\Delta/2}} = 0, \quad [a , b^\dagger]_{q^{\Delta/2}} = 0\,.
\end{malign}Now let us consider the action of $d$ on $\ket{0}$. The fact that $d$ creates a dashed chord ending at $P_\omega$ suggests that $\ket{0}$ can be realized as a `condensate' of dashed chords which satisfies\begin{malign}
    d \ket{0} = \ket{0}.
\end{malign}
Finally, we have the rules for intersection between the dashed chords and $H$- and $M$-chords which are encoded in the following commutation relations: 
\begin{malign}
[d, b^\dagger]_{q^{\Delta^2/2}} &= [b, d]_{q^{\Delta^2/2}} = 0\,, \\
    [d, a^\dagger]_{q^{\Delta/2}} &= [a, d]_{q^{\Delta/2}} = 0.
\end{malign}The representation of the Hamiltonian and matter operators as linear combinations of creation and annihilation operators suggests that the Hilbert space relevant to the dynamics around $\ket{\omega}$ is spanned by monomials of $a^\dagger$ and $b^\dagger$ acting on the zero chord state $\ket{0}$:
\begin{malign}\label{eq:hilbert_space}
    \cH_{d} = \text{span}\big\{ \prod_{i = 0}^k a^{\dagger\, n_i} b^{\dagger\, m_i} \ket{0}: n_i , m_i, k \in \mathbb{Z}_{\geq 0} \big\}.
\end{malign}The definition of the chord Hilbert space is complete with an inner product. Since the basis states in $\cH_d$ are constructed by acting with $a^\dagger$ and $b^\dagger$ on the zero chord state $\ket{0}$, a natural definition of the inner product is obtained by demanding that $a$ and $b$ are the Hermitian conjugates of $a^\dagger$ and $b^\dagger$ respectively. Thus, we end up with the following definition of the inner product on $\cH_d$:
\begin{malign}
    \langle f(a^\dagger,b^\dagger) \ket{0}, g(a^\dagger, b^\dagger) \ket{0}\rangle = \bra{0} (f(a^\dagger,b^\dagger))^\dagger g(a^\dagger, b^\dagger) \ket{0}, \quad \langle 0 | 0 \rangle = 1. 
\end{malign}for arbitrary polynomials $f(a^\dagger,b^\dagger)$ and $g(a^\dagger,b^\dagger)$. Here $(f(a^\dagger, b^\dagger))^\dagger$ is the Hermitian conjugate of $f(a^\dagger, b^\dagger)$. 

With the above definition of the inner product, the commutation relations and the definition of the zero chord state $\ket{0}$, it follows that two basis states in \eqref{eq:hilbert_space} have a non-zero inner product iff the states have an equal number of $a^\dagger$ (and $b^\dagger$) acting on $\ket{0}$. This observation motivates the definition of chord number operators $\hat n_H$ and $\hat n_M$ which count the number of $a^\dagger$ and $b^\dagger$ acting on the zero chord state:
\begin{malign}
   & \hat n_H \prod_{i = 0}^k a^{\dagger\, n_i} b^{\dagger\, m_i} \ket{0} =\left( \sum_{j = 0}^k n_i \right) \prod_{i = 0}^k a^{\dagger\, n_i} b^{\dagger\, m_i} \ket{0}, \\    & \hat n_M \prod_{i = 0}^k a^{\dagger\, n_i} b^{\dagger\, m_i} \ket{0} = \left( \sum_{j = 0}^k m_i \right) \prod_{i = 0}^k a^{\dagger\, n_i} b^{\dagger\, m_i} \ket{0} \,.
\end{malign}It is also useful to define a ``total chord number" $\hat n_{\text{tot}}$ operator as a weighted sum of $\hat n_H$ and $\hat n_M$:
\begin{malign}
  \hat n_{\text{tot}}  = \hat n_H + \Delta \hat n_M .
\end{malign}From this definition,  it follows that $ \hat n_{\text{tot}}$ has the following commutation relations with the chord creation and annihilation operators:
\begin{malign}
    [\hat n_{\text{tot}}, a^\dagger] = a^\dagger\,, \quad  [\hat n_{\text{tot}}, b^\dagger] = \Delta b^\dagger, \\ 
    [\hat n_{\text{tot}}, a] = -a\,, \quad  [ \hat n_{\text{tot}}, b] = -\Delta b \,.
\end{malign}The above commutation relations imply that $q^{\frac{\Delta}{2} \hat n_{\text{tot}}}$ and $d$ have the exactly the same commutation relations with other creation and annihilation operators. Moreover, they have identical action on the zero chord state:
\begin{malign}
    q^{\frac{\Delta}{2} \hat n_{\text{tot}}} \ket{0} = \ket{0} = d \ket{0}.
\end{malign}Therefore, $q^{\frac{\Delta}{2} \hat n_{\text{tot}}}$ and $d$  act identically  on $\cH_{d}$. This identification gives us another useful representation of $M$ and $M^\dagger$ on the chord Hilbert space:
\begin{malign}
    M^\dagger \rightarrow b^\dagger   q^{\frac{\Delta}{2} \hat n_{\text{tot}}},\quad M \rightarrow q^{\frac{\Delta}{2}  \hat n_{\text{tot}}} b\,.
\end{malign}To conclude this section, we have found that correlation functions of the Hamiltonian and the matter operators in the KM state in the double-scaled limit can be computed by the following formula:
\begin{malign}
    \bra{\omega} f(M, M^\dagger, H)\ket{\omega} = \bra{0}f(q^{\frac{\Delta}{2}  \hat n_{\text{tot}}} b,b^\dagger   q^{\frac{\Delta}{2} \hat n_{\text{tot}}}, a + a^\dagger) \ket{0}
\end{malign}
where $f$ is an arbitrary polynomial of $M, M^\dagger$ and $H$.

\section{Emergent Operator Algebras and Characterization of Purity} \label{sec:vn_algebras}
In this section, we analyze the Murray--von Neumann type of the two classes of double-scaled operator algebras introduced above, distinguished by whether or not the dressed chord operators $M$ and $M^\dagger$ are included. We begin with the algebra generated by the two microscopic operators
\be \label{eq:op1}
H_N = \sum_{I} J_{I} \Psi_{I},
\qquad
O_N = \sum_{I^\prime} K_{I^\prime} \Psi_{I^\prime},
\ee
where $|I|=p$, $|I^\prime|=p^\prime=\Delta p$ with $\Delta>0$, and the subscript $N$ emphasizes that these operators are defined in the SYK model at finite $N$. We will show that the GNS representation of the corresponding ensemble-averaged double-scaled limit gives rise to the familiar Type II$_1$ von Neumann algebra studied in~\cite{Lin:2022rbf,Xu:2024von}.

By contrast, if one incorporates operators adapted to the Kourkoulou–Maldacena (KM) state,
\be \label{eq:op2}
M_N = \sum_{I} K_I C_{I}, \quad M_{N}^\dagger = \sum_{I} K_I^* C^{\dagger}_I,
\ee
then the limiting algebra becomes  Type~I$_\infty$. In this case, access to the full algebra allows one to reconstruct all information in the dual bulk theory, and the corresponding state appears pure. The distinction between the two von Neumann types therefore provides a precise characterization of whether the emergent state in the double-scaling limit is mixed or pure. Importantly, in both cases the underlying sequence of states is the same; what differs is the GNS Hilbert space that emerges in the limit, which depends on the choice of operators generating the algebra.

The fact that the large-$N$ limit of the same sequence of pure states can yield either a pure or a mixed state, depending on the choice of algebra, is not in itself surprising. It simply reflects whether the limiting algebra is capable of probing the entire system. A familiar analogy is provided by a lattice regularization of a local quantum field theory on a global Cauchy slice, with the sequence of states taken to be the vacuum of the lattice theory at each $N=1/\delta$, where $\delta$ is the lattice spacing. In the continuum limit, the vacuum remains pure when considered with respect to the algebra generated by operators acting on all lattice sites, corresponding to the full Cauchy slice, but appears mixed when restricted to an algebra generated by operators supported on a proper subregion.

What is nontrivial in the present setting is that there is no fundamental distinction, at the level of microscopic operators, between \eqref{eq:op2} and \eqref{eq:op1}. Both can be expanded in the basis $\Psi_I$ with appropriate disorder-dependent coefficients, and both may be viewed as simple operators that perturb the state of $N$ fermions by an amount of order $p$, which constitutes an infinitesimal fraction of the degrees of freedom in the double-scaling limit. 

The essential difference is instead encoded in the correlations among the coefficients appearing in \eqref{eq:op2} when expressed as superpositions of the $\Psi_I$ basis. These correlations carry information about the KM state and translate, in the chord Hilbert space, into a nonlocal operator that measures the wormhole length from the insertion point to the insertion of projection on the boundary. Recent work has explored the interpretation of the wormhole length in this setup as a measure of the complexity of the corresponding state~\cite{susskind2014computationalcomplexityblackhole,Rabinovici:2023yex,Rabinovici:2025otw,Xu:2025Complexity,Heller:2024ldz,Ambrosini:2024sre,Aguilar-Gutierrez:2025hty}. This naturally suggests the possibility that correlations among different terms in the operator are geometrized as a distance from the operator insertion to the boundary projection. While we leave a detailed exploration of this interpretation for future work, in the following discussion we establish, independently of any complexity-theoretic considerations, that the algebra generated by double-scaled operators in~\eqref{eq:op2} together with the ordinary chord Hamiltonian is of Type~I$_\infty$, based on the way the double-scaled operators in \eqref{eq:op2} act on the associated chord Hilbert space.

\subsection{The Type II$_1$ Double-Scaled Algebra}

In this section, we formulate the double-scaling limit in the language of operator algebras, focusing first on the emergent algebra generated only by generic operators. The cleanest approach is to begin with an abstract $*$-algebra whose correlation functions in the limiting state $\omega$ are defined by the chord-diagram expansion. One may then construct the corresponding Hilbert space, following~\cite{Lin:2022rbf}, and complete the algebra to a von Neumann algebra with GNS construction from these data.

Let
\be
\mathfrak A =\mathbb C\langle H,O\rangle
\ee
be the unital free $*$-algebra generated by two self-adjoint symbols $H$ and $O$, with
\be
H^*=H,\qquad O^*=O,
\ee
and with the involution extended anti-linearly and anti-multiplicatively. Thus $\mathfrak A$ consists of finite noncommutative polynomials in $H$ and $O$ with complex coefficients. At this stage, $H$ and $O$ are only formal generators.

The double-scaling limit of KM states defines a linear functional
\be
\omega:\mathfrak A \to \mathbb C
\ee
by
\be
\omega(f):=
\lim_{\substack{N,p\to\infty\\ p^2/N=\lambda}}
\overline{\langle \omega_N|\,f (H_N,O_N)\,|\omega_N\rangle},
\qquad \forall f\in \mathfrak A,
\ee
where we have made the $N$-dependence explicit in the states and operators, so that $H_N$ and $O_N$ act on the SYK Hilbert space with $2N$ fermions. Here we use the same symbol $f$ both for an abstract polynomial $f(H,O)\in\mathfrak A$ and for the corresponding operator $f(H_N,O_N)$ obtained by substituting $H\mapsto H_N$ and $O\mapsto O_N$. It is clear that $f(H_N,O_N)$ is then a well-defined operator on the microscopic Hilbert space at each $N$.

Since for every $f\in\mathfrak A$ one has
\be
\overline{\langle \omega_N|\,f(H_N,O_N)^\dagger f(H_N,O_N)\,|\omega_N\rangle} \ge 0,
\ee
after taking the double-scaling limit, it follows that
\be
\omega(f^* f)\ge 0.
\ee
Hence $\omega$ is a positive linear functional, and therefore a state on $\mathfrak A$.

We may now apply the GNS construction to the pair $(\mathfrak A,\omega)$. Let
\be
\mathcal N_\omega:=\{A\in \mathfrak A :\omega(A^*A)=0\}
\ee
be the null left ideal. The quotient $\mathfrak A/\mathcal N_\omega$ carries the inner product
\be
\langle [A],[B]\rangle=\omega(A^*B),\qquad A_1 \in [A] \ \text{if}\ A_1-A\in \mathcal N_\omega,
\ee
whose completion defines the GNS Hilbert space $\mathcal H_\omega$. The class of the identity  $[\mathbf{1}]$  defines a cyclic vector $|0\ra_\omega$, and left multiplication gives a representation
\be
\pi_\omega:\mathfrak A \to B(\mathcal H_\omega)
\ee
such that
\be
\omega(f)={}_\omega\langle 0|\,\pi_\omega(f)\,|0\rangle_\omega , \qquad \forall f \in \mathfrak{A}.
\ee
The double-scaled algebra is then defined to be the von Neumann algebra
\be
\mathcal A:=\pi_\omega(\mathfrak{A})^{\pp\pp}
= W^*(\pi_\omega(H),\pi_\omega(O))\subseteq B(\mathcal H_\omega),
\ee
where the double-commutant is taken with respect to $\mathcal{B}(\mathcal{H}_\omega)$, and $W^*(S)$ denotes the smallest von Neumann algebra containing the set $S$ together with the identity. This formulation makes precise the sense in which the double-scaled algebra is an emergent object reconstructed from the averaged limiting correlators, rather than an \emph{a priori} input. In practice, one may regard double-scaled SYK as an autonomous theory defined directly by its chord rules, independent of finite-$N$ SYK. In this formulation, the microscopic fermionic degrees of freedom are no longer explicit.

The combinatorial analysis of the previous sections shows that states of the form $\pi_\omega(A)|0\rangle_\omega$ are represented by open chord configurations, with operator insertions determined by $\pi_\omega(A)$, and with inner products computed by reflecting and gluing the corresponding chord diagrams. It follows that the GNS Hilbert space $\mathcal H_\omega$ associated with the limiting state is precisely the chord Hilbert space defined by the chord rules\footnote{For simplicity we focus on the case with two types of chords. The extension to any finite number of chord species is straightforward.}. Explicitly,
\be
\mathcal{H}_\omega = \mathcal{H},\qquad |0\ra_\omega =|0\ra,
\ee
and
\be
\pi_\omega (H) = a+ a^\dagger, \qquad \pi_\omega (O) = \tilde a + \tilde{a}^\dagger, \qquad \omega (f) = \la 0 | f(a+a^\dagger,\tilde{a}+\tilde{a}^\dagger) |0\ra. 
\ee
This completes the operator-algebraic formulation of the statement that the ensemble-averaged double-scaling limit of sequence of correlators in KM states converges to the pair $(\mathcal A,\omega)$ determined by the chord rules. Since the chord rules for generic operators in KM states coincide with those of the tracial state, the analysis of~\cite{Xu:2024von,Cao:2025pir} applies directly and implies that $\mathcal A$ is a Type II$_1$ factor.

\subsection{The Type I$_\infty$ Double-Scaled Algebra}

In this section, we study the operator algebra generated by the double-scaled limits of $M_N$, $M_N^\dagger$, and the SYK Hamiltonian $H_N$. The setup parallels that of the previous subsection, except that we now begin with the $*$-algebra
\be
\mathfrak{A}_{d}=\mathbb{C}\la H,M,M^{\da}\ra.
\ee
Here $H$, $M$, and $M^\dagger$ are treated as independent generators. Only after passing to the GNS representation does $M^\dagger$ become the Hermitian adjoint of $M$ with respect to the inner product on the GNS Hilbert space. At the algebraic level, it is enough to specify the involution on the generators by
\be
H^*=H,\qquad M^*=M^{\da},\qquad (M^{\da})^*=M,
\ee
and then extend it to $\mathfrak{A}_d$ anti-linearly and anti-multiplicatively. The subscript $d$ indicates that this algebra contains the dressed operators.

The analysis of section~\ref{sec:sum-cds} shows that, under the GNS representation, the generators act as
\be
\pi_\omega (H) = a + a^\dagger, \qquad \pi_\omega (M) = q^{\frac{\Delta}{2} \nt} b,\qquad \pi_\omega (M^\dagger)  = b^\dagger q^{\frac{\Delta}{2} \nt},
\ee
where we identify the GNS Hilbert space $\mathcal H_\omega$ with $\mathcal H_d$ defined in~\eqref{eq:hilbert_space}. The corresponding double-scaled algebra is therefore
\be \label{eq:def-Ad}
\mathcal{A}_{d}
=
\{\pi_\omega(H), \pi_\omega(M),\pi_\omega(M^{\da})\}^{\pp\pp}
=
\{a+a^{\da},\, q^{\frac{\Delta}{2}\nt}b,\, b^{\da}q^{\frac{\Delta}{2}\nt}\}^{\pp\pp}
\subseteq \mathcal{B}(\mathcal{H}_d).
\ee
We will show that $\mathcal{A}_d$ is a Type I$_\infty$ factor, and in fact coincides with the full algebra of bounded operators on the GNS Hilbert space:
\be
\mathcal{A}_d = \mathcal{B}(\mathcal{H}_d).
\ee
From this point on, whenever no confusion can arise, we will identify an abstract generator with its GNS image and write $H$, $M$, and $M^\dagger$ in place of $\pi_\omega(H)$, $\pi_\omega(M)$, and $\pi_\omega(M^\dagger)$. We adopt this convention in all subsequent sections and appendices.

It is straightforward to verify that $q^{\Delta \hat{n}_{{\rm tot}}}$ belongs to the algebra $\mathcal{A}_d$:
\be
[q^{\frac{\Delta}{2}\nt}b , b^\dagger q^{\frac{\Delta}{2}\nt} ]_{q^{\Delta^2} } 
= q^{\frac{\Delta}{2}\hat{n}_{{\rm tot}}}[b,b^{\da}]q^{\frac{\Delta}{2}\hat{n}_{{\rm tot}}}
= q^{\Delta \hat{n}_{{\rm tot}}},
\ee
where the first equality follows from the commutation relations
\be
[\hat{n}_{{\rm tot}},b]=-\Delta b,\qquad
[\hat{n}_{{\rm tot}},b^{\da}]=\Delta b^{\da},\qquad
[b,b^{\da}]=1.
\ee
One might be tempted to conclude that $\hat{n}_{{\rm tot}}$ itself belongs to $\mathcal{A}_d$, since it can formally be expressed in terms of $q^{\Delta\hat{n}_{{\rm tot}}}$ as
\be
\hat{n}_{{\rm tot}}=\frac{\ln q^{\Delta\hat{n}_{{\rm tot}}}}{\Delta \ln q}.
\ee
However, the function $f(x)=\ln(x)/(\Delta\ln q)$ is unbounded on the spectrum $\sigma(q^{\Delta\hat{n}_{{\rm tot}}})$ of $q^{\Delta \hat{n}_{{\rm tot}}}$, which consists of the discrete set of eigenvalues $q^{\Delta^2 n_M+\Delta n_H}$ with $n_M,n_H\in\mathbb{Z}_{\ge0}$ and has $0$ as its only accumulation point. Consequently, $\hat{n}_{{\rm tot}}$ does not belong to $\mathcal{A}_d$ as a bounded operator, and the appropriate way to proceed is instead to exploit the spectral properties of $q^{\Delta\hat{n}_{{\rm tot}}}$.

Indeed, $q^{\Delta\hat{n}_{{\rm tot}}}$ is a compact self-adjoint operator, whose only accumulation point in the spectrum is $0$. For any compact self-adjoint operator $T$, the projection onto each eigenspace is a spectral projection and belongs to the $\mathbb{C}^*$-algebra generated by $T$. Concretely, for any $\lambda\in\sigma(q^{\Delta\hat{n}_{{\rm tot}}})$, let $P_\lambda$ denote the projection onto the corresponding eigenspace. This projection can be constructed explicitly using the Riesz functional calculus,
\be \label{eq:Risez}
P_{\lambda }=\frac{1}{2\pi i}\oint_{\gamma_{\lambda}}(zI-q^{\Delta\hat{n}_{{\rm tot}}})^{-1}\dd z,
\ee
where the contour $\gamma_\lambda$ is a small circle enclosing $\lambda$ but no other points in the spectrum. Such a contour exists because $0$ is the only accumulation point of $\sigma(q^{\Delta\hat{n}_{{\rm tot}}})$. For any $z\in\mathbb{C}$ with $z\notin\sigma(q^{\Delta\hat{n}_{{\rm tot}}})$, the resolvent appearing in \eqref{eq:Risez} is a bounded function of $q^{\Delta\hat{n}_{{\rm tot}}}$, and therefore the projection operator $P_\lambda$ defined above belongs to $\mathcal{A}_d$.

Applying the spectral decomposition to $q^{\Delta \hat{n}_{{\rm tot}}}$ yields the following resolution of the identity on $\mh_d$,
\be
\mathbf{1}_{\mh_d}= \sum_{\lambda\in\sigma(q^{\Delta\hat{n}_{{\rm tot}}})} P_\lambda,
\qquad
P_{\lambda_1}P_{\lambda_2}=\delta_{\lambda_1,\lambda_2}P_{\lambda_1},\quad \lambda_{1,2}\in\sigma(q^{\Delta\nt}).
\ee
In particular, one can show that the creation operator $a^\dagger$ itself belongs to $\mathcal{A}_d$~\cite{Cao:2025pir}, and can be written as
\be
a^\dagger=\sum_{\lambda\in\sigma(q^{\Delta\hat{n}_{{\rm tot}}})} P_{q^\Delta\lambda}\,H\,P_{\lambda}.
\ee
It also follows that the annihilation operator $a$ belongs to $\mathcal{A}_d$.

We now prove that $\mathcal{A}_d=\mathcal{B}(\mathcal{H}_d)$, which establishes that $\mathcal{A}_d$ is a von Neumann algebra of Type~I$_\infty$. It suffices to show that the commutant $\mathcal{A}_d^{\pp}$ in $\mathcal{B}(\mathcal{H}_d)$ consists only of scalar multiples of the identity. It follows from the  GNS construction that the double-scaled Kourkoulou--Maldacena state $\omega$ is cyclic for $\mathcal{A}_d$. Now with the chord rules developed in~\ref{sec:sum-cds}, we can derive an explicit operator basis of $\mh_d$ in terms of  chord configurations generated by successively acting on $\ket{\omega}$ with an ordered set of the two types of creation operators,
\be \label{eq:cyclic}
|n_{0},n_{1},\dots,n_{k}\rangle
=q^{-\frac{\Delta}{2}\left(\sum_{j=1}^{k}jn_{j}+\binom{k}{2}\Del\right)}
a^{\da n_{0}}(b^{\da}q^{\frac{\Delta}{2}\hat{n}_{{\rm tot}}})a^{\da n_{1}}\cdots(b^{\da}q^{\frac{\Delta}{2}\hat{n}_{{\rm tot}}})a^{\da n_{k}}\ket{\omega}.
\ee
The cyclic property of $\ket{\omega}$ for $\mathcal{A}_d$ can therefore be stated as follows: for any state $\ket{\Psi}\in\mh_d$, there exists an operator $\hat{\Psi}\in\mathcal{A}_d$ such that $\ket{\Psi}=\hat{\Psi}\ket{\omega}$. A notable difference from the previous algebra is that $\ket{\omega}$ is not separating for $\mathcal{A}_d$, since both $M=q^{\Delta\hat{n}_{{\rm tot}}/2}b$ and $a$ annihilate $\ket{\omega}$ despite being nonzero operators. As a result, for a given state $\ket{\Psi}$ the operator $\hat{\Psi}$ is not unique, and may differ by the addition of operators that annihilate $\ket{\omega}$. A convenient choice is to take $\hat{\Psi}$ to involve only $a^\dagger$ and $b^\dagger q^{\Delta\hat{n}_{{\rm tot}}/2}$, which is always possible due to \eqref{eq:cyclic}. In the following discussion, we therefore assume $\hat{\Psi}=\hat{\Psi}(a^\dagger,b^\dagger q^{\Delta\hat{n}_{{\rm tot}}/2})$.

Now suppose $\mathcal{O}^{\pp}\in\mathcal{A}_d^{\pp}$ acts on $\ket{\omega}$ as
\be \label{eq:def-Op}
\mathcal{O}^{\pp}\ket{\omega}
=c_{\mathcal{O}^{\pp}}\ket{\omega}+\ket{\eta},
\qquad
\mathcal{O}^{\pp\da}\ket{\omega}
=d_{\mathcal{O}^{\pp\da}}\ket{\omega}+\ket{\xi},
\ee
where $\ket{\eta}$ and $\ket{\xi}$ are orthogonal to $\ket{\omega}$. It follows from \eqref{eq:def-Op} that $d_{\mathcal{O}^{\pp\dagger}}=c^{*}_{\mathcal{O}^{\pp}}$. We now show that $\ket{\eta}$ and $\ket{\xi}$ must vanish. It suffices to show that their inner product with any state $\ket{\Psi}$ in a dense subspace of $\mh_d$ vanishes. Indeed,
\be
\begin{aligned}
\langle\eta | \Psi\rangle
&=\langle\eta| \hat{\Psi}\ket{\omega}
=\langle\omega|\left(\mathcal{O}^{\pp\da}-c_{\mathcal{O}^{\pp}}^*\mathbf{1}\right)\hat{\Psi}\ket{\omega} \\
&=\langle\omega|\hat{\Psi}\left(\mathcal{O}^{\pp\da}-c_{\mathcal{O}^{\pp}}^*\mathbf{1}\right)\ket{\omega}
=0,
\end{aligned}
\ee
where the first equality uses the cyclicity of $\ket{\omega}$, the second follows from \eqref{eq:def-Op}, and the third uses the fact that $\mathcal{O}^{\pp\da}\in\mathcal{A}_d^{\pp}$ commutes with all operators in $\mathcal{A}_d$, together with the assumption that $\hat{\Psi}$ annihilates $\bra{\omega}$ since it is composed only of $a^\dagger$ and $b^\dagger q^{\hat{n}_{{\rm tot}}/2}$. Hence $\ket{\eta}$ is orthogonal to every state in $\mh_d$ and must therefore vanish. The same argument shows that $\ket{\xi}=0$.

It follows that $\mathcal{O}^{\pp}$ acts on $\ket{\omega}$ as multiplication by a constant. Applying the cyclic property of $\ket{\omega}$ once more, we conclude that $\mathcal{O}^{\pp}$ acts on any state $\ket{\Psi}\in\mh_d$ as
\be
\mathcal{O}^{\pp}\ket{\Psi}
=\mathcal{O}^{\pp}\hat{\Psi}\ket{\omega}
=\hat{\Psi}\mathcal{O}^{\pp}\ket{\omega}
=c_{\mathcal{O}^{\pp}}\hat{\Psi}\ket{\omega}
=c_{\mathcal{O}^{\pp}}\ket{\Psi}.
\ee
Therefore, $\mathcal{O}^{\pp}=c_{\mathcal{O}^{\pp}}\mathbf{1}_{\mh_d}$, and we conclude that
\be
\mathcal{A}_d^{\pp}=\mathbb{C}\mathbf{1}_{\mh_d} 
\Longrightarrow
\mathcal{A}_d=\mathcal{B}(\mathcal{H}_d).
\ee
This completes the proof that $\mathcal{A}_d$ is a Type~I$_\infty$ von Neumann algebra.

The analysis above reveals an interesting and somewhat surprising structure. Once the dressed chord operators $M_N$ and $M_N^\dagger$ are adjoined to the SYK Hamiltonian, the resulting emergent algebra contains all operators that survive the ensemble-averaged double-scaling limit. In particular, the Type I$_\infty$ algebra $\ma_d$ contains a distinguished proper subalgebra
\be
\ma_{0}=\mathbb{C}\la a+a^{\da},b+b^{\da}\ra_{\mathcal{B}(\mathcal{H}_{d})}^{\pp\pp},
\ee
generated by the two chord operators
\be
H=a+a^\dagger,\qquad \tilde H=b+b^\dagger.
\ee
This subalgebra is proper because its commutant is nontrivial:  there are chord operators that create or annihilate $H$ and $\tilde{H}$ chords on the other side of an open chord diagram.

A notable feature of $\tilde H$ is that self-intersections of the corresponding chord carry unit penalty, while intersections between an $H$-chord and a $\tilde H$-chord still carry the nontrivial factor $q^\Delta\in(0,1)$. Such operators were studied previously in~\cite{Gao_2025Dcommute,Berkooz:2024Integrable-short,Berkooz:2024integrable-long,Sergio2026integrable}, where $\tilde H$ was referred to as an \emph{integrable chord}, in contrast with the \emph{chaotic chord} $H$. Microscopically, $\tilde H$ admits the finite-$N$ realization
\be
\tilde{H}_{N}=(-1)^{p^\prime} \sum_{1\leq i_{1}<\cdots<i_{p^{\pp}}\leq N} B_{i_{1}\cdots i_{p^{\pp}}}Z_{i_{1}}\cdots Z_{i_{p^{\pp}}},
\ee
with properly normalized disorder coefficients $B_{i_1,\dots i_{p^\prime }}$ drawn from a Gaussian ensemble. Accordingly, $\ma_0$ may be viewed as the GNS representation of the algebra generated by $H_N$ and $\tilde H_N$ in the ensemble-averaged double-scaling limit,
\be\label{eq:suba0}
\mathfrak{A}_{0}=\mathbb C\langle H,\tilde H\rangle,
\qquad
\ma_0=\pi_\omega(\mathfrak A_0)^{\pp\pp}.
\ee
What makes this structure especially interesting is that, at finite $N$, there is essentially no direct relation between the two families of operators $\tilde H_N$ and $M_N,M_N^\dagger$. They are constructed in very different ways and are adapted to quite different microscopic data. Nevertheless, after passing to the ensemble-averaged double-scaling limit, the algebra generated by the dressed operators is rich enough to reconstruct the integrable chord operator $\tilde H$. In other words, two operator families that appear unrelated in the microscopic theory become deeply connected in the emergent algebraic description. This is a striking example of how the large-$N$ limit reorganizes operator structure: relations that are invisible at finite $N$ can become exact in the emergent algebra.

A natural reason to expect that $\mathcal{A}_d$ probes a pure state is the self-averaging property of SYK correlators in the regime considered here. Because the dressed operators $M_N$ and $M_N^\dagger$ are already well-defined at each finite $N$ and their correlators self-average in the double-scaling limit, their limiting GNS representation continues to encode the microscopic purity of the KM state, rather than washing it out into a thermal description. In this sense, the purity of the emergent state is not restored by hand, but is inherited from the existence of a self-averaging sequence of state-adapted operators in the underlying theory.

By contrast, the algebra generated directly by $H_N$ and $\tilde H_N$ still has a nontrivial commutant in the double-scaled limit, which suggests that it describes a mixed state, as in the previous discussion.\footnote{It is not clear whether $\mathcal A_0$ is itself a Type II$_1$ factor. Since the $\tilde H$-chord carries no self-intersection penalty, arbitrarily many such chords may accumulate at no cost, leading to a qualitative change in the analytic structure relative to the standard double-scaled algebra.} In the bulk interpretation, this nontrivial commutant naturally suggests the presence of operators behind a horizon that cannot be reconstructed from the algebra itself. Once the dressed operators $M_N$ and $M_N^\dagger$ are included, however, the commutant becomes trivial.  Equivalently, the fact that $\mathcal A_d = \mathcal{B}(\mathcal{H}_d)$  means that the algebra sees no horizon: its entanglement wedge is expected to contain the full Cauchy slice of the emergent bulk geometry.

\section{Correlation Functions of Dressed Chord Operators} \label{sec:correlators}

In this section, we evaluate correlation functions involving the charged chord operators $M$ and $M^\dagger$. In contrast to neutral operators such as $O$, whose GNS representation in the state $\omega$ acts as the standard chord operators on the empty chord state $|0\ra$, the operators $M$ and $M^\dagger$ are represented by dressed chord creation and annihilation operators. As a result, their correlators are not the conventional thermal correlators, and they depend on the location of the KM projection on the boundary. We derive closed-form expressions for these correlators and give an intuitive interpretation of them in terms of chord diagrams.

The derivations in the following subsections rely heavily on matrix elements of chord operators in the energy eigenbasis, which are often expressed in terms of $q$-special functions such as the $q$-Gamma function $\Gamma_q$. For completeness, we review the bulk Hilbert space description of these operators in conventional double-scaled SYK, together with the relevant formulas and standard notation used throughout this section, in appendix~\ref{app:review-sec}. Readers already familiar with these results and conventions may proceed directly, as we follow the standard notation in the literature.

\subsection{Two-point Functions}
We begin by computing the two-point function
\be \label{eq:G2-def3}
G_{2}(\tau_{1},\tau_{2},\tau_{3})
=\la\omega|\wick{ e^{-\tau_{1}H}\c1 Me^{-\tau_{2}H}\c1 M^{\da}e^{-\tau_{3}H} } |\omega\ra.
\ee
Under the GNS representation in the emergent state $\omega$, this correlator is mapped to an expectation value on the dressed chord Hilbert space $\mathcal H_d$:
\be \label{eq:G2-def30}
G_{2}(\tau_{1},\tau_{2},\tau_{3})
=\la 0 | e^{-\tau_{1}H}  q^{\frac{\Delta}{2}\nt } b\, e^{-\tau_{2}H} b^\dagger q^{\frac{\Delta}{2}\nt} e^{-\tau_{3}H} |0\ra.
\ee
Here $H$ is represented by the standard chord Hamiltonian acting on the tracial chord Hilbert space, while $M$ and $M^\dagger$ are represented by dressed chord annihilation and creation operators. Since the empty chord state $|0\ra$ is tracial with respect to the Hamiltonian $H$, it is natural to evaluate $G_2$ by resolving the identity in the energy eigenbasis of $H$. Expanding the Euclidean evolution along the $\tau_1$ and $\tau_3$ segments then gives
\be \label{eq:G2-def4}
G_{2}
=\int_{0}^{\pi}\dd\mu(\te_{1})\dd\mu(\te_{3})
\,e^{-\tau_{1}E_{1}-\tau_{3}E_{3}}
\la\te_{1}|q^{\frac{\Delta}{2}\nt} b e^{-\tau_2 H} b^\dagger q^{\frac{\Delta}{2}\nt}  |\te_{3}\ra,
\ee
where $|\te\rangle$ denotes the energy eigenbasis of the standard chord Hamiltonian, and we have used
\be
H |\te_i \ra = E_i |\te_i \ra , \quad E_i =\frac{2\cos\theta_i}{\sqrt{1-q}}. 
\ee
Since the $|\theta\ra$ states are in zero-particle sector with respect to $M$-chords,  the total chord number operator $\nt=\Delta \hat{n}_M + \hat{n}_H$ acts on them effectively as $\hat{n}_H$:
\be
q^{\frac{\Delta}{2}\hat{n}_{{\rm tot}}}\ket{\te_{3}}
=q^{\frac{\Del}{2}\hat{n}_{H}}\ket{\te_{3}}
=\int_{0}^{\pi}\dd\mu(\te_{5})
|\theta_5\ra \la\te_{5}|q^{\frac{\Del}{2}\hat{n}_{H}}|\te_{3}\ra.
\ee
Inserting complete sets of energy eigenstates then gives
\be \label{eq:G2-1}
\la\te_{1}|q^{\frac{\Delta}{2}\nt} b e^{-\tau_2 H} b^\dagger q^{\frac{\Delta}{2}\nt}  |\te_{3}\ra 
=\int_{0}^{\pi}\dd\mu(\te_{4})\dd\mu(\te_{5})
\,\la\te_{1}|q^{\frac{\Del}{2}\hat{n}_{H}}|\te_{4}\ra
\,\la\te_{4}|b e^{-\tau_2 H} b^{\da}|\te_{5}\ra
\,\la\te_{5}|q^{\frac{\Del}{2}\hat{n}_{H}}|\te_{3}\ra.
\ee
To evaluate the middle factor, we insert a complete energy eigenbasis between $b$ and $b^\dagger$, which diagonalizes the Hamiltonian in that interval. Using the matrix elements \eqref{eq:bd-components} and their conjugates, we obtain
\be \label{eq:G2-2}
\la\te_{4}|b e^{-\tau_{2}H} b^{\da}|\te_{5}\ra
=\frac{\delta(\te_{4}-\te_{5})}{\mu(\te_{4})}
\int_{0}^{\pi}\dd\mu(\te_{2})
\,e^{-\tau_{2}E_{2}}
\la \te_2 |q^{\frac{\Delta}{2} \hat{n}_H}|\te_4\ra.
\ee
Substituting \eqref{eq:G2-1} and \eqref{eq:G2-2} into \eqref{eq:G2-def4}, we arrive at
\be \label{eq:G2-result}
G_{2}(\tau_{1},\tau_{2},\tau_{3})
=\mathcal{N}_{\Delta/2}^{3}
\int_{0}^{\pi}\prod_{i=1}^{3}\dd\mu(\te_{i})
\,e^{-\sum_{i=1}^{3}\tau_{i}E_{i}} \int^\pi_0 \dd\mu(\phi)
\prod_{i=1}^{3}
\frac{\Gamma_{q}(\frac{\Del}{2}\pm\frac{i\te_{i}}{\lambda}\pm\frac{i\phi}{\lambda})}
{\Gamma_{q}(\Del)},
\ee
where we used \eqref{eq:m-density} to express the integrand in terms of matter densities, closely resembling analogous expressions in JT gravity. The alternating signs in $\Gamma_q$ mean taking the product over all choices of signs in the argument.

It is obvious from \eqref{eq:G2-result} that $\omega$ is not tracial for $\mathcal{A}_d$, where
\be
G_2 (\tau_1,\tau_2,\tau_3)
\not=
G_2 (\tau_1+\tau_3,\tau_2,0).
\ee
Nevertheless, the dependence on the three Euclidean time intervals is symmetric, which becomes manifest in the chord-diagram representation of the correlator:
\be
G_2(\tau_1,\tau_2,\tau_3)
=
\int_0^\pi \prod_{i=1}^3 d\mu(\theta_i)\,
e^{-\sum_{i=1}^3 \tau_i E_i}
\,\int^\pi_0 \dd\mu(\phi)\,
\begin{tikzpicture}[baseline=-0.5ex, scale=1.2]
  \draw[thick] (0,0) circle (1);
  \fill (90:1) circle (0.05);
  \node[above=1pt] at (90:1) {\small $P_{\omega}$};

  \coordinate (A1) at (90:1);
  \coordinate (B1) at (205:1);
  \coordinate (C1) at (325:1);
  \coordinate (A2) at (90:1);
  \coordinate (B2) at (215:1);
  \coordinate (C2) at (335:1);

  \coordinate (O) at (0,-0.08);

  \draw[dashed, thick] (A1) .. controls (30:0.25) .. (C2);
    \draw[dashed, thick] (A2) .. controls (150:0.25) .. (B1);
    \draw[blue, thick] (B2) .. controls (270:0.25) .. (C1);

  \node at (-0.65,0.42) {\small $\theta_3$};
  \node at (0.65,0.42) {\small $\theta_1$};
  \node at (0,-0.75) {\small $\theta_2$};
  \node at (0,0) {\small $\phi$};
\end{tikzpicture}
\ee
From this perspective, the projector $P_\omega=\ket{\omega}\!\bra{\omega}$ can be viewed as a projection inserted at ``infinity.'' After compactifying the horizontal diagram representation of the correlator \eqref{eq:G2-def3} as a transition amplitude into a chord-diagram picture, the distance from an $M$-chord insertion to infinity is naturally represented by a pair of ordinary matter chords with weight $\Delta/2$, represented as $O$ below, with one of the pair effectively ending on the point of the circle that represents infinity. In this representation, \eqref{eq:G2-def3} may be equivalently written as
\be \label{eq:G2-def-wick}
G_2 (\tau_1,\tau_2,\tau_3)
=
\langle 0 |
\wick{ \c1 O e^{-\tau_1 H} \c1 O \c1 O e^{-\tau_2 H} \c1 O \c1 O e^{-\tau_3 H} \c1 O }
|0\ra,
\ee
which makes it clear why $G_2$ takes the form of a particular uncrossed six-point function of ordinary matter operators in conventional double-scaled SYK.

\subsection{Uncrossed $2n$-point Functions}

It is straightforward to generalize the preceding analysis to uncrossed $2n$-point functions of dressed chord operators separated by Euclidean evolution. We define
\be\label{eq:G2n-def}
G_{2n}(\tau_{1},\tau_{2},\cdots,\tau_{2n+1})=\la\omega|\wick{ e^{-\tau_{1}H}\c1 M e^{-\tau_{2}H}\c1 M^{\da}\cdots e^{-\tau_{2n-1}H}\c1 Me^{-\tau_{2n}H} \c1 M^{\da}e^{-\tau_{2n+1}H} }|\omega\ra.
\ee
Projecting onto the KM state effectively produces a condensate of matter excitations: each dressed insertion $M$ or $M^\dagger$ on the boundary is accompanied by a contracted matter line that runs from the insertion point to the condensate. As a result, $G_{2n}$ can be rewritten as a correlator of ordinary matter chord operators $O$ evaluated in the tracial state,
\be 
G_{2n}= \la0|\wick{\dots \c3 O\c2 O\c1 O e^{-\tau_{1}H}\cdots \c1O \c1 O e^{-\tau_{2i}H}\c1 O\c2 O e^{-\tau_{2i+1}H}\c3 O \c1 O ..\c1 .. e^{-\tau_{2n+1}H} }|0\ra,
\ee 
where the initial string of $2n$ matter operators $O$ represents the condensate induced by the projection onto the KM state. These operators are contracted with the $O$ insertions adjacent to each factor of $e^{-\tau_{2i+1}H}$ for $i=1,2,\dots,n$, while the two $O$'s adjacent to each $e^{-\tau_{2i}H}$ are contracted with one another. In the chord-diagram representation, the resulting correlator is illustrated by
\be\label{fig:G2n}
\begin{tikzpicture}[scale=2,line cap=round,line join=round]

  \def\R{1.0}

  \coordinate (P) at (90:\R);

  \draw[line width=1.2pt]
    (180:1) arc[start angle=180,end angle=0,radius=\R];
  \draw[line width=1.2pt,dashed]
    (0:\R) arc[start angle=0,end angle=-30,radius=\R];
    \draw[line width=1.2pt]
    (-30:\R) arc[start angle=-30,end angle=-150,radius=\R];
  \draw[line width=1.2pt,dashed]
    (-150:\R) arc[start angle=-150,end angle=-180,radius=\R];

  \fill (P) circle (0.035);
  \node[above=1pt] at (P) {$P_{\omega}$};

  \draw[line width=.75pt,dashed] (P) -- (47:1);
  \draw[line width=.75pt,dashed] (P) -- (0:1);
  \draw[line width=.75pt,dashed] (P) -- (-32:1);
  \draw[line width=.75pt,dashed] (P) -- (-78:1);
  \draw[line width=.75pt,dashed] (P) -- (-102:1);
  \draw[line width=.75pt,dashed] (P) -- (-148:1);
  \draw[line width=.75pt,dashed] (P) -- (180:1);
  \draw[line width=.75pt,dashed] (P) -- (133:1);

    \draw[blue, line width=1.2pt] (45:1) .. controls (25:0.8) .. (5:1);
    \draw[blue,line width=1.2pt] (-35:1) .. controls (-55:0.75) .. (-75:1);
    \draw[blue,line width=1.2pt] (-105:1) .. controls (-125:0.75) .. (-145:1);
    \draw[blue,line width=1.2pt] (135:1) .. controls (155:0.8) .. (175:1);

    \node at (65:1.15) {$\theta_1$};
    \node at (25:1.15) {$\theta_2$};
    \node at (-55:1.15) {$\theta_{2i}$};
    \node at (-90:1.15) {$\theta_{2i+1}$};
    \node at (-125:1.15) {$\theta_{2i+2}$};
    \node at (155:1.2) {$\theta_{2n}$};
    \node at (120:1.2) {$\theta_{2n+1}$};

    \node at (50:0.85) {$\phi_1$};
    \node at (-32:0.45) {$\phi_i$};
    \node at (-150:0.45) {$\phi_{i+1}$};
    \node at (130:0.82) {$\phi_{n}$};
    \node at (-8:0.8) {$\dots$};
    \node at (-172:0.8) {$\dots$};

\end{tikzpicture}
\ee
where the $\theta_i$'s and $\phi_i$'s are energy parameters associated with the chord diagram: $\theta_i$ labels each bulk region adjacent to a boundary segment, while $\phi_i$ labels the additional regions introduced by contractions with the condensate through the projector $P_\omega$. Although $P_\omega$ induces extra contractions between boundary insertions and the condensate, in the uncrossed configuration the resulting diagram still contains no chord crossings in the bulk. As a result, the analytic expression for \eqref{eq:G2n-def} can be written directly as
\be
G_{2n}=\mathcal{N}_{\Del/2}^{3n}\int_{0}^{\pi}\left(\prod_{i=1}^{2n+1}\dd\mu(\theta_{i})e^{-\tau_{i}E(\theta_{i})}\right)\int_{0}^{\pi}\left(\prod_{i=1}^{n}\dd\mu(\phi_{i})\right)\prod_{i=1}^{n}\prod_{j=-1,0,1}\frac{\Gamma_{q}(\frac{\Del}{2}\pm\frac{i\phi_{i}}{\lambda}\pm\frac{i\te_{2i+j}}{\lambda})}{\Gamma_{q}(\Del)},
\ee

\subsection{Crossed Four-Point Function}
In this section we evaluate the crossed four-point function, defined by the correlator in which the two charged matter chords are contracted in a crossing configuration:
\be\label{eq:G4-cr-def}
G_{4}^{({\rm crossed})}
=\la\omega|
\wick{ e^{-\tau_{1}H}\c1 Me^{-\tau_{2}H}\c2 Me^{-\tau_{3}H}\c1 M^{\da}e^{-\tau_{4}H}\c2 M^{\da}e^{-\tau_{5}H} }
|\omega\ra.
\ee
Our strategy is to decompose this correlator into three pieces. The bra and ket are prepared by acting with $M$ and $M^\dagger$, dressed by Euclidean evolution, while the central segment isolates the bulk crossing kernel. Using the chord rules, we first rewrite the ket preparation as
\be
e^{-\tau_{4}H}M^{\da}e^{-\tau_{5}H}|\omega\ra\longrightarrow e^{-\tau_{4}H}O q^{\frac{\Del}{2}\hat{n}_{{\rm tot}}}e^{-\tau_{5}H}|0\ra.
\ee
Here the operator $q^{\frac{\Delta}{2}\hat{n}_{\rm tot}}$ acts within the zero-particle Hilbert space $\mh_0$. We expand the thermal state $e^{-\tau_5 H}|0\ra$ in the energy basis and insert the matrix element of $q^{\frac{\Delta}{2}\hat{n}_{\rm tot}}$ in terms of the vertex function~\eqref{eq:vertex-func} with weight $\Delta/2$:
\be
q^{\frac{\Del}{2}\hat{n}_{{\rm tot}}}e^{-\tau_{5}H}|0\ra=\int_{0}^{\pi}\dd\mu(\te_{5})e^{-\tau_{5}E_{5}}q^{\frac{\Del}{2}\hat{n}_{{\rm tot}}}|\te_{5}\ra
=\int_{0}^{\pi}\prod_{i=3,5}\dd\mu(\te_{i})e^{-\tau_{5}E_{5}}\gamma_{\te_{3}\te_{5}}^{2}|\te_{3}\ra.
\ee
Next, the matter chord $O$ maps a zero-particle energy eigenstate into a superposition of one-particle energy eigenstates:
\be
O|\te_{3}\ra=\int_{0}^{\pi}\dd\mu(\te_{4})|\frac{\Delta}{2};\te_{4},\te_{3}\ra,
\ee
The definitions of the states appearing in the equation above are reviewed in~\eqref{eq:0p-state-def} and~\eqref{eq:def-1p-state}. We choose the ordering $(\theta_4,\theta_3)$ so that the subsequent Euclidean evolution by $e^{-\tau_4 H}$ diagonalizes with energy eigenvalue $E_4$. Putting these ingredients together, we obtain
\be \label{eq:1p-states-1}
e^{-\tau_{4}H}M^{\da}e^{-\tau_{5}H}|\omega\ra\longrightarrow\int_{0}^{\pi}\prod_{i=3}^{5}\dd\mu(\te_{i})\thinspace e^{-\tau_{4}E_{4}-\tau_{5}E_{5}}\gamma_{\te_{3}\te_{5}}^{2}|\frac{\Delta}{2};\te_{4},\te_{3}\ra.
\ee
An analogous derivation applies to the preparation of the bra state, yielding
\be \label{eq:1p-states-2}
\la\omega|e^{-\tau_{1}H}Me^{-\tau_{2}H}
    \longrightarrow \int_{0}^{\pi}\prod_{i=0}^{2}\dd\mu(\te_{i})
    e^{-\tau_{1}E_{1}-\tau_{2}E_{2}}
    \gamma_{\te_{0}\te_{1}}^{2}
    \la \frac{\Del}{2};\te_{2},\te_{0}|.
\ee
The remaining part of the correlator is the crossed matter region separated by Euclidean time $\tau_3$, namely the part involving crossed $M e^{-\tau_3 H} M^\dagger$ in \eqref{eq:G4-cr-def}. After the bra and ket preparations in \eqref{eq:1p-states-1} and \eqref{eq:1p-states-2}, this middle segment reduces to a matrix element of the one-particle crossing kernel:
\be \label{eq:def-I}
I
=\la\frac{\Del}{2};\te_{2},\te_{0}|
q^{\Del\nt/2}
\chi_{{\rm \frac{\Del}{2}}}(\tau_{3})
q^{\Del\nt/2}
|\frac{\Del}{2};\te_{4},\te_{3}\ra,
\ee
where $\chi_\Delta(\tau)$ implements the exchange of two matter insertions separated by Euclidean time $\tau$:
\be \label{fig:chi}
\chi_{\Delta}(\tau)=
\begin{tikzpicture}[baseline=-0.6ex, line width=0.9pt,scale=.8]
  \draw (0,0) -- (5,0);

  \draw[blue] (1,0) -- (1,.4) -- (4.5,.4);

  \draw[blue] (3.6,0) -- (3.6,.8) -- (0.4,.8);

  \node at (1,-0.35) {$\Delta$};
  \node at (3.6,-0.35) {$\Delta$};
  \node at (2.5,-0.35) {$e^{-\tau H}$};
\end{tikzpicture}
\ee 
The insertions of $q^{\frac{\Delta}{2}\hat{n}_{\rm tot}}$ weight the total number of incoming or outgoing chords. Diagrammatically, they can be represented by straight lines adjacent to the crossing region that extend to infinity. Projecting onto one-particle energy eigenstates then corresponds to annihilating the matter chords with fixed energy boundary conditions on the two sides of the matter line. Combining these elements with \eqref{fig:chi}, we depict $I$ as
\be
I =
\begin{tikzpicture}[baseline=-0.5ex, line width=1pt]

\draw (0,0) -- (6,0);

\draw[blue]
(0.5,0) -- (0.5,0.8) -- (3.6,0.8) -- (3.6,0);

\draw[blue]
(2.0,0) -- (2.0,0.5) -- (5.2,0.5) -- (5.2,0);

\draw[dashed] (1.8,0) -- (1.8,1.3);
\draw[dashed] (3.9,0) -- (3.9,1.3);

\node at (3,-0.25) {$e^{-\tau_3 H}$};
\node at (0.5,-0.35) {$\frac{\Delta}{2}$};
\node at (5.2,-0.35) {$\frac{\Delta}{2}$};
\node at (1,0.25) {$\theta_2$};
\node at (1,1.2) {$\theta_0$};
\node at (4.5,0.25) {$\theta_4$};
\node at (4.5,1.2) {$\theta_3$};

\end{tikzpicture}
\ee
In summary, after using the chord rules to rewrite the original expectation value of dressed operators in the KM state in terms of ordinary chord operators in the tracial state, the crossed four-point function can be expressed as
\be
G_{4}^{({\rm crossed})}
=\int_{0}^{\pi}
\prod_{i=0}^{5}\dd\mu(\te_{i})
e^{-\tau_{1}E_{1}-\tau_{2}E_{2}-\tau_{4}E_{4}-\tau_{5}E_{5}}
\gamma_{\te_{0}\te_{1}}^{2}
\gamma_{\te_{3}\te_{5}}^{2}
\times I.
\ee
In the following, we evaluate $I$ explicitly using the matrix elements of the crossing operator $\chi_{\Delta/2}(\tau)$.

The analytic structure of the crossing kernel $\chi_{\Delta/2}(\tau)$, viewed as an operator on the one-particle Hilbert space $\mathcal{H}_1$, was studied in~\cite{Cao:2025pir}. In the chord-number basis, its matrix elements are controlled by how the outgoing $H$-chords are distributed as they traverse the crossing region. For our purposes, we will use its matrix elements in the one-particle energy eigenbasis, which we derive in Appendix~\ref{app:useful}. They take the form
\be
\begin{aligned}
\left\langle\Delta ; \phi_1, \phi_2\right|
\chi_{\Delta}\left(\tau\right)
\left|\Delta ; \phi_4, \phi_3\right\rangle
& =q^{\Delta^2}
\frac{\delta\left(\phi_2-\phi_3\right)}{\mu\left(\phi_2\right)}
\int \mathrm{d} \mu(\phi)
\,e^{-\tau E(\phi)} \\
& \quad \times
\gamma_{\phi_1 \phi_2}
\gamma_{\phi_2 \phi_4}
\gamma_{\phi_4 \phi}
\gamma_{\phi \phi_1}
\left\{
\begin{array}{ccc}
\frac{\Delta}{2} & \phi_1 & \phi_2 \\
\frac{\Delta}{2}  & \phi & \phi_4
\end{array}
\right\}_q.
\end{aligned}
\ee
We now evaluate the middle factor $I$ in \eqref{eq:def-I}. Inserting the resolution of identity \eqref{eq:resolve-1}, and using the fact that matrix elements of $q^{\Delta\nt}$ produce quantum $6j$ symbols as in \eqref{eq:slicing-4pt-cr}, we obtain
\be
\begin{aligned}
I
 =q^{\Delta^2}
\int_0^\pi
\prod_{i=1}^4 \mathrm{~d} \mu\left(\phi_i\right)
& e^{-\tau_3 E\left(\phi_3\right)}  \gamma_{\phi_2 \theta_2}
\gamma_{\theta_2 \theta_0}
\gamma_{\theta_0 \phi_1}
\gamma_{\phi_1 \phi_3}
\gamma_{\phi_3 \phi_4} 
\gamma_{\phi_4\theta_4}
\gamma_{\theta_4 \theta_3}
\gamma_{\theta_3\phi_2} \times 
\\
& \quad \times
\left\{
\begin{array}{ccc}
\frac{\Delta}{2} & \phi_1 & \phi_3 \\
\frac{\Delta}{2}  & \phi_2 & \phi_4
\end{array}
\right\}_q
\left\{
\begin{array}{ccc}
\frac{\Delta}{2} & \theta_4 & \theta_3 \\
\frac{\Delta}{2} & \phi_4 & \phi_2
\end{array}
\right\}_q
\left\{
\begin{array}{ccc}
\frac{\Delta}{2} & \phi_2 & \theta_2 \\
\frac{\Delta}{2} & \phi_1 & \theta_0
\end{array}
\right\}_q.
\end{aligned}
\ee
A chord-diagram representation of this decomposition is shown in Figure~\ref{fig:G4-crossed}:
\be\label{fig:G4-crossed}
I =\int_0^\pi \prod_{i=1}^4 d \mu\left(\phi_i\right) e^{-\tau_3 E\left(\phi_3\right)}\,
\begin{tikzpicture}[scale=1.2,line cap=round,line join=round,baseline=-0.5ex]

  \def\R{1.0}

  \coordinate (P) at (90:\R);
  
  \draw[line width=1.2pt]
    (1,0) arc[start angle=0,end angle=360,radius=\R];

  \fill (P) circle (0.045);
  \node[above=1pt] at (P) {$P_{\omega}$};

    \draw[line width=.75pt,dashed] (P) -- (-30:1);
    \draw[line width=.75pt,dashed] (P) -- (210:1);

    \draw[blue, line width=1.2pt] (45:1) -- (225:1);
    \draw[blue,line width=1.2pt]  (135:1) -- (315:1);

    \node at (65:1.2) {$\theta_2$};
    \node at (5:1.25) {$\theta_0$};
    \node at (175:1.25) {$\theta_4$};
    \node at (115:1.2) {$\theta_3$};
    
    \node at (90:0.45) {$\phi_2$};
    \node at (-90:0.45) {$\phi_3$};
    \node at (0:0.35) {$\phi_1$};
    \node at (180:0.35) {$\phi_4$};

\end{tikzpicture}
\ee
Substituting this result for $I$ into the expression for $G_4^{(\text{crossed})}$ obtained in the previous subsection, we arrive at 
\be \label{eq:G4-cr-answer}
\begin{aligned}
G_4^{\text {(crossed) }}
& =q^{\Delta^2}
\int_0^\pi \prod_{i=0}^5 \mathrm{~d} \mu\left(\theta_i\right)
\int^\pi_0\prod_{i=1}^4 \mathrm{~d} \mu\left(\phi_i\right)
e^{-\tau_3 E\left(\phi_3\right)}
e^{-\sum_{i=1,2,4,5}\tau_i E\left(\theta_i\right)} \\
& \quad \times
\gamma_{\theta_0 \theta_1}^2
\gamma_{\theta_3 \theta_5}^2
\gamma_{\phi_2 \theta_2}
\gamma_{\theta_2 \theta_0}
\gamma_{\theta_0 \phi_1}
\gamma_{\phi_1 \phi_3}
\gamma_{\phi_3 \phi_4} 
\gamma_{\phi_4\theta_4}
\gamma_{\theta_4 \theta_3}
\gamma_{\theta_3\phi_2} \\
& \quad \times\left\{
\begin{array}{ccc}
\frac{\Delta}{2} & \phi_1 & \phi_3 \\
\frac{\Delta}{2} & \phi_2 & \phi_4
\end{array}
\right\}_q
\left\{
\begin{array}{ccc}
\frac{\Delta}{2} & \theta_4 & \theta_3 \\
\frac{\Delta}{2} & \phi_4 & \phi_2
\end{array}
\right\}_q
\left\{
\begin{array}{ccc}
\frac{\Delta}{2} & \phi_2 & \theta_2 \\
\frac{\Delta}{2} & \phi_1 & \theta_0
\end{array}
\right\}_q.
\end{aligned}
\ee
Overall, we have used the chord rules to rewrite $G_4^{(\text{crossed})}$ as an expectation value of ordinary chord operators in the tracial state:
\be
\begin{split}
  G_4^{(\text{crossed})} = & \la\omega|
\wick{ e^{-\tau_{1}H}\c1 Me^{-\tau_{2}H}\c2 Me^{-\tau_{3}H}\c1 M^{\da}e^{-\tau_{4}H}\c2 M^{\da}e^{-\tau_{5}H} }
|\omega\ra  \\
= & 
\la 0 |
\wick{\c3 O \c1 O e^{-\tau_1 H} \c1 O \c2 O e^{-\tau_2 H}
\c3  O \c1 O e^{-\tau_3 H}
\c2 O \c3 O e^{-\tau_4 H}
\c1 O  \c1 O e^{-\tau_5 H}
\c1 O \c3 O}
|0\ra,
\end{split}
\ee
where in the second line we used the tracial property of $|0\ra$ to cyclically permute two of the condensate operators to the far right, adjacent to the ket state, to make the expression more symmetric. The essential step throughout these sections is purely kinematical: we rewrite KM expectation values of dressed chord operators as expectation values of ordinary chord operators in the tracial state. Concretely, the projection onto the KM state is replaced by a condensate of matter operators, and each dressed insertion $M$ or $M^\dagger$ is represented by a pair of ordinary matter chords, one of which contracts with this condensate in an appropriate way. With these replacements, the original KM correlator reduces to a standard tracial correlator with a fixed pattern of Wick contractions dictated by the chord diagram (uncrossed or crossed), and the remaining nontrivial dynamics is entirely captured by the corresponding chord diagrams.

\section{Exploring the Semiclassical Regime} \label{sec:classical}
In this section, we study the semiclassical $\lambda \rightarrow 0$ limit of the finite-temperature correlators and show that they factorize into products of thermal two-point functions. We then consider the Schwarzian regime near the spectral edge and show that the dressed two-point function reduces to a vertex function in JT gravity coupled to matter.

\subsection{Semi-classical Limit of Euclidean Correlators of Dressed Operators}

In this section, we study the semi-classical limit of correlation functions involving the dressed operators $M$ and $M^\dagger$. We will show that, although these correlators are not thermal correlators in the original formulation, their semi-classical limit factorizes into a product of thermal correlators of the ordinary matter chord operators. 

We begin with the following two-point function of dressed operators:
\be \label{eq:G2-def0}
G_{2}(\beta_1,\beta_2,\beta_3)=\la\omega|\wick{ e^{-\beta_{1}H} \c1 Me^{-\beta_{2}H}\c1 M^{\da}e^{-\beta_{3}H} }|\omega\ra.
\ee
In the previous section, we derived an exact representation for $G_2$ by diagonalizing the Boltzmann factors in the fixed-energy basis and integrating against the matter density. The result is
\be
G_2= \int\prod_{i=1}^{4}\dd\mu(\theta_{i})e^{-\sum_{i=1}^{3}\beta_{i}E_{i}}\mn_{\frac{\Delta}{2}}^{3}\prod_{i=1}^{3}\frac{\Gamma_{q}(\frac{\Delta}{2}\pm\frac{i\theta_{i}\pm i\te_{4}}{\lambda})}{\Gamma_{q}(\Delta)}.
\ee
Here, the variables $\theta_i$ are the spectral parameters associated with the boundary energy basis, and the matter-dependent factors couple $\theta_4$ to $\theta_1,\theta_2,\theta_3$.

In the $\lambda\to0$ limit with the inverse temperatures $\beta_i$ held fixed, the matter density strongly suppresses configurations with large differences $\theta_i-\theta_4$. This suggests that the dominant configurations satisfy
\be
\te_{i}=\te+\lambda\alpha_{i},\quad i=1,2,3,4,
\ee
with $\theta$ and $\alpha_i$ of order $\mo(\lambda^0)$. Using this parametrization, we can rewrite the integral as
\be \label{eq:G2-def1}
\begin{aligned}
G_2=\lambda \mathcal{N}_q^4 \mathcal{N}_{\frac{\Delta}{2}}^3 & \int_0^\pi \frac{\mathrm{d} \theta}{\prod_{j=1}^{4} \Gamma_q( \pm (\frac{2 i \theta}{\lambda} + 2i \alpha_j ))} e^{-\frac{2 \beta \cos \theta}{\lambda}} \prod_{j=1}^3 \Gamma_q\left(\frac{\Delta}{2} \pm\left(\frac{2 i \theta}{\lambda}+i \alpha_{j+4}\right)\right) \\
& \times \int_{-\frac{\pi}{\lambda}}^{\frac{\pi}{\lambda}} \mathrm{d} \alpha_4 \prod_{i=1}^3 \mathrm{~d} \alpha_i \frac{\Gamma_q\left(\frac{\Delta}{2} \pm i \alpha_{i 4}\right)}{\Gamma_q(\Delta)} e^{\sum_{i=1}^3(2 \sin \theta) \beta_i \alpha_i} \delta\left(\sum_{i=1}^4 \alpha_i\right),
\end{aligned}
\ee
where, for convenience, we defined
\be
\beta= \sum_{j=1}^3 \beta_j, \quad \alpha_{i4}= \alpha_i -\alpha_4, \quad \alpha_{j+4}= \alpha_j+\alpha_4,
\ee
and rewrote the spectral measure in terms of $\Gamma_q$ functions as
\be
\dd\mu(\theta)=\frac{\mathcal{N}_{q}}{\Gamma_{q}(\pm\frac{2i\theta}{\lambda})}\dd\theta,\quad\mathcal{N}_{q}=\frac{(1-q)^{2}\left(q;q\right)_{\infty}^{3}}{2\pi}.
\ee
The organization of \eqref{eq:G2-def1} is useful for the saddle-point analysis. For $\theta,\Delta,\alpha_i=\mo(\lambda^0)$, the first line contributes to the exponentially large factor of the form $\exp(\mo(\lambda^{-1}))$, and therefore determines the saddle point of the $\theta$-integral. By contrast, the second line contributes only at order $\mo(\lambda^0)$ and does not affect the saddle-point location.

To extract the leading semi-classical behavior, we use the $\lambda\to0$ asymptotics of the $q$-Gamma functions:
\be\label{eq:gamma-ratio}
\frac{\Gamma_{q}(\pm(\frac{2i\theta}{\lambda}+2i\alpha))}{\Gamma_{q}(\pm\frac{2i\theta}{\lambda})}\simeq e^{-2(\pi-2\theta)\alpha},\quad\frac{\Gamma_{q}(\frac{\Delta}{2} \pm(\frac{2i\theta}{\lambda}+i\alpha_{i+4}))}{\Gamma_{q}(\pm\frac{2i\theta}{\lambda})}\simeq\left(\frac{2\sin\theta}{\lambda}\right)^{\Delta}e^{-(\pi-2\theta)(\alpha_{i}+\alpha_{4})},
\ee
where $\simeq$ means equality at leading order in $\lambda$. These relations allow us to rewrite the dominant part of the integrand in terms of the standard spectral-density factor $\Gamma_q(\pm 2i\theta/\lambda)$, times an overall factor of order $\mo(\lambda^0)$.

Substituting \eqref{eq:gamma-ratio} into \eqref{eq:G2-def1}, the correlator simplifies to
\be \label{eq:G2-def2}
\begin{aligned}
G_2 & =\lambda \mathcal{N}_q^4 \mathcal{N}_{\frac{\Delta}{2}}^3 \int_0^\pi \frac{\mathrm{d} \theta}{\Gamma_q\left( \pm \frac{2 i \theta}{\lambda}\right)} e^{-\frac{2 \beta \cos \theta}{\lambda}} \\
& \times \int_{-\frac{\pi}{\lambda}}^{\frac{\pi}{\lambda}} \delta\left(\sum_{i=1}^4 \alpha_i\right)  \mathrm{d} \alpha_4  \left[ \prod_{i=1}^3 \mathrm{~d} \alpha_i \frac{\Gamma_q\left(\frac{\Delta}{2} \pm i \alpha_{i 4}\right)}{\Gamma_q(\Delta)}   \right] e^{2 \sin \theta\sum_{i=1}^3 \beta_i \alpha_i-  2(\pi - 2\theta) \alpha_4},
\end{aligned}
\ee
where the first line is precisely the standard double-scaled SYK partition-function integrand at inverse temperature $\beta$. Its saddle-point analysis is well known~\cite{Goel:2023svz,Baur:2023tcg,Sergio:2025Complexity-Dictionary}. Introducing
\[
\theta = \frac{\pi}{2} (1+v),
\]
the saddle-point equation of  $\theta$ becomes
\be \label{eq:saddle-beta}
\beta = \frac{\pi v}{ \cos \frac{\pi v}{2}},
\ee
which determines $v$ as a function of the total inverse temperature $\beta$. Evaluating the $\theta$-integral at the saddle and substituting back into \eqref{eq:G2-def2}, we obtain
\be
G_2 \simeq Z(\beta)\int_{-\infty}^{\infty}\dd\nu_{4}\prod_{i=1}^{3}\dd\nu_{i}\frac{\Gamma_{q}(\frac{\Delta}{2}\pm i\nu_{i})}{\Gamma_{q}(\Delta)}e^{ 2\cos\frac{\pi v}{2} \sum_{i=1}^{3}\beta_{i}\alpha_i+2 \beta\cos\frac{\pi v}{2} \alpha_{4}}\delta(\nu_{1}+\nu_{2}+\nu_{3}+4\nu_{4}),
\ee
where we introduced the variables
\be
\nu_{i}=\alpha_{i}-\alpha_{4},\quad\nu_{4}=\alpha_{4}.
\ee
This change of variables isolates the relative modes $\nu_i$ from the common shift $\nu_4$. The $\nu_4$ integral is then fixed by the delta function, which imposes
\[
\nu_4=-\frac{1}{4}\sum_{i=1}^3 \nu_i.
\]
After substituting this back, the remaining integral factorizes manifestly:
\be \label{eq:G2-result2}
Z(\beta)^{-1} G_{2}=\prod_{i=1}^{3}\left[\int\dd\nu_{i}\frac{\Gamma_{q}(\frac{\Delta}{2}\pm i\nu_{i})}{\Gamma_{q}(\Delta)}e^{(2\cos\frac{\pi v}{2})\tau_{i}\nu_{i}}\right] = G_{\frac{\Delta}{2}}(\tau_1)G_{\frac{\Delta}{2}}(\tau_2)G_{\frac{\Delta}{2}}(\tau_3),
\ee
where the Euclidean separations $\tau_i$ are
\be
\tau_{1}=\beta_{1}-\frac{\beta}{2},\quad\tau_{2}=\beta_{2}-\frac{\beta}{2},\quad\tau_{3}=\beta_{3}-\frac{\beta}{2}.
\ee
This result shows that the dressed two-point function $G_2$ factorizes in the semi-classical limit into a product of three thermal two-point functions of the ordinary matter operators, multiplied by the thermal partition function $Z(\beta)$.

As a simple consistency check, if any $\beta_i=0$, then the corresponding $\tau_i=-\frac{\beta}{2}$, and the standard normalization implies $G_{\frac{\Delta}{2}}(\tau_i)=1$, as expected from the standard normalization of fermion operators.

The above analysis generalizes to arbitrary higher-point functions with arbitrary crossing patterns. In conventional double-scaled SYK, this is the regime in which one recovers large-$p$ SYK~\cite{Baur:2023tcg}. For crossed matter lines in the bulk, a careful analysis of the semi-classical limit of the quantum $6j$ symbol around the saddle~\eqref{eq:saddle-beta} shows that it reduces to a delta function imposing conservation of the incoming and outgoing energy fluctuations, defined by the $\nu_i$ above. This is reminiscent of the generalized free-field limit of JT gravity coupled to matter~\cite{jafferis2023}. Since we have shown that the $2n$-point functions of dressed operators can be mapped to $6n$-point functions of ordinary correlators in double-scaled SYK with a particular configuration, it follows that the $2n$-point function of any fixed Wick-contraction pattern of dressed operators factorizes into a product of $3n$ thermal two-point functions $G_{\Delta/2}$ with the corresponding  Wick pairing configuration.

\subsection{Schwarzian limit of Correlators of Dressed Operators}

It is also instructive to study the Schwarzian limit of correlators of dressed operators, obtained by zooming into the edge of the spectrum as $\lambda\to0$. In this subsection, we analyze the two-point function~\eqref{eq:G2-def0} in this regime.

We introduce the following variables:
\be
\te_{i}=\pi-\lambda s_{i},\quad\tilde{\beta}_{i}=\lambda\beta_{i},
\ee
and expand the energy around the ground state:
\be
E_{i}=\frac{2\cos\theta_i}{\lambda}\simeq-\frac{2}{\lambda}+\lambda s^{2}_i,
\ee
where the first term is the constant (negative) ground-state energy. In what follows, we subtract this constant shift, i.e. we redefine the zero of energy so that the spectral density starts at $s=0$. The remaining quadratic term captures the nontrivial $\mo(\lambda)$ energy fluctuations above the ground state.

With this scaling, the two-point function~\eqref{eq:G2-def0} reduces to
\be
\begin{aligned}
G_2 & \simeq   \int_0^{\infty} e^{-\sum_{i=1}^3 \tilde{\beta}_i s_i^2} \prod_{i=1}^4 \frac{\mathrm{~d} s_i}{\Gamma\left( \pm 2 i s_i\right)} \prod_{i=1}^3 \frac{\Gamma\left(\frac{\Delta}{2} \pm i s_i \pm i s_4\right)}{\Gamma(\Delta)} \\
& =\int_0^{\infty} \prod_{i=1}^3 \rho\left(s_i\right) \mathrm{d} s_i e^{-\tilde{\beta}_i s_i^2} V_{\Delta}\left(s_1, s_2, s_3\right),
\end{aligned}
\ee
where the spectral density becomes the standard Schwarzian spectral density,
\be
\rho(s)=\frac{1}{\Gamma(\pm2is)}=\frac{2 s \sinh 2\pi s }{\pi},
\ee
and, more importantly, the remaining factor is the vertex function
\be \label{eq:V-def-limit}
V_{\Delta}(s_{1},s_{2},s_{3})=\int\dd s_{4}\rho(s_{4})\prod_{i=1}^{3}\frac{\Gamma(\frac{\Delta}{2}\pm is_{i}\pm is_{4})}{\Gamma(\Del)}.
\ee
This vertex function originates from the Schwarzian-dressed three-point function of primary operators. It can be represented as
\be \label{eq:V-def}
\begin{aligned}
V_{\Delta_1, \Delta_2, \Delta_3}\left(s_1, s_2, s_3\right) & =\int \mathrm{d} l_{12} \mathrm{~d} l_{23} \mathrm{~d} l_{31} 2 K_{2 i s_3}\left(4 e^{-l_{12} / 2}\right) 2 K_{2 i s_1}\left(4 e^{-l_{23} / 2}\right) 2 K_{2 i s_2}\left(4 e^{-l_{31} / 2}\right) \\
& \times \Upsilon\left(l_{12}, l_{23}, l_{31}\right)\left(2 e^{-l_{12} / 2}\right)^{\Delta_{1+2-3}}\left(2 e^{-l_{23} / 2}\right)^{\Delta_{2+3-1}}\left(2 e^{-l_{31} / 2}\right)^{\Delta_{3+1-2}},
\end{aligned}
\ee 
where $\Delta_{i+j-k}\equiv \Delta_i + \Delta_j -\Delta_k$. The structure of \eqref{eq:V-def} has a clear geometric interpretation. The boundary three-point function depends on the relative separations between operator insertions on the physical boundary, encoded by the weight-dependent exponentials in the second line. In the nearly AdS$_2$ gravity, the insertion locations are promoted to dynamical variables of the boundary particle, and one integrates over all boundary trajectories modulo $SL(2,\mathbb{R})$ redundancy, weighted by the Euclidean propagation amplitude of the boundary particle \cite{Yang:2018gdb}. 

One can use invariant variables such as the renormalized geodesic distances $l_{ij}$, as in \eqref{eq:V-def}, and integrate them against the gravitational wavefunction \cite{penington2020replica,Kolchmeyer:2023gwa}. This wavefunction can be viewed as consisting of two parts. The first is an exterior part, represented by the product of three $2K_{2is}$ functions in the first line of \eqref{eq:V-def}, which corresponds to the overlap between boundary-particle energy eigenstates and fixed geodesic-length states in the bulk Hilbert space. The second is an interior part, which computes the hyperbolic area contribution of the triangular region:
\be
\Upsilon\left(l_{12},l_{23},l_{31}\right)=\int_{0}^{\infty}\frac{\dd s}{\Gamma(\pm2is)}(2K_{2is}(4e^{-l_{12}/2}))(2K_{2is}(4e^{-l_{23}/2}))(2K_{2is}(4e^{-l_{31}/2})).
\ee
The equivalence between \eqref{eq:V-def} and \eqref{eq:V-def-limit} is seen by setting $\Delta_1=\Delta_2=\Delta_3=\Delta$, together with the identity
\be
\frac{\Gamma\left(\frac{\Delta}{2}\pm is_{1}\pm is_{2}\right)}{\Gamma(\Delta)}=\int_{-\infty}^{\infty}d\ell\, ( 2^{\Delta}e^{-\Delta\ell/2} ) \left(2K_{2is_{1}}(4e^{-\ell/2})\right) \left(2K_{2is_{2}} (4e^{-\ell/2} ) \right).
\ee
Therefore, the Schwarzian-limit expression of the dressed two-point function in the KM state is precisely the three-point function of boundary primaries dressed by Schwarzian modes, with all three primaries carrying the same conformal weight $\Delta$.

This match should not be surprising. As we already showed, the two-point function of dressed chord operators in the KM state can be written as three pairs of contracted ordinary matter chords, as suggested by~\eqref{eq:G2-def-wick}. In that representation, the corresponding chord diagram can be interpreted as the expectation value of three paired chords, sliced between an exterior state prepared by three fixed-energy states and the interior state $|\Upsilon\ra$ studied in \cite{Xu:2025Complexity}, which is viewed as a $q$-deformed version of \eqref{eq:V-def}.

It is natural to ask whether this correspondence extends to higher-point
functions, i.e. whether the \(2n\)-point functions of dressed operators in the
KM state reduce, in the Schwarzian regime, to the corresponding \(3n\)-point
functions of primaries in JT gravity coupled to matter for \(n>1\).  We do not
expect such an identification to hold in general.  The reason is that the higher-point JT+matter correlator contains more dynamical OPE data than are
present in the DSSYK dressed chord expansion.  In the double-scaled SYK correlator, generic operators are Gaussian random \(p\)-fermion operators, and
their double-scaling limit is governed by chord combinatorics: Wick contractions and crossing weights.  By contrast, a gravitational
\(3n\)-point function of matter primaries admits many inequivalent OPE channels,
with internal representations and OPE coefficients associated with different fusion trees.  These data are not captured by the universal chord rules of the
dressed SYK correlator.  Therefore the matching observed at the level of the
basic dressed two-point function should be viewed as a special kinematic
correspondence, rather than as evidence for a general equality of higher-point
functions. Thus, the match at $n=1$ should be regarded primarily as a kinematic correspondence, and it does not directly generalize to higher-point functions that probe the dynamical content of Schwarzian interactions.

This does not mean, however, that higher-point functions of dressed chord operators are irrelevant for gravity. On the contrary, the $2n$-point functions of these chord operators may capture particular OPE channels of the corresponding $3n$-point gravitational correlators, accounting for the chaotic nature of gravitational dynamics. It is suggested by the earlier results that crossed chord operators involve the quantum $6j$-symbol in the energy basis, which is known to reduce to the Dray-'t Hooft $S$-matrix element relevant for the near-horizon gravitational scattering~\cite{Maldacena:2016upp,Lam_2018,Penington:2025hrc}. It is therefore an interesting open question to understand the Schwarzian limit of higher-point correlators of dressed chord operators and to clarify their precise relation to gravitational correlators. We leave this for future work.

We conclude this section by recording a Mellin--Barnes representation for~\eqref{eq:V-def}:
\be
\begin{aligned}
& V_{\Delta}\left(s_1, s_2, s_3\right)=\frac{\Gamma\left(\Delta+i s_2 \pm i s_3\right) \Gamma\left(\Delta-i s_2 \pm i s_1\right)}{\Gamma(\Delta)} \\
& \quad \times \int_{-i \infty+c}^{i \infty+c} \frac{\mathrm{~d} u}{2 \pi i} \frac{\Gamma( \Delta+u) \Gamma\left(\Delta+i s_2 \pm i s_1+u\right) \Gamma\left(-i s_2 \pm i s_3-u\right) \Gamma(-u)}{\Gamma(2 \Delta+u) \Gamma( \Delta-u)},
\end{aligned}
\ee
where the contour is chosen with $-\Delta<c<0$. For a derivation, see equation (2.3) of \cite{Yurii2002}, and discussion in JT gravity in~\cite{Kolchmeyer:2023gwa}. Evaluating this contour integral by the residue theorem yields a sum of three ${}_4F_3(1)$ functions.

\section{Creating Wormhole-Brane State with Chords} \label{sec:branes}

It is natural to ask whether a small perturbation of the chord Hamiltonian can probe the thermal KM state. In this section we consider a deformation by the total length operator $q^{2\nt}$,
\be \label{eq:Hk-def}
H_{\ka}=a+a^{\da}+ \frac{\kappa}{\sqrt{1-q^2}} q^{2\nt},
\ee
and study how the spectrum depends on $\kappa$. Throughout this section, we adopt the convention $[a,a^\dagger]_{q^2}= 1$ which differs from that used elsewhere in the paper but simplifies the formulas below. The factor $1/\sqrt{1-q^2}$ is chosen so that $\kappa$ coincides directly with the deformation parameter of the continuous big $q$-Hermite polynomials, thereby giving it a simple interpretation.  Since we have already shown that $q^{2\nt}$ belongs to the (dressed) chord algebra, the deformed Hamiltonian $H_\kappa$ remains an element of the same algebra.

Moreover, since the generators of the dressed algebra arise as the double-scaling limit of disorder-averaged SYK operators, the deformation~\eqref{eq:Hk-def} may be viewed as the double-scaling limit of a corresponding microscopic deformation of the SYK Hamiltonian:\footnote{A different microscopic deformation that leads to similar double-scaled Hamiltonian has been considered in~\cite{Berkooz:2025ydg}. }
\be
H_{N, \kappa} = \sum_I J_I \Psi_I + \frac{\kappa}{\sqrt{1-q^2}} \sum_{I^\prime,J^\prime} M_{I^\prime} M^{*}_{J^\prime} \big(C_{I^\prime} C_{J^\prime}^{\dagger} + C_{J^\prime}^\dagger C_{I^\prime}\big). 
\ee
Here we have chosen the size of the dressed operators to be $p^\prime = 2p$, so that their emergent GNS representation in the double-scaling limit carries weight $\Delta=1$. With this choice, the GNS representation of the deformed Hamiltonian indeed takes the form~\eqref{eq:Hk-def}.

We analyze the spectrum of $H_\kappa$ in the chord Hilbert space $\mh_0$, the zero-particle sector of $\mathcal{H}_d$ that does not contain any $\tilde{H}$ chords. Relative to the undeformed case $\kappa=0$, the only change is that the recursion relation in the chord-number basis acquires a $\kappa$-dependent diagonal term that weights the number of chords:
\be\label{eq:recur}
H_\kappa |n\ra=\sqrt{[n+1]_{q^{2}}}|n+1\ra+
\frac{\kappa q^{2n}}{\sqrt{1-q^2}}|n\ra+ \sqrt{[n]_{q^{2}}}|n-1\ra,\quad n\geq0.
\ee
This recursion immediately reveals a $\mathbb{Z}_2$ redundancy: flipping the sign of $\kappa$ produces an equivalent operator, since the redefinition $|n\ra\mapsto(-1)^n|n\ra$ maps \eqref{eq:recur} with $\kappa$ to the same recursion with $-\kappa$. We therefore restrict to $\kappa>0$ in what follows.

The spectral structure depends qualitatively on the value of $\kappa$. For $0<\kappa<1$, the spectrum of $H_\kappa$ remains purely continuous and coincides with the $\kappa=0$ case. In particular,
\be
H_\kappa |\te\ra_{\ka}=E(\theta)|\te\ra_{\ka},\quad E(\theta)=\frac{2\cos\theta}{\sqrt{1-q^{2}}},
\ee
and the corresponding (deformed) energy eigenfunctions are
\be
\Psi_{n}(\theta)=\la n|\te\ra_{\kappa}=\frac{H_{n}(\cos\te,\kappa |q^{2})}{\sqrt{\left(q^2;q^2\right)_{n}}},
\ee
where $H_n$ denotes the $n$-th continuous big-$q$ polynomial. The spectral density is also deformed~\cite{okuyama2023end}:
\be
\dd\mu_{\kappa}(\theta)=\rho_{\kappa}(\theta)\dd\theta=\frac{\left(e^{\pm2i\theta},q^{2};q^{2}\right)_{\infty}}{2\pi(\kappa e^{\pm i\theta};q^{2})_{\infty}} \dd \theta,
\ee
and the wavefunctions satisfy the orthonormality relation
\be \label{eq:orth-kappa}
\int^\pi_0 \dd\mu_\kappa (\theta) \Psi^{*}_n (\theta) \Psi_m (\theta) =\delta_{mn}. 
\ee
Equivalently, the continuous spectrum provides a resolution of the identity in $\mh_0$:
\be
\int^{\pi}_0 \dd\mu_\kappa (\theta) {}_\kappa|\te\ra \la \te|_\kappa = \mathbf{1}_{\mh_0}.
\ee

The more interesting regime is $\kappa>1$, where a discrete component of the spectrum emerges. This discrete sector will later be interpreted as bound states once we take the Schwarzian limit. The bound states obey
\be
H_\ka |u_{l}\ra=-\frac{2\cosh u_{l}}{\sqrt{1-q^{2}}}|u_{l}\ra,\quad u_{l}= \ln(\ka q^{2l}),\quad l=0,1,\dots\lfloor\frac{\ln\ka}{-\ln q^2}\rfloor
\ee
with $\lfloor x\rfloor$ the integer part of $x$. The spectral decomposition of the identity now involves, in addition to the continuum, a discrete sum over these bound states:
\be \label{eq:id-bound}
\begin{aligned}
\delta_{m n} & =\int_0^\pi \mathrm{d} \mu_\kappa(\theta) \frac{H_n\left(\cos \theta, \kappa \mid q^2\right) H_m\left(\cos \theta, \kappa \mid q^2\right)}{\left(q^2 ; q^2\right)_m} \\
& +\sum_{0 \leq l< \ln |\kappa|/(2\lambda)} \omega_\kappa(l) \frac{H_m\left(\cosh u_l ,\kappa \mid q^2\right) H_n\left(\cosh u_l , \kappa \mid q^2\right)}{\left(q^2 ; q^2\right)_m},
\end{aligned}
\ee
where the bound-state weights are
\be \label{eq:discrete-weight}
\omega_{\kappa}(l)=\left(\frac{-1}{\kappa^{4}}\right)^{l} q^{-3l^{2}-l} \left(\kappa^{-2};q^{2}\right)_{\infty} \frac{(1-\kappa^{2}q^{4l})(\kappa^{2};q^{2})_{l}}{(1-\kappa^{2})(q^{2};q^{2})_{l}}.
\ee
The appearance of the discrete sum in \eqref{eq:id-bound} can be derived by analytically continuing the orthogonality relation \eqref{eq:orth-kappa} in $\kappa$. Concretely, the $\theta$ integral can be rewritten as a contour integral in $z=e^{i\theta}$ around the unit circle. For $\kappa<1$, all poles from the spectral density lie inside the unit disk. As $\kappa$ increases past the critical value $\kappa=1$, some poles move outside the unit disk. If we nevertheless keep the unit-circle contour prescription for all $\kappa$, then analytically continuing across $\kappa=1$ requires adding the residue contributions of the poles that exit the contour. These residues produce precisely the discrete contribution in the second line of \eqref{eq:id-bound}. A detailed derivation is provided in Appendix~\ref{app:spectral}.

Having explicit eigenfunctions and spectral measures is useful because they provide analytic control over the wormhole length operator $q^{ \hat{n}}$. In conventional SYK, its matrix elements in the energy basis compute the corresponding two-point function, and its relation to Krylov complexity and scrambling has been explored extensively in recent work~\cite{Sergio:2025Complexity-Dictionary, Ambrosini:2024sre,Heller:2024ldz,Xu:2024gfm,Rabinovici:2023yex,Xu:2025Complexity,Fu:2025kkh,Rabinovici:2025otw}. In the present deformed setup, which we will later interpret as describing wormhole states with an End-of-the-World brane insertion,\footnote{The interpretation of~\eqref{eq:recur} as the insertion of an End-of-the-World brane on one side of the boundary in double-scaled SYK was first suggested by Okuyama~\cite{okuyama2023end}. Its realization in sine-dilaton gravity, proposed as a bulk dual of double-scaled SYK, was subsequently explored in~\cite{Blommaert:2024puzzle,Blommaert:2025avl,Cui:2025sgy}.} the operator $\hat n$ measures the distance from the opposite boundary to the brane.\footnote{The matrix elements of $\hat n$ may be obtained by differentiating the corresponding matrix elements of $q^{\Delta \hat n}$ with respect to $\Delta$ and then setting $\Delta=0$.} A key qualitative distinction between the continuous and discrete parts of the spectrum is that the former is delta-function normalizable, while the latter consists of normalizable bound states. For this reason, the physically meaningful comparison on the continuous side is really in terms of wave packets rather than individual plane-wave eigenstates. Nevertheless, since the energy eigenstates in the continuous spectrum behave asymptotically like plane waves, one expects the corresponding length to diverge, whereas for bound states, which are localized, the expectation value of the length operator remains finite. Motivated by this distinction, we will derive exact expressions for matrix elements of $q^{2 \Delta \hat n}$ and then explore their physical implications, especially in the Schwarzian regime.

The matrix element of $q^{2\Delta \hat{n}}$ between continuous energy eigenstates is
\be\label{eq:kernel-1}
{}_\kappa \la\te_{1}|q^{2\Del\hat{n}}|\te_{2}\ra_\kappa=\sum_{n=0}^{\infty}q^{2\Del n}\Psi_{n}(\te_{1})\Psi_{n}(\te_{2})=\frac{\left(\kappa^2  q^{2\Del}e^{\pm i\theta_{1}},\kappa^2  q^{2\Del}e^{\pm i\theta_{2}};q^2 \right)_{\infty}}{\left(\kappa^2  q^{2\Del},q^{2\Del}e^{i\left(\pm\theta_{1}\pm\theta_{2}\right)};q^2 \right)_{\infty}}\mathcal{S}\left(\theta_{1},\theta_{2}\right),
\ee
where the symmetric kernel $\mathcal{S}(\theta_1,\theta_2)$ is
\be \label{eq:kernel-S}
\mathcal{S}(\theta_{1},\theta_{2})=\sum_{n=0}^{\infty}\frac{1-\ka^{2}q^{2\Del-2}q^{4n}}{1-\ka^{2}q^{2\Del-2}}\frac{\left(\ka^2 q^{2\Delta -2 },\ka e^{\pm i\theta_{1}},\ka e^{\pm i\theta_{2}};q^{2}\right)_{n}}{\left(q^{2\Del}\ka e^{\pm i\theta_{1}},q^{2\Del}\ka e^{\pm i\theta_{2}},q^2;q^{2}\right)_{n}}(-1)^{n}q^{n(n-1)}q^{4\Del n}.
\ee
This expression can be derived by noting that the continuous big-$q$ Hermite polynomials arise from the Al-Salam--Chihara polynomials upon setting one of their parameters to zero, which we derive in appendix~\ref{app:derivation-S}. Consequently, $\mathcal{S}$ follows from the Poisson kernel of the Al-Salam--Chihara polynomials in the corresponding limit.  Note that \eqref{eq:kernel-1} is manifestly symmetric under $\theta_1\leftrightarrow\theta_2$. We will therefore refer to $\mathcal{S}$ as the symmetric kernel, and distinguish it from the asymmetric kernel that appears in the subsequent discussion.


As a quick consistency check, when $\kappa=0$ one finds
\be
\lim_{\kappa\to0}\mathcal{S}(\theta_1,\theta_2)= \sum_{n=0}^{\infty} \frac{(-1)^n}{(q^2;q^2)_n} q^{n(n-1)}q^{4\Delta n} =\sum_{n=0}^{\infty} \frac{(-1)^n}{(q^2;q^2)_n}  q^{2\binom{n}{2}}(q^{4\Delta})^{n} = (q^{4\Delta};q^2)_\infty,
\ee
and ${}_\kappa \la \te_1 |q^{2\Delta \hat{n}}|\te_2\ra_\kappa$ reduces to the standard DSSYK result~\cite{Berkooz:2018qkz,Berkooz:2018jqr}.

The expression \eqref{eq:kernel-1} also applies to bound states, after analytically continuing $\theta_{1,2}$ to imaginary values, $\theta_{1,2}\to i u_{l_{1,2}}$. A notable simplification then occurs: for the bound-state parameters $u_l$, the symmetric kernel truncates because $(q^{-2l},q^2)_n\equiv0$ for $n>l$. As a result, $\mathcal{S}$ becomes a finite sum,
\be \label{eq:kernel-Su}
\begin{aligned}
\mathcal{S}\left(u_{l_1}, u_{l_2}\right) & =\sum_{n=0}^{\min \left(l_1, l_2\right)} \frac{1-\kappa^2 q^{2 \Delta-2+4 n}}{1-\kappa^2 q^{2 \Delta-2}}(-1)^n q^{n(n-1)+4 \Delta n} \\
& \times \frac{\left(\kappa^2 q^{2 \Delta-2}, \kappa^2 q^{2 l_1}, \kappa^2 q^{2 l_2}, q^{-2 l_1}, q^{-2 l_2} ; q^2\right)_n}{\left(q^2, \kappa q^{2 \Delta+2 l_1}, \kappa q^{2 \Delta+2 l_2}, q^{2 \Delta-2 l_1}, q^{2 \Delta-2 l_2} ; q^2\right)_n}.
\end{aligned}
\ee
The corresponding matrix element of $q^{2\Delta \hat{n}}$ between bound states is then
\be \label{eq:kernel-u1}
\la  u_{l_{1}}|q^{2\Delta\hat{n}}|u_{l_{2}}\ra=\frac{\left(\ka^{2}q^{2\Del+2l_{1}},\ka^{2}q^{2\Del+2l_{2}},q^{2\Del-2l_{1}},q^{2\Delta-2l_{2}};q^2\right)_{\infty}}{\left(\ka q^{2\Del},\ka^{2}q^{2\Del+2l_{1}+2l_{2}},\ka^{-2}q^{2\Del-2l_{1}-2l_{2}},q^{2\Del\pm(2l_{1}-2l_{2})};q^2\right)_{\infty}}\mathcal{S}\left(u_{l_{1}},u_{l_{2}}\right).
\ee
An alternative representation of \eqref{eq:kernel-1} is known in terms of a basic hypergeometric series ${}_3\phi_2$. This expression is often referred to as the non-symmetric Poisson kernel~\cite{askey1994qFourier}:
\be \label{eq:kernel-nons}
\sum_{n=0}^{\infty}q^{2\Del n}\Psi_{n}(\te_{1})\Psi_{n}(\te_{2})=\frac{\left(q^{4\Del},\ka q^{2\Del}e^{i\te_{2}},\ka e^{-i\te_{2}};q^{2}\right)_{\infty}}{\left(q^{2\Del}e^{\pm i\te_{1}\pm i\te_{2}};q^{2}\right)_{\infty}}\thinspace_{3}\phi_{2}\left(\begin{array}{c}
q^{2\Del},q^{2\Del}e^{\pm i\theta_{1}+i\te_{2}}\\
q^{4\Del},\ka q^{2\Del}e^{i\te_{2}}
\end{array};q^{2},\ka e^{-i\te_{2}}\right).
\ee
The term ``non-symmetric'' emphasizes that the right-hand side is not manifestly symmetric under $\theta_1\leftrightarrow\theta_2$, even though the left-hand side is. Despite this, \eqref{eq:kernel-nons} is particularly useful in the JT/Schwarzian analysis, where its asymptotics match expressions derived in JT gravity with end-of-the-world branes.

It is also worth noting a practical distinction at finite $q$: as a series expansion (for generic allowed parameters), the symmetric kernel \eqref{eq:kernel-S} typically converges much faster than the non-symmetric expansion \eqref{eq:kernel-nons}, due to the additional $q^{n(n-1)}$ suppression in \eqref{eq:kernel-S}. In what follows, we will find the bound-state form \eqref{eq:kernel-Su} especially convenient for analyzing the behavior of the wormhole length operator $\hat{n}$ in the discrete sector.

\subsection{Distance to the End of the World in the Quantum Regime}

The expectation value of $\hat{n}$ in both the continuous and bound-state sectors can be extracted from the matrix elements of $q^{2\Delta \hat{n}}$ in the corresponding energy eigenstates. We will show that for finite $0<q<1$ the expectation value diverges in fixed-energy (continuous) states, while it remains finite in bound states. Concretely, we diagnose this behavior by studying the $\Delta\to0$ limit of $\la q^{2\Delta \hat{n}}\ra$.

For a fixed-energy state, we use \eqref{eq:kernel-nons} with $\theta_1=\theta_2=\theta$. In this case the ${}_3\phi_2$ function is regular as $\Delta\to0$:
\be
{ }_3 \phi_2\binom{q^{2 \Delta}, q^{2 \Delta} e^{ \pm i \theta_1+i \theta_2}}{q^{4 \Delta}, \kappa q^{2 \Delta} e^{i \theta_2} ; q^2, \kappa e^{-i \theta_2}} \stackrel{\Delta\to0}{=} 1 + \mathcal{O}(\Delta),
\ee
which follows directly from its series expansion in the argument $\kappa e^{-i\theta_2}$: the leading term is $1$, and for $n\in\mbz_{n>0}$ the remaining terms carry a factor $(q^{2\Delta},q^2)_n\propto(1-q^{2\Delta})$, providing the $\Delta$ suppression. The singular behavior therefore arises entirely from the prefactor in \eqref{eq:kernel-nons}, giving
\be \label{eq:expectation-1}
{}_\kappa \la\te|q^{2\Del\hat{n}}|\te\ra_{\ka}=\frac{1}{\Delta}\frac{\left(\ka q^{2\Del}e^{\pm i\te};q^{2}\right)_{\infty}}{\left(q^{2},e^{\pm2i\theta};q^{2}\right)_{\infty}}+\mo(\Delta^{0}).
\ee
The divergence as $\Delta\to0$ therefore implies that $\la\te|\hat{n}|\te\ra$ diverges. Indeed, using \eqref{eq:expectation-1} and extracting $\la\hat{n}\ra$ by differentiating with respect to $\Delta$ at $\Delta=0$ yields an infinite result. By contrast, for bound states with $l_1=l_2=l$, the expectation value of $q^{2\Delta \hat{n}}$ is regular as $\Delta\to0$. This can be seen by analyzing both the kernel \eqref{eq:kernel-Su} and the prefactor in \eqref{eq:kernel-u1}. In particular, the truncated kernel admits the expansion
\be
\mathcal{S}(u_l, u_l)= \sum_{n=0}^{l}\frac{1-\ka^{2}q^{4n-2}}{1-\ka^{2}q^{-2}}\frac{\left(\ka^{2}q^{-2};q^{2}\right)_{n}}{\left(q^{2};q^{2}\right)_{n}}\left(-1\right)^{n}q^{n(n-1)} + \mo(\Delta) \equiv S_0(l) +\mo(\Delta),
\ee
while the prefactor remains finite because
\be
\frac{\left(q^{2\Del-2l};q^{2}\right)_{\infty}}{\left(q^{2\Del};q^{2}\right)_{\infty}}=\left(q^{2\Del-2l};q^{2}\right)_{l}\stackrel{\Del\to0}{=}(-1)^{l}q^{-l(l+1)}\left(q^{2};q^{2}\right)_{l}.
\ee
Combining these, we obtain
\be
\la u_l | q^{2\Delta\hat{n}}|u_l\ra = (-1)^l q^{-2l(l+1)}\frac{\left(q^{2};q^{2}\right)_{l}^{2}\left(\ka^{2}q^{2l};q^{2}\right)_{\infty}^{2}}{\left(\ka,\ka^{2}q^{4l},\ka^{-2}q^{-4l};q^{2}\right)_{\infty}} S_0(l)+\mo(\Delta). 
\ee
Because the $\Delta\to0$ limit is finite, the expectation value of $\hat{n}$ can then be extracted by differentiating with respect to $\Delta$ at $\Delta=0$. This can be done explicitly with the truncated result \eqref{eq:kernel-Su}, though we don't find the final result expressible as a compact closed form formula. We will find such a formula for the length expectation value in the Schwarzian regime in the following discussion.

\subsection{The Wormhole-Brane state in the Schwarzian Regime}

The Schwarzian limit of the eigenstates of the deformed Hamiltonian $H_\kappa$ with $|\kappa|<1$ was proposed in~\cite{okuyama2023end} to describe JT gravity quantized with both an asymptotic nearly AdS$_2$ boundary and an end-of-the-world (EOW) brane boundary. In this section, we instead focus on the regime $\kappa>1$, where bound states appear. These states will be interpreted as bulk geometries in which an EOW brane becomes naked on one side. 

Our main goal is to evaluate the expectation value of $e^{-\Delta \hat{l}}$ in both the continuous and bound-state sectors. These matrix elements control two-point functions in the semi-open channel. We also present an explicit expression for the expectation value of the renormalized length operator $\hat{l}$ in the bound states, where $\hat{l}$ is the length variable that emerges naturally in the Schwarzian limit. Finally, we briefly comment on two kinematical regimes corresponding to infinite brane tension and vanishing brane tension.

The Schwarzian limit of the wavefunction $\Psi_n(\theta)$ is implemented by taking $\lambda\to0$ together with the scaling
\be \label{eq:Sch-limit}
l=\lambda n+\ln\lambda,\quad\theta=\lambda k,\quad\kappa=q^{2\nu+1}.
\ee
Note that in the $\kappa=0$ case the renormalized length is usually defined as $\tilde{l}=\lambda n+2\ln\lambda$, where each $\ln\lambda$ subtracts an IR divergence associated with approaching one of the AdS boundaries~\cite{Lin:2022rbf}. In the present setup, the brane terminates at a finite proper distance in the bulk, and $l$ should be interpreted as the geodesic length that starts at the asymptotic boundary and ends orthogonally on the brane. Accordingly, only a single $\ln\lambda$ subtraction is needed to remove the divergence near the asymptotic boundary. This interpretation is consistent with canonically quantized JT gravity with a brane of tension parameterized by $\nu$, which motivates the proposal of~\cite{okuyama2023end}. We will adopt this viewpoint in our analysis of expectation values of the operator $\hat{l}$ in the gravitational states obtained from the Schwarzian limit of the finite-$q$ results.

Under the scaling~\eqref{eq:Sch-limit}, the recursion relation \eqref{eq:recur} for the continuous sector becomes the Liouville-type equation
\be\label{eq:brane-dynamics}
(-\partial_{l}^{2}+\nu e^{-l}+\frac{1}{4}e^{-2l})\psi_{k,\nu}(l)=k^{2}\psi_{k,\nu}(l),
\ee
where $\psi_{k,\nu}(l)$ arises as the Schwarzian limit of $\Psi_n(\theta)$. It describes a continuum eigenstate with positive energy $E_k=k^2$, and may be written in terms of Whittaker functions as
\be
\psi_{k,\nu}(l)=e^{l/2}W_{-\nu,ik}(e^{-l}).
\ee
We choose the normalization
\be
\int_{-\infty}^{\infty}\dd l\,\psi_{k_{1},\nu}(l)\psi_{k_{2},\nu}(l)=\frac{\delta(k_{1}-k_{2})}{\rho_{\nu}(k_{1})},\quad\rho_{\nu}(k)=\frac{\Gamma(\nu+\frac{1}{2}\pm ik)}{2\pi\Gamma(\pm2ik)}.
\ee

For $\kappa>1$, which corresponds to $\nu<-\frac{1}{2}$, the potential admits normalizable bound states. These arise from the Schwarzian limit of the discrete spectrum discussed above. Denoting the $n$-th bound state by $\psi_{n,\nu}(l)$, it satisfies
\be
(-\partial_{l}^{2}+\nu e^{-l}+\frac{1}{4}e^{-2l})\psi_{k_n,\nu}(l)=-k_n^{2}\psi_{k_n,\nu}(l),\quad \nu<-\frac{1}{2},
\ee
with discrete labels
\be
\psi_{k_n,\nu}(l)= e^{l/2} W_{-\nu, k_n}(e^{-l}),\quad k_n=-(\nu+n+\frac{1}{2}),\quad n=0,1,\dots \lfloor -\nu-\frac{1}{2}\rfloor,
\ee
and with the following normalization: 
\be \label{eq:discrete-weight-q1}
\int_{-\infty}^{\infty} \dd l \psi_{k_n, \nu}(l) \psi_{k_{m},\nu}(l) = \frac{\delta_{mn}}{w_{n,\nu}},\quad w_{n,\nu}=\frac{-2\nu -2n -1}{n! \Gamma(-2\nu -n)}.
\ee
The normalization factor above can of course be obtained directly from the standard normalization of Whittaker $W$-functions. Here we emphasize a complementary perspective: the same relations follow straightforwardly from taking the $q\to1$ limit of \eqref{eq:discrete-weight}, using the Schwarzian scaling \eqref{eq:Sch-limit} for the relevant $q$-functions. In particular, one finds
\be\label{eq:limit-1}
(\frac{-1}{\ka^{4}})^{l}q^{-3l^{2}-l}\to(-1)^{l},\quad\frac{1-\ka^{2}q^{4l}}{1-\ka^{2}}\to\frac{2\nu+2l+1}{2\nu+1},\quad\frac{\left(\ka^{2};q^{2}\right)_{l}}{(q^{2};q^{2})_{l}}\to\frac{(2\nu+1)_{l}}{l!},
\ee
together with the slightly more nontrivial limit
\be\label{eq:limit-2}
\frac{(q^{-4\nu-2};q^{2})_{\infty}}{(q^{2};q^{2})_{\infty}}=\frac{(1-q^{2})^{2+2\nu}}{\Gamma_{q^{2}}(-2\nu-1)}\to\frac{(2\lambda)^{2\nu+2}}{\Gamma(-2\nu-1)},
\ee
where in the first equality we used the definition of the $q$-Gamma function. Substituting \eqref{eq:limit-1} and \eqref{eq:limit-2} into \eqref{eq:discrete-weight} reproduces \eqref{eq:discrete-weight-q1}, up to an overall state-independent factor $\lambda^{4+4\nu}$. This factor can be absorbed into the definition of the Schwarzian limit of the wavefunctions when passing from continuous big $q$-Hermite polynomials to Whittaker functions.

\paragraph{The Distance to the Brane}

It is known that the expectation value of $\hat{l}$ in a fixed continuous eigenstate diverges~\cite{Maldacena-Qi:2018lmt}. Nevertheless, boundary correlation functions in the presence of an EOW brane are finite and are controlled by matrix elements of $e^{-\Delta\hat{l}}$ in the energy basis. These matrix elements can be obtained by taking the Schwarzian limit of our finite-$q$ expressions derived above. Using the basic limit
\be
\lim_{q\to1}\thinspace_{3}\phi_{2}\left(\begin{array}{c}
q^{a_{1}},q^{a_{2}},q^{a_{3}}\\
q^{b_{1}},q^{b_{2}}
\end{array};q,q^{x}\right)=\thinspace_{3}F_{2}\left(\begin{array}{c}
a_{1},a_{2},a_{3}\\
b_{1},b_{2}
\end{array};1\right),
\ee
the Schwarzian limit of \eqref{eq:kernel-nons} yields
\be \label{eq:expl-exact}
\begin{aligned}
\int^{\infty}_{-\infty} \mathrm{d} l e^{-\Delta l} \psi_{k, \nu}(l)^* \psi_{k^{\prime}, \nu}(l) & =\frac{\Gamma\left(\Delta \pm i k \pm i k^{\prime}\right)}{\Gamma(2 \Delta) \Gamma\left(\Delta+\nu+\frac{1}{2}-i k^{\prime}\right) \Gamma\left(\nu+\frac{1}{2}+i k^{\prime}\right)} \\
& \times { }_3 F_2\left(\begin{array}{c}
\Delta, \Delta \pm i k-i k^{\prime} \\
2 \Delta, \nu+\frac{1}{2}+\Delta-i k^{\prime}
\end{array} ; 1\right).
\end{aligned}
\ee
Setting $k'=k$ then exhibits the expected divergence as $\Delta\to0$ in the continuous sector:
\be
\int^{\infty}_{-\infty} \dd l e^{-\Delta l} \psi_{k,\nu}(l)^*  \psi_{k,\nu}(l) = \frac{2}{\Delta} \cdot \frac{\Gamma(\Delta \pm 2ik)}{\Gamma(\nu+\frac{1}{2}\pm ik)} +\mo(\Delta^0).
\ee
In contrast, for bound states the length expectation value is finite. It can be computed as:
\be
\la \hat{l}\ra_{n}=\frac{\int_{-\infty}^{\infty}\dd l\thinspace l\thinspace|\psi_{\nu,n}(l)|^{2}}{\int_{-\infty}^{\infty}\dd l\thinspace|\psi_{\nu,n}(l)|^{2}},
\ee
and one finds after some algebra that
\be
\la l\ra_{n}= - \psi(-2n-2\nu -1)  - \psi(-2n-2\nu)+ \psi(-n-2\nu) .
\ee
where $\psi(x)=(\ln \Gamma(x))^\prime$ is the digamma function. We include the detailed calculation of this result in appendix~\ref{app:distance}.

\paragraph{Infinite Brane Tension Limit}

Setting $\kappa=0$ turns off the deformation in \eqref{eq:Hk-def}, and the finite-$q$ wavefunctions reduce to those of the pure-gravity sector of double-scaled SYK. In relating the finite-$\kappa$ wavefunctions to their Schwarzian counterparts, we used the identification $\kappa=q^{2\nu+1}$ and interpreted $\nu$ as the brane tension. If the $\kappa\to0$ limit commutes with the Schwarzian limit~\eqref{eq:Sch-limit}, then $\kappa\to0$ corresponds to $\nu\to+\infty$, i.e. an infinite-tension brane that decouples from the dynamics. In that case one should recover the pure JT gravity sector. We now verify this expectation explicitly.

To compare more directly with wormhole dynamics in pure gravity, it is convenient to introduce a shifted length variable $\tilde{l}$ by
\be
l\to\tilde{l}+\ln\nu,\quad e^{-l}\to\nu^{-1}e^{-\tilde{l}}.
\ee
Under this redefinition, the Schr\"odinger equation \eqref{eq:brane-dynamics} becomes
\be
(-\partial_{\tilde{l}}^{2}+e^{-\tilde{l}}+\frac{1}{4\nu^{2}}e^{-2\tilde{l}})\psi_{k,\nu}(\tilde{l}+\ln\nu)=k^{2}\psi_{k,\nu}(\tilde{l}+\ln\nu).
\ee
In the $\nu\to\infty$ limit the $e^{-2\tilde{l}}$ term is suppressed, so the wavefunction should reduce to the pure-gravity solution. At the level of functional form, this suggests
\be\label{eq:expectation-2}
\psi_{k,\nu}(\tilde{l}+\ln\nu)\stackrel{\nu\to\infty}{\simeq}K_{2ik}(2e^{-\tilde{l}/2}).
\ee
To make this reduction precise, we start from the Mellin--Barnes representation
\be \label{eq:integral-psik1}
\psi_{k,\nu}(l)=\frac{e^{-e^{-l}/2}}{\Gamma(\nu+\frac{1}{2}\pm ik)}\int_{c-i\infty}^{c+i\infty}\frac{\dd u}{2\pi i}\thinspace e^{ul}\Gamma(u\pm ik)\Gamma(\nu-u),
\ee
where the contour satisfies $0<\Re(c)<\Re(\nu)$ so that it separates the poles of $\Gamma(u\pm ik)$ from those of $\Gamma(\nu-u)$. Expressing this in terms of $\tilde{l}$ and taking $\nu\to\infty$, the prefactors simplify as
\be
e^{-e^{-l}/2}\simeq e^{-e^{-\tilde{l}}/(2\nu)}=1+\mo(\nu^{-1}), \quad e^{u l} \Gamma(\nu-u)= e^{u\tilde{l}} \nu^u \Gamma(\nu -u)=e^{u\tilde{l}}\Gamma(\nu)(1+\mo(\nu^{-1})),
\ee
where we used the large-$\nu$ behavior of $\Gamma(\nu-u)$ for fixed $u=\mo(1)$. Substituting these estimates into \eqref{eq:integral-psik1} yields
\be
\psi_{k,\nu}(\tilde{l}+\ln\nu)\stackrel{\nu\to\infty}{\simeq}\frac{\Gamma(\nu)}{\Gamma(\nu+\frac{1}{2}\pm ik)}\int_{c-i\infty}^{c+i\infty}\frac{\dd u}{2\pi i}\thinspace e^{u\tilde{l}}\Gamma(u\pm ik)=\frac{\Gamma(\nu)}{\Gamma(\nu+\frac{1}{2}\pm ik)}\left(2K_{2ik}(e^{-\tilde l/2})\right),
\ee
which fixes the overall normalization left implicit in \eqref{eq:expectation-2}. Knowing this normalization is useful when taking the $\nu\to\infty$ limit of observables computed in the brane setup and comparing them directly to the pure-gravity results.

A physical interpretation of this limit was discussed in~\cite{Maldacena-Qi:2018lmt,penington2020replica} and later in the Lorentzian, canonically-quantized JT+EOW-brane framework of~\cite{Gao:2021uro}.~\footnote{The wormhole-brane states have also been explored in sine-dilaton gravity in \cite{Blommaert:2025avl,Blommaert:2024puzzle,Cui:2025sgy}} In the infinite-tension limit, the brane is forced to follow the minimal Euclidean geodesic anchored at the asymptotic boundary endpoints. Geometrically, this effectively pinches off the brane in the Euclidean saddle and thereby reproduces the pure JT gravity answer. 

Here, the identification $\kappa=q^{2\nu+1}$ provides a direct map between the brane tension $\nu$ in the JT description and the coupling constant $\kappa$ in the microscopic SYK deformation. In particular, the decoupling limit of the brane ($\nu\to+\infty$) is translated into turning off the deformation in the SYK Hamiltonian ($\kappa\to0$).  This relation provides a microscopic viewpoint on wormhole-brane states: the bulk configurations with an EOW brane can be understood as arising from specific eigenstates of the deformed (double-scaled) SYK Hamiltonian. Moreover, because the brane follows a geodesic whose position is controlled by the parameter $\nu$, the correspondence $\kappa=q^{2\nu+1}$ suggests that one can continuously move the brane in the bulk by tuning $\kappa$ on the SYK side.

\paragraph{Zero Brane Tension Limit}

It is also instructive to consider the limit $\nu\to0$, corresponding to a zero-tension EOW brane in the JT+brane description. In this limit, \eqref{eq:brane-dynamics} again reduces to the pure-gravity form, but with an effective Liouville potential proportional to $e^{-2l}$ rather than the more familiar $e^{-l}$. At first sight one might suspect that this is merely a reparametrization of the pure-JT dynamics. However, the geometric interpretation is different from the infinite-tension limit: when $\nu=0$ the brane does \emph{not} decouple, and $l$ should still be interpreted as the renormalized distance from the asymptotic boundary to the EOW brane. From the microscopic point of view this is also nontrivial, since $\nu=0$ corresponds to a specific finite value of the deformation parameter, $\kappa=q$, rather than the undeformed chord Hamiltonian.

The reason the resulting dynamics nevertheless matches the pure-JT length mode can be understood directly from the classical phase-space description of JT gravity with a boundary brane~\cite{Gao:2021uro}. The brane embedding is determined by the boundary conditions
\begin{align}
\nabla_a n^a = 0,
\qquad
n^a \partial_a \Phi = \mu,
\end{align}
where $n^a$ is the unit normal vector to the brane. On the other hand, the general dilaton solution can be written in embedding coordinates as
\begin{align}
\Phi = V^\mu Y_\mu .
\end{align}
Using the $SO(2,1)$ gauge symmetry, one may choose the dilaton profile to take the canonical form
\begin{align}
\Phi = \Phi_h \frac{\cos T}{\sin \sigma},
\end{align}
which is equivalent to fixing $V^\mu=(0,\Phi_h,0)$. This leaves a residual $SO(1,1)$ subgroup unbroken. The geodesic brane solutions may then be parameterized by
\begin{align}
U^\mu Y_\mu =0,
\qquad
U^\mu \propto (r\sin\theta,r\cos\theta,1),
\end{align}
and the remaining $SO(1,1)$ may be used to set $\theta=0$, so that the brane trajectory becomes
\begin{align}
\cos \sigma = r \cos T,
\qquad
r\in[-1,1].
\end{align}
Substituting this trajectory into the boundary condition $n^a\partial_a\Phi=\mu$ gives
\begin{align}
n^a\partial_a\Phi
=
\frac{r\Phi_h}{\sqrt{1-r^2}}
=
\mu,
\qquad\Longrightarrow\qquad
r=\frac{\mu}{\sqrt{\Phi_h^2+\mu^2}}.
\end{align}
Thus the brane position is not arbitrary: once the dilaton profile is fixed, the remaining gauge symmetry only relates equivalent representatives, while the physical location of the brane is determined by $\mu$ through the above relation.

In particular, when $\mu=0$ one finds
\begin{align}
r=0,
\end{align}
so the brane lies on the timelike geodesic through the center of AdS$_2$. Since in our gauge the center is precisely the extremum of the dilaton, the zero-tension brane is forced to pass through the dilaton saddle. This is the precise sense in which the $\nu=0$ boundary condition selects a distinguished timelike geodesic, rather than allowing a tensionless brane to wander freely throughout AdS$_2$.

Once the brane is fixed to this symmetric geodesic, the geometry acquires a natural $\mathbb{Z}_2$ reflection symmetry exchanging the two sides of the brane. In this symmetric sector, the renormalized distance $l$ from an asymptotic boundary point to the brane, defined using the geodesic that meets the brane orthogonally, can be reflected across the brane to produce the corresponding two-sided geodesic in pure JT gravity. The total wormhole length is therefore
\begin{align}
L = 2l,
\end{align}
so the operator $e^{-\Delta \hat l}$ may equivalently be regarded as $e^{-\frac{\Delta}{2}\hat L}$. This explains why the effective dynamics takes the pure-JT form despite the fact that the microscopic theory remains at a finite deformation $\kappa=q$.

Moreover, the one-sided boundary evolution preserves this $\mathbb{Z}_2$-symmetric sector, and therefore agrees with the symmetric two-sided evolution generated by $(H_L+H_R)/2$ in the pure-gravity description. In this sense, the zero-tension brane does not disappear, but rather picks out the special symmetric subsector in which the boundary-to-brane distance mode is identified with half of the standard wormhole length mode.

As a technical remark, one may also wonder how the matrix elements of $e^{-\Delta \hat{l}}$ reduce to the pure-gravity expressions at $\nu=0$. The reason is that at $\nu=0$ the hypergeometric function ${}_3F_2$ in~\eqref{eq:expl-exact} becomes well-poised and can be simplified using Watson's formula:
\begin{align}
{}_3F_2\!\left(
\begin{matrix}
\Delta,\ \Delta\pm ik-ik'\\
2\Delta,\ \Delta+\frac12-ik'
\end{matrix};1
\right)
=
\frac{\sqrt{\pi}\,
\Gamma\!\left(\frac{2\Delta+1}{2}\right)
\Gamma\!\left(\Delta+\frac12-ik'\right)
\Gamma\!\left(\frac12+ik'\right)}
{\Gamma\!\left(\frac{1+\Delta\pm ik\pm ik'}{2}\right)}.
\end{align}
Substituting this into~\eqref{eq:expl-exact} gives
\begin{align}
\int_{-\infty}^{\infty} \dd l\,
e^{-\Delta l}\,
\psi_{k,0}(l)^*\psi_{k',0}(l)
=
\frac{2^{2\Delta-3}}{\pi}
\frac{
\Gamma\!\left(\frac{\Delta}{2}\pm\frac{ik}{2}\pm\frac{ik'}{2}\right)
}{
\Gamma(\Delta)
}.
\end{align}
The effective conformal weight is therefore $\Delta/2$, in agreement with the geometric identification $l=L/2$.

\begin{figure}
    \centering
    \includegraphics[width=0.3\linewidth]{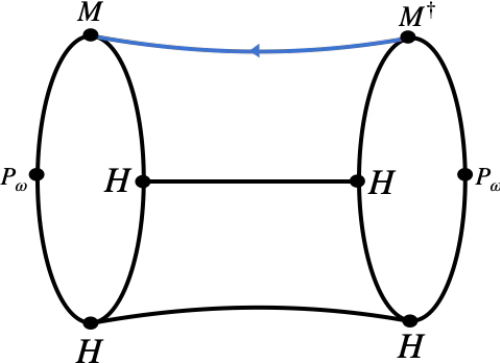}
    \caption{A wormhole-like contribution to the variance of the one-point function of $M$ in the KM state.}
    \label{fig:wormhole}
\end{figure}
\section{Global Symmetry Violation by Double-trace Wormholes} \label{sec:symmetry}
In section \ref{sec:KM}, we found that the dynamics of DSSYK about $\ket{\omega}$ preserves $U(1)$ symmetry. The $U(1)$ symmetry is only emergent in the double-scaled limit as we know that the SYK Hamiltonian has no such symmetry at finite $N$. In this section, we will analyze the violation of this symmetry at the leading order in $1/N$. A simple probe of symmetry violation is the one-point function of $M$ in the KM state. While the ensemble average of one-point function of $M$ vanishes, its variance is non-zero at finite $N$. Our goal in this section is to estimate the order of magnitude of the leading contribution to the variance in $1/N$ expansion. We will show that for a matter operator $M$ defined in  equation \eqref{eq:dressed_operators} as linear combination of products of $p'$ annihilation operators, its variance goes as
\begin{malign}
   \overline{|\bra{\beta}  M \ket{\beta}|^2} \sim N^{-(\text{max}(\lceil p'/p \rceil,2) p + p')/4} 
\end{malign}where $\lceil p'/p \rceil$ is the smallest integer greater than $p'/p$ or equal to $p'/p$.

Let us begin by analyzing the variance of $\bra{\omega}H^n M H^m\ket{\omega}$:
\begin{malign}
 \overline{|\bra{\omega} H^{n} M H^{m} \ket{\omega}|^2} &= \overline{  \bra{\omega} H^{n} M H^{m} \ket{\omega} \bra{\omega} H^{m} M^\dagger H^{n} \ket{\omega}}\\ &= \binom{N}{p'}^{-1} \sum_{J} \overline{ \bra{\omega} H^{n} C_J H^{m} \ket{\omega} \bra{\omega} H^{m} C_J^\dagger H^{n} \ket{\omega} } 
\end{malign}The variance gets a non-trivial contribution from terms in which $C_J H^m \ket{\omega}$ and $C_J^\dagger H^n \ket{\omega}$ are non-zero. $ C_J H^m \ket{\omega}$ is non-zero only if $H^m$ flips all spins of $\ket{\omega}$ that are annihilated by $C_J$. On the other hand, $ C_J^\dagger H^n\ket{\omega} $ is non-zero only if $H^n$ flips no spin annihilated by $C_J$. When each term in the sum over $J$ is expanded as a sum over Wick contractions of $H$, the non-zero terms in the sum must contain crossed contractions between $\bra{\omega} H^n C_J H^m\ket{\omega} $ and $\bra{\omega} H^m C_J^\dagger H^n\ket{\omega}$. Since each crossed contraction in $H$ is suppressed by the factor $\binom{2N}{p}^{-1}$, the leading contribution to the variance comes from terms with the least number of crossed contractions (see figure \ref{fig:wormhole} for an example). To ensure that $C_J H^m \ket{\omega}$ is non-zero, the crossed contractions must come from the $H$'s in the product $H^m$ at the leading order in $1/N$. 

Let us first consider the case $p' < p$. In this case, the leading contribution to the one-point function is from terms with two crossed contractions of $H$. An example of such a contribution is the following:\begin{malign}\label{eq:2pfun_suppression}
    &\bra{\omega} H^n C_J H^{k_1} \wick{\c1{H} H^{k_2} \c2{H} H^{k_3}\ket{\omega} \bra{\omega} H^{l_1} \c1{H} H^{l_2} \c2{H} } H^{l_3} C_J^\dagger H^n\ket{\omega}  \\ &= \binom{2N}{p}^{-2}  \sum_{I,K} \bra{\omega} H^n C_J H^{k_1} \Psi_I H^{k_2} \Psi_K H^{k_3}\ket{\omega} \bra{\omega} H^{l_1} \Psi_I H^{l_2}  \Psi_K H^{l_3} C_J^\dagger H^n\ket{\omega} 
\end{malign}where $k_1 + k_2 + k_3 = l_1 + l_2+ l_3 = m-2$ and no other insertions of $H$ have crossed contractions. As we discussed before, non-trivial contribution to the sum comes from terms in which the product $\Psi_I \Psi_K$ flips all spins of $\ket{\omega}$ that are annihilated by $C_J$. Since we have restricted the sum to terms with two crossed contractions, $\Psi_I \Psi_K$ must flip no other spin.  At leading order in $1/N$, this is achieved when $\Psi_I \Psi_K$ is a product of $p'$ Majorana fermions and each Majorana fermion in the product belongs to a unique annihilation operator in $C_J$. There are $2^{p'}$ choices for such a product. The number of ways in which $\Psi_I$ and $\Psi_K$ can be chosen to form a fixed product of $p'$ Majorana fermions is \begin{malign}
    \binom{p'}{p'/2} \binom{2N - 2 p'}{p-p'/2}\,.
\end{malign}Multiplying these factors with the binomial coefficient in equation \eqref{eq:2pfun_suppression}, we find that the leading contribution to the variance of one-point function is suppressed by the following factor
\begin{malign}
    \binom{2N}{p}^{-2} \times 2^{p'} \times \binom{p'}{p'/2} \binom{2N - 2 p'}{p - p'/2} \sim   N^{-p/2-p'/4}.
 \end{malign}
Now let us consider the case $(k-1) p< p' < kp$ and $k \geq 2$. In this case, leading contribution to the variance comes from terms with $k$ crossed contractions of $H$. Once again, such crossed contractions of $H$ must be from the product $H^m$.  The resulting sum over products of Majorana fermions from the crossed contractions contains a product of the form $\prod_{j = 1}^k \Psi_{I_j}$. At leading order in $1/N$, this must be equal to a product of $p'$ Majorana fermions which must flip all the spins of $\ket{\omega}$ annihilated by $C_J$. 

For a typical product of Majorana fermions $\prod_{j = 1}^k \Psi_{I_j}$ which flips $p'$ spins of $\ket{\omega}$, each $\Psi_{I_j}$ flips $p/k$ spins of $\ket{\omega}$.  Moreover, any pair of products of fermions $\Psi_{I_j}$ and $\Psi_{I_{j'}}$ share $(p - p'/k)(k-1)^{-1}$ common fermions and there are $k(k-1)/2$ such pairs.  Thus, the number of typical products of fermions grows roughly as \begin{malign}
     2^{p'} \times \frac{p'!}{(p'/k)!^k } \times \frac{(2N -2p')!}{[(p - p'/k)(k-1)^{-1}]!^{k(k-1)/2} \times (2N - (p k - p')/2)!}  \sim N^{(pk-p')/4}
\end{malign}Since each crossed contraction is suppressed by $\binom{N}{p}^{-1} \sim N^{-p/2}$, the overall suppression of the variance of the one-point function of $M$ containing $p'$ annihilation operators goes as
\begin{malign}
    N^{-(kp + p')/4}, \quad k = \lceil p'/p \rceil.
\end{malign}

Combining the results for the cases $p' > p$ and $p' < p$, we have found that the leading contribution to the variance of one-point function of $M$ in the KM state goes as\begin{malign}
    N^{- ({\max(\lceil p'/p \rceil,  \,2) p + p'})/4}.
\end{malign}

\section{Conclusion and Discussions}

In this work, we study the double-scaling limit of Kourkoulou--Maldacena states and fermionic operators acting on them. We show, with mathematical precision, that a generic class of such operators generates a von Neumann algebra of Type II$_1$, governed by the standard chord-diagram description, while the corresponding sequence of correlators in the KM state converges to correlation functions of chord operators in the tracial state. We interpret this as saying that generic operators are insensitive to the internal correlations of the underlying pure state, and therefore cannot distinguish it from the maximally mixed state. In this sense, the averaged large-$N$ limit of the KM states is mixed with respect to the algebra generated by generic operators.

More surprisingly, we identify a class of state-adapted operators that retains information about the KM state at every finite $N$. In the double-scaling limit, these operators become dressed chord creation and annihilation operators, whose action on the thermal circle probes the distance to the boundary projection onto the KM state through the Euclidean bulk wormhole. We derive the corresponding chord rules and show that, once these dressed operators are adjoined to the algebra generated by the generic chord operators, the resulting algebra becomes a Type I$_\infty$ factor. Equivalently, its commutant is trivial, so there are no operators behind the horizon that lie outside the algebra generated by these observables. Accordingly, the sequence of KM states now converges to a pure state with respect to the enlarged algebra containing the dressed chord operators.

We then studied correlation functions of dressed chord operators in the KM state at Euclidean separations. At finite DSSYK coupling $q$, we used the modified chord rules to derive exact expressions for uncrossed $2n$-point functions as well as crossed four-point functions. Compared with correlators of ordinary chord operators, these expressions are more intricate because the dressing introduces additional crossings associated with lines connecting the operator insertions to the KM projector. Nevertheless, the final results can still be written in terms of the familiar $q$-special functions that appear in standard DSSYK correlators. We also analyzed two classical regimes: the semi-classical limit at finite temperature, and the Schwarzian regime that zooms in on the spectral edge. In the former, we found a novel factorization property: a general $2n$-point correlator of dressed chord operators factorizes into $3n$ thermal two-point functions of large-$p$ SYK operators. In the latter, we found that the two-point function of dressed chord operators reduces to a  three-point function of boundary primaries dressed to Schwarzian modes in JT gravity on a Euclidean disk.

Furthermore, since the dressed chord operators belong to the observable algebra, we deformed the chord Hamiltonian by a Hermitian bilinear combination of them with the special weight $\Delta=2$, and studied the spectrum of the deformed Hamiltonian, particularly in the regime above the critical deformation where bound states begin to appear. At finite $q$, we obtained analytic expressions for the energy eigenfunctions and showed how the discrete spectrum emerges as the deformation parameter is tuned through the threshold. This was achieved by formulating the orthogonality relations of the wavefunctions as a contour integral in the complex plane of the energy parameter, and then tracking how the relevant residues move as the deformation coupling is varied.  With the explicit bound-state solutions in hand, we were able to show that the expectation value of the length operator is finite in the bound states, in contrast to its divergence in the continuum states. In the Schwarzian limit, we showed that the spectrum of the deformed Hamiltonian reduces to that of JT gravity in the presence of an End-of-the-World brane~\cite{penington2020replica,Gao:2021uro}, extending the earlier observations of~\cite{okuyama2023end} to general brane tension. We also studied the zero- and infinite-tension limits, and gave a geometric explanation of why in both cases the gravitational dynamics reduces to that of pure JT, albeit by different mechanisms. 

Finally, we identify an emergent global $U(1)$ symmetry in the bulk description that exists only in the strict double-scaling limit, and show how it is violated by finite-$N$ effects. In particular, we interpret the leading $1/N$ corrections as wormhole contributions generated by double-trace insertions. As a concrete diagnostic of this symmetry violation, we compute the leading contribution to the variance of one-point functions in the dressed chord algebra.

In the following, we discuss several lessons of the present analysis and outline possible directions for future work:

\subsection{On Holographic Description of Black Hole Interiors and Closed Universes}

We now draw an analogy with the boundary algebra that arises in the holographic description of black hole interior and closed universes, arising as different phases of large $N$ limit of two dimensional CFT~\cite{Antonini:2024mci}, and has been further analyzed in algebraic formulation recently in~\cite{Liu:2025close}, and discuss what lessons from the present work may be relevant in that setting.\footnote{See also recent discussions~\cite{Antonini:2023hdh,Antonini:2025ioh,Engelhardt:2025vsp,Kudler-Flam:2025cki,Belin:2025ako,VanRaamsdonk:2026tnv}.}

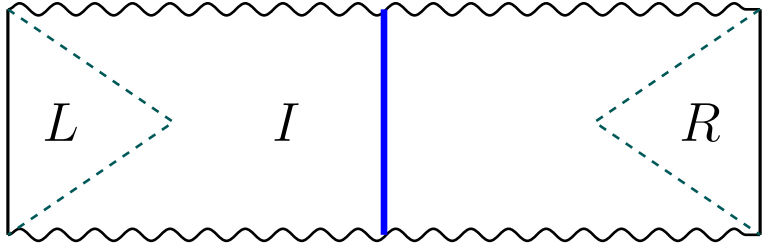
\begin{figure} \label{fig:longBH}
\centering
\begin{tikzpicture}[scale=1.0]

    \coordinate (TL) at (0,3);
    \coordinate (BL) at (0,0);
    \coordinate (TR) at (10,3);
    \coordinate (BR) at (10,0);

    \draw[line width=1.2pt] (BL) -- (TL);
    \draw[line width=1.2pt] (BR) -- (TR);

    \draw[line width=1.0pt, decorate, decoration={snake, amplitude=.8mm, segment length=5mm}] (TL) -- (TR);
    \draw[line width=1.0pt, decorate, decoration={snake, amplitude=.8mm, segment length=5mm}] (BL) -- (BR);

    \draw[line width=2.5pt, blue] (5,0) -- (5,3);

    \draw[dashed, line width=1pt, teal!70!black] (0,3) -- (2.2,1.5) -- (0,0);

    \draw[dashed, line width=1pt, teal!70!black] (10,3) -- (7.8,1.5) -- (10,0);

    \node[font=\fontsize{20}{25}\selectfont] at (0.7,1.5) {$L$};
    \node[font=\fontsize{20}{25}\selectfont] at (9.2,1.5) {$R$};
    \node[font=\fontsize{20}{25}\selectfont] at (3.7,1.5) {$I$};
\end{tikzpicture}
\caption{\small{The Lorentzian geometry dual to the large-$N$ limit of $|\omega_N\ra$ in the long-black-hole phase. The thick blue line in the middle represents the insertion of a heavy operator, and supports a long Einstein--Rosen bridge. The left and right triangular regions, bounded by the dashed lines representing the horizon and the vertical lines representing the asymptotic boundaries, correspond to the entanglement wedges of the algebra generated by the GNS representations of the left and right single-trace operators in the emergent state $\omega$. This algebra sees a mixed state and forms a proper subalgebra of the full algebra of operators that survive the limit. The latter is expected to probe a pure state.      } }
\end{figure}

The setup of~\cite{Antonini:2024mci} can be explained from the boundary side as the large-$N$ limit of the sequence of states in two-dimensional holographic CFT:
\be \label{eq:sequence-omega}
|\omega_{N}\ra=\frac{1}{\sqrt{Z}}e^{-\frac{\beta_{R}}{2}H}\mathbb{O}^{(N)}e^{-\frac{\beta_{L}}{2}H}|\Omega_{N}\ra
=\frac{1}{\sqrt{Z}}\sum_{m,n}e^{-\frac{1}{2}\beta_{L}E_{m}-\frac{1}{2}\beta_{R}E_{n}}\mathbb{O}_{nm}^{(N)}|n\rangle_{R}|\tilde{m}\rangle_{L},
\ee
where the energy eigenstates are those of a two-dimensional CFT, with $N$ identified with its central charge. The heavy operator $\mathbb{O}^{(N)}$ has conformal dimension of order $\mo(N)$, so that after analytic continuation to Lorentzian signature, and for sufficiently large $\beta_L$ and $\beta_R$, it supports a baby-universe component in the bulk. For sufficiently small $\beta_L$ and $\beta_R$, the corresponding geometry is instead that of a long black hole, with the matter operator $\mathbb{O}$ supporting a long Einstein--Rosen bridge, as illustrated in Fig~\ref{fig:longBH}. 

In both phases, the emergent, possibly averaged, large-$N$ algebra of single-trace operators cannot probe the internal purity of the limiting state. In the baby universe phase, it probes only local excitations in the bulk region connected to the asymptotic boundaries; in the long-black-hole phase, it probes only exterior excitations. In either case, the same basic question arises: what boundary operators, if any, can access the internal purity of the limiting state defined by the sequence~\eqref{eq:sequence-omega}? The SYK discussion in the current paper is closely analogous to the long black hole phase, which we explain the extent to which they are analogous below, and then comment on possible extension to the baby universe phase.

In the long black hole phase,  the  large-$N$ limit of correlation functions of single-trace operators acting on the left and right CFTs converge to the thermal correlators on each side:
\be \label{eq:thermal-2pt}
\lim _{N \rightarrow \infty}\left\langle\omega_N\right| \mathcal{O}^{(N)}_{L / R}\left(t_1\right) \mathcal{O}^{(N)}_{L / R}\left(t_2\right)\left|\omega_N\right\rangle
=\frac{1}{Z_{\beta_{L / R}}} G_{\beta_{L/R}} (t_1,t_2),
\ee
where $G_{\beta}(t_1,t_2)$ is the thermal two-point function characterized by KMS condition at inverse temperature $\beta$.  Let  $\mathcal{S}_{L/R}$ denote the algebra of single-trace operators of the left/right boundary CFT that survives at infinite $N$, and $\mathcal{S}=\mathcal{S}_L\otimes \mathcal{S}_R$.  Let $\mathcal{A}$ denote the full algebra of CFT operators that survives the large-$N$ limit. The expectation values of the states $|\omega_N\ra$ define a limiting state $\omega$, with GNS Hilbert space $\mathcal{H}^{\text{GNS}}_{\omega}$. The algebras $\mathcal{S}$ and $\mathcal{A}$ then act on $\mathcal{H}^{\text{GNS}}_{\omega}$ through the GNS representation $\pi_\omega$, giving
\be
\mathcal{X} = \pi_\omega (\mathcal{A})^{\prime \prime},\qquad \mathcal{Y}_{L/R} = \pi_\omega (\mathcal{S}_{L/R})^{\prime \prime },\qquad  \mathcal{Y} = \pi_\omega (\mathcal{S})^{\prime\prime},
\ee
where the double-commutant completes the GNS representation to a von Neumann algebra. From~\eqref{eq:thermal-2pt}, one concludes that operators in $\mathcal{Y}_{L}$ and $\mathcal{Y}_R$ see the state $\omega$ as being in exact thermal equilibrium. In this sense, the large-$N$ limit of the pure states $\omega_N$ appears mixed when probed only by each one-sided algebra $\mathcal{Y}_{L/R}$.

By construction, $\mathcal A$ contains all boundary operators that survive the large-$N$ limit, and $\mathcal X$ is the corresponding represented algebra on $\mathcal H_\omega^{\rm GNS}$. If one further assumes that these surviving operators furnish a complete semiclassical description of the bulk quantum field theory on a global Cauchy slice of the fixed long-black-hole saddle, then one may identify
\begin{align}
\mathcal H_\omega^{\rm GNS}\simeq \mathcal H_\omega^{\rm Fock},
\qquad
\mathcal X=\mathcal B(\mathcal H_\omega^{\rm Fock}).
\end{align}
Using the holographic dictionary, the algebra generated by the surviving left and right single-trace sectors $\mathcal{Y}_L$ and $\mathcal{Y}_R$ should then be identified with the local algebra of the two exterior regions $\mathcal{A}_L$ and $\mathcal{A}_R$, and 
\begin{align}
\mathcal Y=\ma_L\vee \ma_R .
\end{align}
In the long-black-hole phase, however, semiclassical bulk effective field theory suggests that the central portion of the elongated Einstein--Rosen bridge supports an additional nontrivial local algebra $\ma_I$. Assuming that $\ma_I$ is associated with a region spacelike separated from both exterior regions on the chosen bulk slice, locality implies
\begin{align}
[\ma_I,\ma_L]=[\ma_I,\ma_R]=0,
\end{align}
and hence
\begin{align}
\ma_I \subset \mathcal Y' \cap \mathcal X .
\end{align}
Since $\ma_I$ is nontrivial, it follows that
\begin{align}
\mathcal Y' \cap \mathcal X \neq \mathbb C .
\end{align}
Equivalently, although the global bulk algebra on the full slice is
\begin{align}
\mathcal X=\mathcal B(\mathcal H_\omega^{\rm Fock})
=\ma_L\vee \ma_R\vee \ma_I,
\end{align}
operators in $\mathcal Y=\ma_L\vee \ma_R$ probe only the two exterior regions and  see the limiting state as thermal. The purification is encoded in operators lying outside $\mathcal Y$, semiclassically associated with the black hole interior region.

This structure is closely analogous to what happens in the SYK analysis of the present paper. There too, one begins with a sequence of pure states $\omega_N$ and studies the algebra of operators that survives an averaged large-$N$ limit. Here the ensemble average should be viewed primarily as a tool for accessing the large-$N$ limit, since for the class of correlators considered in the KM state the result is expected to be self-averaging: the averaged correlators should agree, at leading order in $1/N$, with those of a typical individual realization, up to fluctuations that vanish at large $N$.

The algebra generated by generic operators is then analogous to the single-trace algebra $\mathcal{Y}_{L/R}$: its GNS representation is in exact thermal equilibrium and does not retain access to the purity of the microscopic state. By contrast, the dressed chord operators are analogous to operators in $\mathcal X\setminus \mathcal Y$ that continue to encode information about the underlying pure state. In this sense, SYK provides an explicit example of the sort of enlarged operator algebra one would like to identify in the long-black-hole phase.

There is, however, an important difference between the SYK algebra and the algebra in the long-black-hole phase. In the latter case, the algebra $\mathcal S$ is a proper subalgebra of $\mathcal A$, and the additional operators may still admit an interpretation in terms of observables associated with a definite bulk region.\footnote{We thank Hong Liu for emphasizing this point.} In the SYK example, by contrast, the dressed operators do not appear to admit a natural interpretation as operators localized in any fixed bulk subregion, since they fail to commute with ordinary chord operators inserted anywhere on the boundary.

More concretely, one may heuristically regard the subalgebra $\mathcal A_0$ defined in~\eqref{eq:suba0} as probing the exterior of a black hole, so that its commutant $\mathcal A_0^\prime$ would play the role of an interior algebra. From this viewpoint, the dressed operators $M$ and $M^\dagger$ belong to neither side of this decomposition: they fail to commute with both $\mathcal A_0$ and $\mathcal A_0^\prime$. Thus, unlike the long-black-hole case, the enlarged algebra in SYK does not seem to arise by simply adjoining operators associated with a new geometric region. Here we are borrowing intuition from the standard large-$N$ SYK picture~\cite{kourkoulou2017purestatessykmodel}; a possible caveat is that the notion of a horizon becomes sharp only in the semiclassical regime of double-scaled SYK, namely as $q\to 1$.

It would therefore be interesting to understand the modular structure of the double-scaled algebra more fully in its semiclassical limit, as well as the way this structure is modified by finite-$q$ corrections. Such an analysis may help clarify how operators that continue to retain information about the purity of the state are organized within the algebra, and could in turn provide useful lessons for the boundary construction of operators probing black-hole interiors.

One might be tempted to draw a similar analogy for the baby-universe phase, namely AS$^2$ cosmology. There is, however, an important difference. In that setting, the single-trace operators do not themselves admit a pointwise large-$N$ limit of the form~\eqref{eq:thermal-2pt}. Instead, \cite{Liu:2025close} proposed a refined notion of large-$N$ limit in which one first averages over theories in a window of $N$, and only then takes the limit. With this prescription, an averaged version of~\eqref{eq:thermal-2pt} defines a convergent algebra analogous to the single-trace algebra, and this algebra sees a thermal state in the limit.

The key difference from the SYK story is that the states relevant for AS$^2$ cosmology are supported in a microcanonical window of width $O(1)$, where self-averaging is not expected to hold in the same way. In the SYK setting studied here, self-averaging is precisely what allows the purity of the microscopic state to survive in suitable operators: the state-adapted operators that probe this purity admit a microscopic realization, in the sense that they arise as the large-$N$ limit of a sequence of operators defined at each finite $N$. By contrast, once this property is lost, the limiting state obtained after averaging is generically mixed with respect to the emergent algebra. Moreover, when the definition of the limit itself involves averaging over $N$, the resulting emergent algebra may contain operators that effectively act across different $N$ sectors. Such operators may participate in a purification of the averaged limiting state within the enlarged emergent algebra, but they need not belong to the microscopic theory at any fixed $N$. Their existence is then tied not only to the underlying theory, but also to the limiting procedure itself.

Despite this difference from AS$^2$ cosmology, recent work~\cite{Sasieta:2025vck} proposed a realization of a baby-universe saddle in the bulk using two SYK models coupled through the Maldacena--Qi interaction~\cite{Maldacena-Qi:2018lmt}. The main difference from AS$^2$ cosmology is that the bulk entanglement in that setup is of order $O(N)$, so that, at this level, the cosmological phase and the black-hole phase are not sharply distinguished.\footnote{We thank Martin Sasieta for bringing this to our attention.} It would be very interesting to apply the operator-algebraic perspective developed here to the corresponding double-scaled algebra in that model.

Finally, it is worth emphasizing that although the dressed chord operators are qualitatively different from ordinary chord operators in the emergent description, at finite $N$ they look remarkably similar to generic SYK operators. When expanded in the $\Psi_I$ basis, both $M$ and $M^\dagger$ are superpositions of fermionic operators of length $\mathcal O(N^{1/2})$. What distinguishes them is not their microscopic size, but the way their coefficients are correlated and adapted to the KM state, allowing them to act as effective annihilation and creation operators. We discuss in section~\ref{ssec:complexity} to what extent this may be interpreted as a notion of operator complexity. The lesson is that information about the internal correlations of the state is already encoded in the operator at finite $N$, and is therefore retained in the double-scaling limit. In the GNS description, this microscopic structure becomes geometrized as a dressing by wormhole distance. While we do not claim that the same form of dressed operators should probe the internal purity of emergent states in general AdS/CFT settings, the SYK construction suggests a useful broader perspective: operators capable of probing purity may have to be state-adapted, even when they look microscopically very similar to ordinary operators.

\subsection{Dynamical Features of Dressed Operators}

It is natural to ask whether one can extract dynamical properties of the bulk dual to the KM state from OTOCs of dressed operators, by Fourier transforming the exact crossed four-point function~\eqref{eq:G4-cr-answer}. Intuitively, since the dressed operators probe the internal correlations of the KM state, one may expect them not to experience the bulk as having a horizon, and hence not to exhibit the scrambling dynamics associated with near-horizon scattering. By contrast, generic operators do probe the horizon, and their scrambling behavior is encoded in the imaginary pole structure of the corresponding  OTOCs in the energy domain, which in the semi-classical limit is governed by the SU$(1,1)$ $6j$ symbol~\cite{Lam_2018}. 

Because the exact result~\eqref{eq:G4-cr-answer} also involves these $6j$ symbols, but in a more intricate way, it is not technically obvious whether the OTOCs of dressed operators exhibit any signature of scrambling at the usual scrambling time $t\sim \log(\lambda^{-1})$ of double-scaled SYK in the semi-classical regime, or instead show no scrambling behavior at all. In either case, it would be interesting to understand the late-time behavior of dressed OTOCs more systematically, and to clarify how it differs from that of generic operators.

\subsection{Implications for Supersymmetric Models}

We briefly comment on possible implications for supersymmetric JT gravity and supersymmetric double-scaled SYK, with particular emphasis on the derivation of out-of-time-ordered correlators in these settings.

In JT gravity, such correlators are related to the crossed four-point function
\be\label{eq:G4-length}
\begin{tikzpicture}[scale=1.4, line cap=round, line join=round,baseline=-0.5ex]

\def\R{1}

\coordinate (T) at (0,\R);
\coordinate (R) at (\R,0);
\coordinate (B) at (0,-\R);
\coordinate (L) at (-\R,0);

\draw[line width=1pt,dashed] (0,0) circle (\R);

\draw[line width=1pt]  (1,0) .. controls (45:0.6) .. (0,1); 
\draw[line width=1pt]  (0,1) .. controls (135:0.6) .. (-1,0); 
\draw[line width=1pt]  (-1,0) .. controls (225:0.6) .. (0,-1); 
\draw[line width=1pt]  (0,-1) .. controls (315:0.6) .. (1,0); 

\draw[blue, line width=1pt] (L) -- (R);
\draw[red,  line width=1pt] (T) -- (B);

\node at (-2.3,0) {$G_4(\ell_1,\ell_2,\ell_3,\ell_4) =$}; 
\node at (45:.85) {$\ell_2$};
\node at (135:0.85) {$\ell_1$};
\node at (225:0.85) {$\ell_4 $};
\node at (315:0.85) {$\ell_3$}; 
\node at (0.4,0.15) {\textcolor{blue}{$\ell$}};
\node at (-0.15,-0.4) {\textcolor{red}{$\ell^\prime$}};

\end{tikzpicture}
\ee
where the red and blue curves represent insertions of $SL(2,\mbr)$-invariant bilocal operators $\mathcal{O}(\tau_1,\tau_2)$ on the Euclidean circle in Schwarzian quantum mechanics. The boundary segments of the star-shaped region are labeled by geodesic lengths $\ell_i$, with $i=1,2,3,4$, corresponding to fixed-length boundary conditions on the adjacent slices. The dashed circle indicates that the theory is defined on a Euclidean disk, on which one may impose fixed-energy or fixed-temperature boundary conditions. Given the overlap between a length eigenstate $|\ell\ra$ and a fixed-energy state $|E\ra$, one can readily convert $G_4$ into a correlator in a microcanonical window. After a further Fourier transform to an appropriate real-time configuration, this yields the out-of-time-ordered correlator.

A useful feature of~\eqref{eq:G4-length} is that its evaluation admits a simple geometric interpretation. For example, one may divide the star-shaped region into upper and lower halves separated by the blue chord; each half then contributes a factor of the interior wavefunction $\Upsilon$ with the corresponding length boundary conditions~\cite{Yang:2018gdb}. Each bulk chord contributes a factor of $e^{-\Delta \ell}$, so that
\be\label{eq:G4-geometric}
G_4 (\ell_1,\ell_2,\ell_3,\ell_4)=\int \dd \ell \, \Upsilon\left(\ell_1, \ell_2, \ell\right) \Upsilon\left(\ell, \ell_3, \ell_4\right) e^{-\Delta \ell} e^{-\Delta^\prime \ell^{\prime}}
\ee
where $\ell^\prime$ is fixed by the geometry as a function of the remaining length variables. This use of $\Upsilon$ in the derivation of crossed four-point functions was extended to double-scaled SYK at finite coupling in~\cite{Xu:2025Complexity}, where a discrete analogue of $\Upsilon$ was identified.

We expect a derivation of crossed four-point functions analogous to~\eqref{eq:G4-geometric} to extend to supersymmetric models, including JT supergravity and $\mathcal{N}=1,2$ supersymmetric double-scaled SYK. In such theories, the bulk wormhole state can be characterized by its length, together with additional quantum numbers associated with the fermionic degrees of freedom~\cite{Lin:2022zxd,Berkooz:2020xne}. What is particularly interesting, however, is that these theories may contain both continuous and discrete sectors, with the latter corresponding to BPS states. In the $\mathcal{N}=2$ case, the two sectors can be separated by a gap, making the situation reminiscent of the effective potential for the length operator in the deformed Hamiltonian above the bound-state threshold. As discussed in section~\ref{sec:branes}, the length eigenstate in such situations is a superposition of both the continuous and discrete sectors. It would therefore be interesting to understand what can be learned about the BPS sector by projecting such correlation functions onto it, and in particular whether one can extract a meaningful notion of chaos, relevant for microstates of supersymmetric black holes,  from the corresponding BPS crossed four-point functions. We leave this as an interesting direction for future work.

\subsection{The Double-Scaling Limit of General Pure States in SYK} 

In section \ref{sec:KM}, we considered the double-scaling limit of a limited class of thermal pure states, namely the KM states, which were constructed by Euclidean evolution of the spin states (see equation \ref{eq:KM_state}). The fact that the dressed operators survive in the large $N$ limit of KM states makes them a special class of pure states. The dressed operators were indeed crucial in showing that the algebra of operators which survive in the large $N$ limit is Type $\text{I}_\infty$. It is interesting to ask about the nature of the algebra for more general pure states. 

Consider the following generalization of the KM state which is obtained by rotating each spin in $\ket{\omega}$ on the Bloch sphere. \begin{malign}
        \ket{\Psi_\alpha(\beta)} = \frac{e^{-\frac{\beta}2 H} \ket{\Psi_\alpha}}{\sqrt{\langle{\Psi_{\alpha}}| e^{-\beta H} |\Psi_{\alpha}\rangle}}, \quad \ket{\Psi_\alpha} = \left(\frac{\ket{-1} + \alpha{\ket{+1}}}{\sqrt{1 + |\alpha|^2}}\right)^{\otimes N} 
\end{malign}where $\ket{\pm 1}$ are the eigenstates of $Z$ with eigenvalues $\pm 1$ respectively.

Unlike the spin states, the expectation value of any polynomial of dressed operators in $\ket{\Psi_\alpha}$ approaches zero in the double-scaling limit. Thus, the dressed operators do not survive in the algebra of operators acting on $\ket{\Psi_\alpha}$ in the double-scaling limit. We may still define an analog of the chord number operator $\hat n_{\text{tot}}$ as follows:\begin{malign}
    \tilde n = \frac{1}{\sqrt{N}} \sum_i \left( Z_i - \langle  Z_i \rangle_{\Psi_\alpha}\right)  
\end{malign}where $\langle \cdot \rangle_{\Psi_{\alpha}}$ denotes the expectation value of an operator in $\ket{\Psi_{\alpha}}$. We have chosen the above normalization to ensure that $ \tilde n$ acts non-trivially on the states of the form $H^n \ket{\Psi_\alpha}$ in the double-scaling limit. It is straightforward to check that $\tilde n$ has $O(1)$ variance in $\ket{\Psi_\alpha}$:\begin{malign} \langle \tilde n^2\rangle_{\Psi_\alpha} &=  \frac{1}{N} \langle \left(\sum_{i} (Z_i - \langle Z_i\rangle_{\Psi_{\alpha}}) \right)^2 \rangle_{\Psi_\alpha} \\ &= 1 -\langle Z_i\rangle_{\Psi_\alpha}^2
\end{malign}Thus, $\tilde n$ does not act on $\ket{\Psi_\alpha}$ like  the conventional chord number operator as $\hat n_{\text{tot}}$ acts on the KM states in the double-scaling limit. 

The partial results above suggest that the dynamics around $\ket{\Psi_{\alpha}}$ in the double-scaling limit is qualitatively different from that around the KM states. It would thus be interesting to study the operator algebra that survives in the double-scaling limit of these general states. 

\subsection{The Complexity of Reconstructing the Dressed Operators}\label{ssec:complexity}

As discussed in section~\ref{sec:KM}, the dressed operators adapted to a KM state are constructed using detailed knowledge of the initial spin configuration. In this respect, they are qualitatively different from state-independent operators such as the SYK Hamiltonian, which is defined as a generic linear combination of products of $p$ Majorana fermions. The dressed operators are able to probe correlations in the KM state that remain inaccessible to such generic operators. In the holographic description of the KM state, correlation functions of the dressed operators depend on the location of the end-of-the-world brane. Equivalently, they are sensitive to data associated with the black-hole interior.

While the size\footnote{By operator size, we mean the average length of the string of Majorana fermions when the operator is expanded in the basis of Majorana fermions.} of a dressed operator scales in the same way as that of the Hamiltonian in the double-scaling limit, the difficulty of reconstructing it depends crucially on how much information the observer has about the state. For an observer who already knows that the state was prepared by Euclidean evolution from a spin state as in equation~\eqref{eq:spin_state}, and who also knows the corresponding choice of spin operators, the reconstruction is straightforward: one may measure those spins in the KM state. This is reliable because KM states with different initial spin configurations are orthogonal up to $1/N$ corrections:
\begin{malign}
    \frac{\bra{\vec{s}\,'} e^{-\beta H} \ket{\vec{s}\,} }{\sqrt{\bra{\vec{s}\,'} e^{-\beta H} \ket{\vec{s}\,'}\bra{\vec{s}\,} e^{-\beta H} \ket{\vec{s}\,}}}
    = \delta_{\vec{s}\,', \vec{s}} + O\!\left(\frac{1}{N}\right).
\end{malign}

For an observer with no prior information about the initial state, however, the reconstruction problem becomes much harder. The reason is that there are $(2N-1)!!$ possible ways of pairing the $2N$ Majorana fermions into $N$ spin operators, which scales as
\begin{align}
    (2N-1)!! \sim \exp\!\left(N\log N + O(N)\right)
\end{align}
at large $N$. Thus the space of possible spin structures entering the definition of the KM state grows factorially with $N$. This does not by itself define a precise computational complexity, but it suggests that the reconstruction of the dressed operators is controlled not by their microscopic size alone, but by the amount of state-specific information required to identify the correct spin structure.

It would therefore be interesting to formulate a sharper notion of reconstruction complexity for the dressed operators, especially in situations where the observer has only partial information about the underlying state. Such a notion may help clarify how state-dependent bulk observables can remain microscopically simple in size while being highly nontrivial to identify.

\section*{Acknowledgments} We thank Ahmed Almheiri, Micha Berkooz, Sergio E. Aguilar-Gutierrez, Chuanxin Cui, Charlie Cummings, Xuchen Cao, Hong Zhe (Vincent) Chen,  Xi Dong, Elliott~Gesteau, Ping Gao,  David Kolchmeyer,   Henry W. Lin, Hong Liu, Sam Leutheusser, Alexey Milekhin, Masamichi Miyaji, 
Vladimir Narovlansky, Onkar Parrikar, Geoffrey R. Penington, Pratik Rath, Martin Sasieta, 
Tim Schuhmann,  Elisa~Tabor, Mykhaylo~Usatyuk  and Herman Verlinde for helpful comments and discussions.  HR acknowledges support from the Department of Atomic Energy,
Government of India, under project identification number RTI 4002, and from the Infosys
Endowment for the study of the Quantum Structure of Spacetime. HR also acknowledges support from grant NSF PHY-2309135 to the Kavli Institute for Theoretical Physics (KITP). The work of J.X. is supported by the U.S. Department of Energy,
Office of Science, Office of High Energy Physics, under Award Number DE-SC0011702.
J.X. acknowledges the support by the Graduate Division Dissertation Fellowship and
the Physics Department Graduate Fellowship at UCSB.

\appendix

\section{Review of the Bulk Hilbert Space of Double-Scaled SYK} \label{app:useful}

In this section, we review the bulk Hilbert-space description of conventional double-scaled SYK. We first introduce the Hilbert space underlying the chord-diagram formulation, and then summarize the main analytic results for inner products and for matrix elements of chord operators in the zero- and one-particle sectors. These results will be used in section~\ref{sec:correlators} to evaluate the correlation functions. 

\subsection{The Bulk Hilbert Space of Double-Scaled SYK in Tracial State} \label{app:review-sec}
The conventional double-scaling limit of SYK concerns correlation functions of operators of the form~\eqref{eq:op1}
\be
H_N = \sum_{I} J_{I} \Psi_{I},
\qquad
O_N = \sum_{I^\prime} K_{I^\prime} \Psi_{I^\prime},
\ee
with properly normalized disorder coefficients $J_I$ and $K_{I^\prime}$, evaluated in the tracial state. It is known that, after disorder averaging and taking the double-scaling limit, these correlators admit the representation
\be \label{eq:dslimit}
\lim_{\stackrel{p,N\to\infty}{p^2/N \text{ fixed}}}\overline{\tr\left(H_{N}^{k_{1}}O_{N}H_{N}^{k_{2}}\cdots O_{N}H_{N}^{k_{n}}\right)}
=
\la 0 |H^{k_{1}}OH^{k_{2}}\cdots OH^{k_{n}}|0\ra,
\ee
where the trace $\tr$ is normalized such that $\tr(\mathbf{1}_{N})=1$ for any $N$. The emergent operators $H$ and $O$ are chord operators of weights $1$ and $\Delta$, respectively, acting on an emergent Hilbert space $\mathcal H$. The state $|0\rangle\in\mathcal H$ is the empty-chord state. Repeated action of $H$ and $O$ on $|0\rangle$ generates the full chord Hilbert space $\mathcal H$, whose basis states may be viewed as configurations of $H$- and $O$-chords.

The chord Hilbert space $\mathcal{H}$ admits a natural decomposition into sectors labeled by the number of $O$-chords,
\be \label{eq:direct-sum}
\mh=\bigoplus_{n=0}^{\infty}\mh_{n},
\ee
where the $k$-particle sector $\mh_k$ is spanned by states with $k$ $O$-chords and an arbitrary number of $H$-chords inserted between them:
\be
\mh_{k}=\text{span}_{\mathbb{C}}
\bigl\{\ket{n_{0},n_{1},\dots,n_{k}} \,\big|\, (n_{0},\dots,n_{k})\in\mathbb{Z}_{\geq0}^{k+1}\bigr\}.
\ee
These states can be illustrated as:
\begin{equation}
\includegraphics[width=0.4\linewidth]{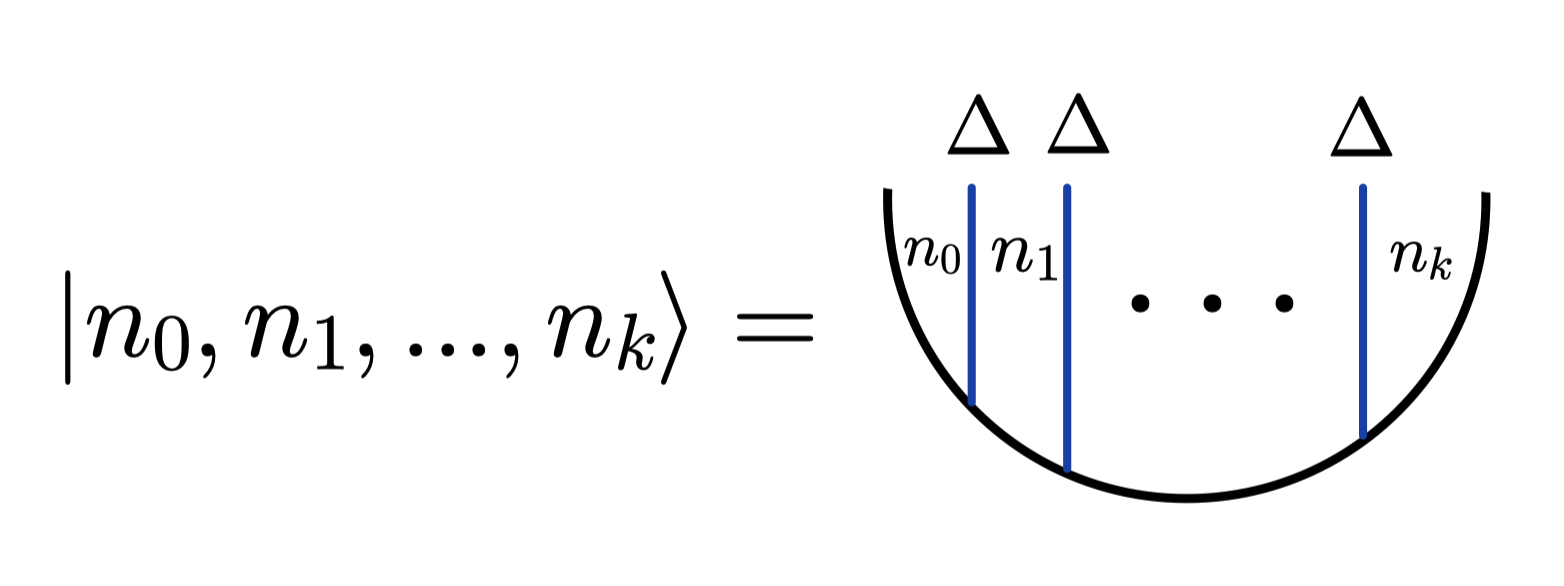}
\end{equation}
Equivalently, one may decompose $\mathcal{H}$ into sectors labeled by the number of $H$-chords, with each sector describing different $O$-chord configurations. In practice, it is convenient to choose $H$ as the reference chord operator of unit weight, corresponding to the double-scaling limit of the SYK Hamiltonian, and to treat all other chords with weight $\Delta\neq1$ as matter chords. The decomposition \eqref{eq:direct-sum} can then be interpreted as a decomposition into sectors with fixed particle number, which facilitates the study of the dynamics generated by $H$ within each sector.

The inner product on $\mathcal{H}$ is defined by the chord contraction rules: the overlap between two states is given by a weighted sum over all admissible contractions of chords in the bra and ket, with the restriction that any two chords may cross at most once. A given contraction carries a weight
\[
q^{\Delta^2\,\text{cr}(O,O)+\Delta\,\text{cr}(O,H)+\text{cr}(H,H)},
\]
where $\text{cr}(O,H)$ counts the number of crossings between $O$- and $H$-chords. As an illustration, the inner product between $\ket{1,1,0}$ and $\ket{0,2,0}$ can be represented as the sum of following chord diagrams:
\be
\langle 1,1,0|0,2,0\rangle
=
\begin{tikzpicture}[baseline=-0.6ex,scale=0.75]
  \draw[thick] (0,0) circle (1);

  \draw[thick] (150:1) -- (270:1);
  \draw[thick] (-60:1) -- (60:1);

  \draw[blue,thick] (115:1) -- (-115:1);
  \draw[blue,thick] (45:1) -- (-45:1);
\end{tikzpicture}
\;+\;
\begin{tikzpicture}[baseline=-0.6ex,scale=0.75]
  \draw[thick] (0,0) circle (1);

  \draw[thick] (150:1) -- (270:1);
  \draw[thick] (-60:1) -- (60:1);

  \draw[blue,thick] (115:1) -- (-45:1);
  \draw[blue,thick] (45:1) -- (-115:1);
\end{tikzpicture}
=
q^{\Delta}+q^{\Delta^2+3\Delta}
\ee
In this setup, the limiting state $\ket{\omega}$ in~\eqref{eq:dslimit} is naturally identified with the zero–chord-number state $\ket{0}$.

As operators on $\mh$, the chord operators $H$ and $O$ act by deleting or inserting chords from a given side. It is important to distinguish left and right actions, as deleting or inserting a chord from different sides generally leads to different crossing factors. For example, if $a$ denotes the annihilation operator for an $H$-chord, then
\be
a_L \ket{0,1} = q^{\Delta} \ket{0,0}, \qquad
a_R \ket{0,1} = \ket{0,0},
\ee
where the difference arises because deleting an $H$-chord to the left of an $O$-chord introduces an $H$--$O$ crossing, whereas deleting it from the right does not.

Accordingly, one introduces left- and right-acting creation and annihilation operators for both $H$- and $O$-chords. The left $H$-chord creation and annihilation operators act within each sector $\mathcal{H}_k$, and are defined as
\be \label{eq:defa1}
\begin{aligned}
a_{1, L}\left|n_0, \ldots, n_k\right\rangle
&=\sum_{j=0}^k q^{j \Delta+\sum_{i=0}^{j-1} n_i}\left[n_j\right]
\left|n_0, \ldots, n_j-1, \ldots, n_k\right\rangle, \\
a_{1, L}^{\dagger}\left|n_0, \ldots, n_k\right\rangle
&=\left|n_0+1, n_1, \ldots, n_k\right\rangle,
\end{aligned}
\ee
while the left $O$-chord creation and annihilation operators shift the particle number by one:
$a_{2,L}:\mathcal{H}_k\mapsto\mathcal{H}_{k-1}$ and
$a_{2,L}^\dagger:\mathcal{H}_k\mapsto\mathcal{H}_{k+1}$, and are defined by
\be \label{eq:defa2}
\begin{aligned}
a_{2, L}\left|n_0, \ldots, n_k\right\rangle
&=\sum_{j=1}^k q^{\Delta^2(j-1)+\Delta \sum_{i=1}^{j-1} n_i}
\left|n_0, \ldots, n_{j-2}, n_{j-1}+n_j, n_{j+1}, \ldots, n_k\right\rangle, \\
a^\dagger_{2, L}\left|n_0, \ldots, n_k\right\rangle
&=\left|0, n_0, \ldots, n_k\right\rangle.
\end{aligned}
\ee
The corresponding left chord operators are
\be
H_L = a_{1,L} + a^{\dagger}_{1,L}, \qquad
O_L = a_{2,L} + a^{\dagger}_{2,L}.
\ee
Right-acting operators $a_{1,R},a_{1,R}^\dagger,a_{2,R},a_{2,R}^\dagger$ are defined analogously, and the corresponding right chord operators are
\be
H_R = a_{1,R} + a^{\dagger}_{1,R}, \qquad
O_R = a_{2,R} + a^{\dagger}_{2,R}.
\ee

The chord contraction rules are equivalently encoded in the commutation relations of these operators. Defining the $q$-commutator by $[A,B]_q=AB-qBA$, one finds the one-sided relations
\be
[a_{i,L},a_{j,L}^{\da}]_{q^{\Del_{ij}}}
=
[a_{i,R},a_{j,R}^{\da}]_{q^{\Del_{ij}}}
=\delta_{ij},
\ee
where
\be
q^{\Del_{ij}}=
\begin{bmatrix}
q & q^{\Del} \\
q^{\Del} & q^{\Del^{2}}
\end{bmatrix}_{ij},
\qquad i,j\in\{1,2\}.
\ee
In addition, the two-sided commutation relations take the form
\be
[a_{i,L},a_{j,R}^{\da}]
=\delta_{ij}\,q^{\Del_i\hat{n}_{H}}\,q^{\Del_i\Del\hat{n}_{O}},
\ee
with
\be
\Del_{i}=
\begin{cases}
1, & i=1,\\
\Del, & i=2.
\end{cases}
\ee
and $\hat{n}_H$ and $\hat{n}_O$ count the number of $H$- and $O$-chords in a state respectively. 

From these relations, it follows immediately that the left chord operators $H_L,O_L$ commute with the right chord operators $H_R,O_R$. Moreover, it can be shown that the algebras generated by the left and right chord operators form a pair of mutually commuting von Neumann algebras, which are related by an isomorphism implemented by the reflection operator $\mathcal{R}$ that reverses the chord ordering.

Previously, we introduced the inner product on $\mathcal{H}$ using the chord contraction rules. Given the algebraic structure described above, the inner product on $\mathcal{H}$ can equivalently be defined by requiring either the left or the right set of chord operators $\{O_{L/R}, H_{L/R}\}$ to be Hermitian, together with the normalization condition $\langle\omega|\omega\rangle=1$.  In~\cite{Lin_2023}, it was further shown that the symmetry algebra $U(J)$, defined as the universal enveloping algebra generated by
\be
J_{ij}=a_{1,i}a_{1,j}^\dagger-q^{\hat{n}_{{\rm tot}}}, \quad 
\hat{n}_{\rm tot}=\hat{n}_H + \Delta \hat{n}_O,\qquad i,j\in\{L,R\},
\ee
and forming a proper subalgebra of the algebra generated by the left and right creation and annihilation operators, acts as the analogue of the near-horizon $\mathfrak{sl}_2$ symmetry that preserves the wormhole length~\cite{Maldacena:2016upp,Lin:2019qwu}. This symmetry algebra can be used to decompose $\mathcal{H}$ into irreducible representations, with the $0$- and $1$-particle sectors providing the basic building blocks for this decomposition. The quantum group origin of such structure has been explored in recent works~\cite{xu:2025qg,Mertens:2025qg1,math:2025qg}. In the following, we therefore focus on the detailed structure of the sectors $\mh_0$ and $\mh_1$, which facilitates the calculations of correlators in the KM states in subsequent sections.

\paragraph{The $0$-particle sector $\mathcal{H}_0$}
The zero-particle Hilbert space $\mathcal{H}_0$ is spanned by states labeled by the number of $H$-chords. The inner product between these basis states is given by
\be
\la m|n\ra=\delta_{mn}[n]!,
\ee
where $[n]=(1-q^n)/(1-q)$ denotes the $q$-integer and $[n]!\equiv\prod_{j=1}^{n}[j]=(q;q)_n/(1-q)^n$ is the corresponding $q$-factorial.

Within $\mathcal{H}_0$, the reflection operator $\mathcal{R}$ acts trivially. As a consequence, the left and right chord Hamiltonians coincide when restricted to this sector and define a single operator $H$:
\be \label{eq:0p-state-def}
H_L|n\ra = H_R|n\ra \equiv H|n\ra = |n+1\ra + [n]|n-1\ra .
\ee
Since $H$ is self-adjoint on $\mathcal{H}_0$, the spectral theorem applies and yields a decomposition of $\mathcal{H}_0$ into its energy eigenstates $|\theta\ra$, satisfying
\be
H|\theta\ra = E(\theta)|\theta\ra, \qquad
E(\theta)=\frac{2\cos\theta}{\sqrt{1-q}}, \qquad \theta\in[0,\pi].
\ee
The overlap between the energy eigenstates and fixed chord-number states is given by
\be
\la\theta|n\ra=\frac{H_n(\cos\theta\,|\,q)}{(1-q)^{n/2}},
\ee
where $H_n(x|q)$ denotes the $q$-Hermite polynomial. The spectral decomposition yields the following resolution of the identity on $\mathcal{H}_0$:
\be
\mathbf{1}_{\mh_0}
=\int_0^\pi \dd\mu(\theta)\,|\theta\ra\la\theta|
=\sum_{n=0}^{\infty}\frac{1}{[n]!}\,|n\ra\la n| ,
\ee
where the measure is
\be
\dd\mu(\theta)=\mu(\theta)\dd\theta
=\frac{(e^{\pm2i\theta},q;q)_\infty}{2\pi}\dd\theta .
\ee
Here the notation $(e^{\pm2i\theta},q;q)_\infty$ indicates the product over both choices of sign in the argument of the $q$-Pochhammer symbol. In particular, the zero-chord state is an equal-weight superposition of all energy eigenstates in $\mathcal{H}_0$:
\be \label{eq:omega-def}
|0\ra = \int_0^\pi \dd\mu(\theta)\,|\theta\ra .
\ee

\paragraph{The $1$-particle sector $\mathcal{H}_1$}
The one-particle sector $\mathcal{H}_1$ is spanned by states with a single matter insertion and arbitrary numbers of $H$-chords to its left and right. 


Within $\mathcal{H}_1$, the left and right chord Hamiltonians $H_L$ and $H_R$ are commuting self-adjoint operators. The spectral theorem therefore applies simultaneously to both, yielding a joint spectral decomposition of $\mathcal{H}_1$ in terms of their common eigenstates,
\be\label{eq:def-1p-state}
H_{L/R}\ket{\Delta;\theta_L,\theta_R}
=E(\theta_{L/R})\ket{\Delta;\theta_L,\theta_R}.
\ee
This leads to the resolution of identity in $\mh_1$:
\be \label{eq:resolve-1}
\mathbf{1}_{\mathcal{H}_{1}}=\int_{0}^{\pi}\dd\mu(\te_{L})\dd\mu(\te_{R})\frac{|\Delta;\te_{L},\te_{R}\ra\la\Delta;\te_{L},\te_{R}|}{\la\te_{L}|q^{\Del\hat{n}_{H}}|\te_{R}\ra}.
\ee
Note that we have included explicit dependence on the matter weight $\Delta$. The normalization of these states is chosen so that they directly reproduce the matter two-point function. Explicitly, their inner product is defined as
\be
\la\Delta;\phi_L,\phi_R|\Delta;\theta_L,\theta_R\ra
=\frac{\delta(\phi_L-\theta_L)\delta(\phi_R-\theta_R)}{\mu(\phi_L) \mu(\phi_R)}
\la\theta_L|q^{\Delta\hat{n}_H}|\theta_R\ra,
\ee
where the matter density factor is given by
\be \label{eq:m-density}
\la\theta_L|q^{\Delta\hat{n}_H}|\theta_R\ra
=\sum_{n=0}^{\infty}q^{\Delta n}
\frac{\la\theta_L|n\ra\la n|\theta_R\ra}{[n]!}
=\frac{(q^{2\Delta};q)_{\infty}}{(q^{\Delta}e^{\pm i\theta_L\pm i\theta_R};q)_{\infty}}
=\mathcal{N}_{\Delta}
\frac{\Gamma_q\!\left(\Delta\pm\frac{i\theta_L}{\lambda}\pm\frac{i\theta_R}{\lambda}\right)}{\Gamma_q(2\Delta)}.
\ee
The second equality follows from the generating function of the $q$-Hermite polynomials, while the third equality uses the definition of the $q$-Gamma function. The numerical normalization constant is
$\mathcal{N}_{\Delta}=(1-q)^{2\Delta-3}(q;q)_{\infty}^{-3}$.
This matter density appears naturally in the computation of two-point correlation functions of the operator $O$ at fixed boundary energies.

It is convenient to introduce the vertex function
\be\label{eq:vertex-func}
\gamma_{\theta_L\theta_R}
=\sqrt{\la\theta_L|q^{\Delta\hat{n}_H}|\theta_R\ra},
\ee
which may be interpreted as the matrix element associated with a single boundary insertion of $O$, resolved in the energy eigenbasis corresponding to the two boundary segments. This interpretation is primarily heuristic: while such vertex factors correctly capture uncrossed matter insertions, they are not sufficient to describe configurations involving crossed matter lines. In those cases, additional structure involving the quantum $R$-matrix is required.
\be
\includegraphics[width=0.25\linewidth]{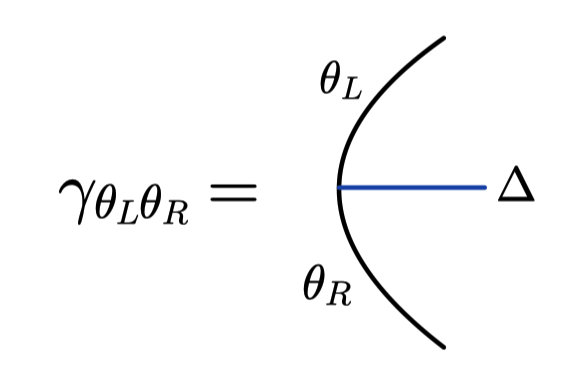}
\ee

The matrix element of the creation operator of an $O$-chord can therefore be derived as:
\be \label{eq:bd-components}
\la\Del;\te_{1},\te_{2}|b^{\da}|\te_{3}\ra=\frac{\delta(\te_{2}-\te_{3})}{\mu(\te_{2})}\gamma_{\te_{1}\te_{2}}^{2},
\ee
we leave the details to Appendix~\eqref{ssec:cross-element}. Application of the exact solutions to these one-particle states involves the derivation of crossed four-point function, relevant for the OTOC calculation of DSSYK.  It is shown in~\cite{Xu:2024gfm} that the matrix element of $q^{\Delta \nt}$ in these one-particle states of given left and right energy leads to the crossed four-point function:
\begin{equation} \label{eq:slicing-4pt-cr}
\begin{aligned}
\langle \Delta; \theta_1, \theta_2 | q^{\Delta_1 \hat{n}_{\rm tot}} | \Delta; \theta_3, \theta_4 \rangle & =
    \left\langle \begin{tikzpicture}[baseline={([yshift=-0.1cm]current bounding box.center)}, line width=1pt, scale=0.6]
\draw (0,1) arc[start angle= 90, end angle = 270, radius=1];

\draw[thick, blue] (-1,0) -- (0,0);
\fill[blue] (-1,0) circle (1.5pt); 

\node at (-0.3,1.2) {\tiny $\te_1$};
\node at (-0.3,-1.2) {\tiny $\te_2$};
\node at (-0.3,0.3) {\tiny $\Delta$};

\end{tikzpicture}
\right| q^{\Delta \nt } \left| 
\begin{tikzpicture}[baseline={([yshift=-0.1cm]current bounding box.center)}, line width=1pt, scale=0.6]

\draw (0,1) arc[start angle= 90, end angle = -90, radius=1];

\draw[thick, blue] (0,0) -- (1,0);
\fill[blue] (1,0) circle (1.5pt); 

\node at (0.3,1.2) {\tiny $\te_3$};
\node at (0.3,-1.2) {\tiny $\te_4$}; 
\node at (0.3,0.3) {\tiny $\Delta$};

\end{tikzpicture}
\right\rangle 
=\begin{tikzpicture}[baseline={([yshift=-0.1cm]current bounding box.center)}, line width=1pt, scale=0.6]

\draw (0,0) circle (1);

\draw[thick, blue] (-1,0) -- (1,0);
\draw[thick,red] (0,-1) -- (0,1);

\node at (-0.6,1.2) {\tiny $\te_1$};
\node at (-0.6,-1.2) {\tiny $\te_2$};
\node at (0.6,1.2) {\tiny $\te_3$};
\node at (0.6,-1.2) {\tiny $\te_4$};
\node at (0.6,-0.3) {\tiny $\Delta$};
\node at (-0.3, 0.6) {\tiny $\Delta$};

\end{tikzpicture} \\
&= \gamma_{\te_1\te_2}\gamma_{\te_2\te_4}\gamma_{\te_4\te_3}\gamma_{\te_3\te_1}\left\{\begin{array}{ccc}
\Delta & \theta_1 & \theta_3 \\
\Delta & \theta_2 & \theta_4
\end{array}\right\}_q
\end{aligned}
\end{equation}

\subsection{The Crossing Element} \label{ssec:cross-element}

In this subsection we study the matrix element of the crossing element $\chi_\Delta(\tau)$. Physically, $\chi_\Delta(\tau)$ encodes the transition amplitude for one-particle states to pass through a region containing crossed matter chords, separated by Euclidean evolution of inverse temperature $\tau$, as illustrated in Fig.~\ref{fig:chi}.

In \cite{Cao:2025pir}, the matrix elements of $\chi_\Delta(\tau)$ in the chord-number basis $\{|\Delta;m,n\ra\}$ were computed explicitly. A key outcome is an identity relating $\chi_\Delta(\tau)$ to the quantum $6j$ symbol:
\be \label{eq:identity-Cao}
\begin{aligned}
\langle 0 | e^{-\tau_4 H} b e^{-\tau_3 H} \chi_{\Delta}\left(\tau_2\right) e^{-\tau_1 H} b^{\dagger}|0\rangle & =\int_0^\pi \prod_{i=1}^4 \mathrm{~d} \mu\left(\theta_i\right) e^{-\tau_i E_i} \times \\
& \times \prod_{i=1}^4 \gamma_{i, i+1}\left\{\begin{array}{ccc}
\Delta & \theta_1 & \theta_2 \\
\Delta & \theta_4 & \theta_3
\end{array}\right\}_q,
\end{aligned}
\ee
where $b$ and $b^\dagger$ are the matter ladder operators, and the product of vertex functions is understood as $\gamma_{\theta_1 \theta_2}\gamma_{\theta_2 \theta_3}\gamma_{\theta_3 \theta_4}\gamma_{\theta_4 \theta_1}$. For our purposes, we would like to extract from this relation the matrix element of $\chi_\Delta(\tau)$ in the one-particle energy basis.

To do so, we evaluate the left-hand side of \eqref{eq:identity-Cao} by inserting resolutions of the identity in terms of one-particle energy eigenstates. This gives
\be \label{eq:LHS}
\begin{aligned}
\mathrm{LHS} & =\int \frac{\prod_{i=1}^4 \mathrm{~d} \mu\left(\theta_i\right)}{\gamma_{\theta_3 \theta_4}^2 \gamma_{\theta_1 \theta_2}^2}\langle0| e^{-\tau_4 H} b e^{-\tau_3 H}\left|\Delta ; \theta_3, \theta_4\right\rangle\left\langle\Delta ; \theta_1, \theta_2\right| e^{-\tau_1 H} b^{\dagger}|0\rangle \\
& \hspace{2.2cm}\times\left\langle\Delta ; \theta_3, \theta_4\right| \chi_\Delta\left(\tau_2\right)\left|\Delta ; \theta_1, \theta_2\right\rangle.
\end{aligned}
\ee
The two overlap factors in the first line can be evaluated directly. First,
\be
\la0|e^{-\tau_{4}H}be^{-\tau_{3}H}|\Del;\te_{3},\te_{4}\ra=\la0|b_{L}e^{-\tau_{3}H_{L}-\tau_{4}H_{R}}|\Delta;\te_{3},\te_{4}\ra=e^{-\tau_{3}E_{3}-\tau_{4}E_{4}}\gamma_{\te_{3}\te_{4}}^{2},
\ee
and similarly,
\be
\la\Del;\te_{1},\te_{2}|e^{-\tau_{1}H}b^{\da}|0\ra=e^{-\tau_{1}E_{1}}\gamma_{\te_{1}\te_{2}}^{2}.
\ee
Substituting these back into \eqref{eq:LHS} and comparing with \eqref{eq:identity-Cao}, we read off the desired energy-basis matrix element:
\be
\la\Del;\te_{3},\te_{4}|\chi_{\Del}(\tau)|\Del;\te_{1},\te_{2}\ra=\frac{\delta(\te_{2}-\te_{4})}{\mu(\theta_{4})}\int\dd\mu(\theta)e^{-\tau E(\theta)}\prod_{i}\gamma_{i,i+1}\left\{ \begin{matrix}\Del & \te_{1} & \te\\
\Delta & \te_{4} & \te_{3}
\end{matrix}\right\} _{q}.
\ee

\section{The Emergence of the Discrete Spectrum}\label{app:spectral}

In this appendix we derive the orthogonality relation \eqref{eq:id-bound} in the regime $\kappa>1$. This includes an explicit derivation of the discrete weights $\omega_\kappa(l)$ and a concrete explanation of how the bound states arise as $\kappa$ is analytically continued from the continuous regime $|\kappa|<1$ to the bound-state regime $|\kappa|>1$. The key mechanism is the analytic continuation of the contour representation of the completeness relation: as $\kappa$ crosses $1$, poles of the integrand cross the unit circle, and their residue contributions generate the discrete part of the spectrum.

We begin with the orthogonality relation for continuous big $q$-Hermite polynomials in the regime $0<\kappa<1$:
\be \label{eq:ortho-z}
\delta_{mn}=\int_{0}^{\pi}\dd\mu_{\kappa}(\theta)\frac{H_{m}(\cos\te,\kappa|q^{2})H_{n}(\cos\theta,\ka|q^{2})}{\left(q^2;q^2\right)_{m}}=\frac{1}{2}\int_{|z|=1}\frac{\dd z}{2\pi i  z}W(z)P_{m}(z)P_{n}(z),
\ee 
where we set $z=e^{i\theta}$ and introduced
\be
W(z)=\frac{(z^{2},z^{-2},q^{2};q^{2})_{\infty}}{\left(\ka z,\ka z^{-1};q^{2}\right)_{\infty}},\quad P_{m}(z)=\frac{H_{m}(\frac{1}{2}(z+z^{-1}),\ka|q^{2})}{\sqrt{\left(q^{2};q^{2}\right)_{m}}}.
\ee
The factor of $\frac{1}{2}$ in \eqref{eq:ortho-z} comes from converting the original integral over $\theta\in[0,\pi]$ into a contour integral over $|z|=1$: the change of variables $z=e^{i\theta}$ maps $[0,\pi]$ to the upper semicircle, while the full unit circle traverses both the upper and lower semicircles. Since the integrand is invariant under $z\leftrightarrow z^{-1}$, the contribution from the lower semicircle duplicates that of the upper semicircle, producing the overall factor $\frac{1}{2}$.

For $0<\kappa<1$, the poles coming from the factor $(\kappa z;q^{2})_{\infty}^{-1}$ in $W(z)$ lie outside the unit disk, while those coming from $(\kappa z^{-1};q^{2})_{\infty}^{-1}$ lie inside. When $\kappa$ is increased past $1$, some poles cross the unit circle: the poles
\[
z=\kappa^{-1}q^{-2l},\qquad l=0,1,\dots
\]
from $(\kappa z;q^{2})_{\infty}^{-1}$ move into the unit disk, while the poles
\[
z=\kappa q^{2l}, \qquad l=0,1,\dots
\]
from $(\kappa z^{-1};q^{2})_{\infty}^{-1}$ move out. If we insist on keeping the contour fixed to be the unit circle for all $\kappa$, then analytic continuation across $\kappa=1$ requires accounting for the poles that cross the contour. Concretely, relative to the $\kappa<1$ expression, we must subtract the residues of poles that enter the unit disk and add the residues of poles that leave it. This gives
\be \label{eq:contour-extended}
\begin{aligned}
\delta_{m n} & =\frac{1}{2}\int_{|z|=1} \frac{\mathrm{~d} z}{2 \pi i z} W(z) P_m(z) P_n(z)+2 \pi i \sum_{k: 1<\kappa q^{2 k} \leq \kappa} \operatorname{Res}_{z=\kappa q^{2 k}}\left(\frac{W(z)}{4 \pi i z} P_m(z) P_n(z)\right) \\
& -2 \pi i \sum_{k: 1<\kappa q^{2 k} \leq \kappa} \operatorname{Res}_{z=\kappa^{-1} q^{-2 k}}\left(\frac{W(z)}{4 \pi i z} P_m(z) P_n(z)\right).
\end{aligned}
\ee
We first compute the residues in the second term of \eqref{eq:contour-extended}. The point $z=\kappa q^{2k}$ is a simple pole of $(\kappa z^{-1};q^{2})_{\infty}^{-1}$, and its residue is
\be
\begin{aligned}
\operatorname{Res}_{z=\kappa q^{2 k}} \frac{1}{\left(\kappa z^{-1} ; q^2\right)_{\infty}} & =\operatorname{Res}_{z=\kappa q^{2 k}}\left(\frac{1}{\left(1-\kappa q^{2 k} z^{-1}\right)} \prod_{j \neq k+1, j \geq 1} \frac{1}{\left(1-\kappa q^{2(j-1)} z^{-1}\right)}\right) \\
& =\kappa q^{2 k} \prod_{j \geq 1, j \neq k+1} \frac{1}{\left(1-q^{2(j-1-k)}\right)}.
\end{aligned}
\ee
The product in the second line can be rewritten using $q$-Pochhammer symbols:
\be
\prod_{j\geq1,j\not=k+1}(1-q^{2(j-1-k)})=\prod_{j=1}^{k}(1-q^{-2j})\prod_{j=1}^{\infty}(1-q^{2j})=(-1)^{k}q^{-2\binom{k+1}{2}}\left(q^{2};q^{2}\right)_{k}\left(q^{2};q^{2}\right)_{\infty},
\ee
and hence
\be \label{eq:residue-factor}
{\rm Res}_{z=\ka q^{2k}}\frac{1}{\left(\ka z^{-1};q^{2}\right)_{\infty}}=\frac{(-1)^{k}\ka q^{k(k+1)+2k}}{\left(q^{2};q^{2}\right)_{k}\left(q^{2};q^{2}\right)_{\infty}}.
\ee
The remaining factors in the integrand of \eqref{eq:contour-extended} are evaluated at $z=\kappa q^{2k}$. First, the polynomial factor becomes
\be \label{eq:Pm-value}
P_{m}(z)|_{z=\ka q^{2k}}= \frac{H_{m}(\cosh u_{k},\ka|q^{2})}{\sqrt{(q^2;q^2)_m}},\quad u_{k}=k\ln\ka q^{2},
\ee
and the remaining part of $W(z)$ gives
\be
\frac{(z^{2},z^{-2},q^{2};q^{2})_{\infty}}{\left(\ka z;q^{2}\right)_{\infty}}|_{z=\ka q^{2k}}=\frac{\left(\ka^{2}q^{4k},\ka^{-2}q^{-4k},q^{2};q^{2}\right)_{\infty}}{\left(\ka^{2}q^{2k};q^{2}\right)_{\infty}}.
\ee
We simplify this ratio using
\be
\begin{aligned}
\left(\kappa^{-2} q^{-4 k} ; q^2\right)_\infty  & =\left(1-\kappa^{-2} q^{-4 k}\right) \cdots\left(1-\kappa^{-2} q^{-2}\right)\left(\kappa^{-2} ; q^2\right)_{\infty} \\
& =\kappa^{-4 k} q^{-2 k(2 k+1)}\left(\kappa^2 q^2 ; q^2\right)_{2 k}\left(\kappa^{-2} ; q^2\right)_{\infty},
\end{aligned}
\ee
and
\be
\left(\kappa^2 q^{4 k} ; q^2\right)_{\infty}=\frac{\left(\kappa^2 ; q^2\right)_{\infty}}{\left(\kappa^2 ; q^2\right)_{2 k}}, \quad\left(\kappa^2 q^{2 k} ; q^2\right)_{\infty}=\frac{\left(\kappa^2 ; q^2\right)_{\infty}}{\left(\kappa^2 ; q^2\right)_k},
\ee 
so that
\be \label{eq:ratio-simplified}
\begin{aligned}
\frac{\left(\kappa^2 q^{4 k}, \kappa^{-2} q^{-4 k}, q^2 ; q^2\right)_{\infty}}{\left(\kappa^2 q^{2 k} ; q^2\right)_{\infty}} & =\kappa^{-4 k} q^{-2 k(2 k+1)} \frac{\left(\kappa^2 q^2 ; q^2\right)_{2 k}}{\left(\kappa^2 ; q^2\right)_{2 k}}\left(\kappa^2 ; q^2\right)_k\left(\kappa^{-2} ,q^2; q^2\right)_{\infty} \\
& =\kappa^{-4 k} q^{-2 k(2 k+1)} \frac{1-\kappa^2 q^{4 k}}{1-\kappa^2}\left(\kappa^2 ; q^2\right)_k \left(\kappa^{-2}, q^2 ; q^2\right)_{\infty}.
\end{aligned}
\ee
Combining \eqref{eq:residue-factor}, \eqref{eq:Pm-value}, and \eqref{eq:ratio-simplified}, we obtain
\be
2\pi i\, {\rm Res}_{z=\ka q^{2k}}\left(\frac{W(z)}{4\pi iz}P_{m}(z)P_{n}(z)\right)=\frac{1}{2}\omega_{\kappa}(k)\frac{H_{m}(\cosh u_{k},\ka|q^{2})H_{n}(\cosh u_{k},\kappa|q^{2})}{\left(q^{2};q^{2}\right)_{m}},
\ee
where the discrete weight is
\be
\omega_{\ka}(k)=(-1)^{k}\ka^{-4k}q^{-3k^{2}-k}\frac{1-\ka^{2}q^{4k}}{1-\ka^{2}}\frac{\left(\ka^{2};q^{2}\right)_{k}}{\left(q^{2};q^{2}\right)_{k}}\left(\ka^{-2};q^{2}\right)_{\infty}.
\ee
Similarly, the pole at $z=\kappa^{-1}q^{-2k}$ contributes
\be
2\pi i\, {\rm Res}_{z=\ka^{-1}q^{-2k}}\left(\frac{W(z)}{4\pi iz}P_{m}(z)P_{n}(z)\right)=-\frac{1}{2}\omega_{\kappa}(k)\frac{H_{m}(\cosh u_{k},\ka|q^{2})H_{n}(\cosh u_{k},\kappa|q^{2})}{\left(q^{2};q^{2}\right)_{m}}.
\ee
Therefore, the residue contributions in \eqref{eq:contour-extended} combine to produce precisely the discrete sum in \eqref{eq:id-bound}, demonstrating how the bound states arise from analytic continuation of the continuous-spectrum completeness relation as $\kappa$ is continued through $\kappa=1$.

Intuitively, the discrete spectrum appears because the analytic structure of the spectral integral changes as $\kappa$ is increased. For $0<\kappa<1$, the completeness relation is fully captured by the contour integral over the unit disk, which encloses the only poles of $(\kappa z^{-1},q^2)_\infty^{-1}$. When $\kappa$ crosses $1$, a finite set of poles of the integrand move across the contour used to represent that integral. If one insists on keeping the same contour prescription while varying $\kappa$, analytic continuation forces one to include the residue contributions from the singularities that cross. These extra residue terms are precisely what reorganize the completeness relation into a continuous part plus a discrete sum, and the latter is naturally interpreted as the contribution of bound states with weights $\omega_\kappa(l)$. This is illustrated in the following figure.

\begin{tikzpicture}[scale=1.2]
\begin{scope}[shift={(0,0)}]

 \coordinate (P) at (3,0);
    \draw[->, line width=1pt] (-2.,0) -- (2,0);
    \draw[->, line width=1pt] (0,-2) -- (0,2);
    \draw (1.8,2.2)--(1.8,1.8) -- (2.2,1.8);
    
    \draw[line width=1pt] (0,0) circle (1.0);

    \draw[->, line width=1pt] (40:1.0) arc[start angle=40,end angle=20,radius=1.0];

    \fill[blue] (0.8,0) circle (0.05);
    \fill[red]  (1.4,0) circle (0.05);

    \node at (2,2) {$z$};
\end{scope}

\node at (3,0) {$\Longrightarrow$};
\node at (3.0,0.3) {Increase $\kappa$};

\begin{scope}[shift={(6,0)}]
    \draw[->, line width=1pt] (-2,0) -- (2,0);
    \draw[->, line width=1pt] (0,-2) -- (0,2);
    \draw (1.8,2.2)--(1.8,1.8) -- (2.2,1.8);
    \draw[line width=1pt] (0,0) circle (1.);

    \draw[->, line width=1pt] (40:1) arc[start angle=40,end angle=20,radius=1];

    \fill[red]  (.7,0) circle (0.05);
    \fill[blue] (1.5,0) circle (0.05);

    \draw[dashed,thin] (0.7, 0) circle (0.2);
    \draw[dashed,thin] (1.5,0) circle (0.2);
 
    \draw[->, line width=.9pt,red] ($(20:.2)+(0.7,0)$) arc[start angle=20,end angle=40,radius=.2];
    \draw[->, line width=.9pt,blue] ($(40:.2)+(1.5,0)$) arc[start angle=40,end angle=20,radius=.2];
    
    \node at (2,2) {$z$};
\end{scope}

\end{tikzpicture}

\section{Detailed Derivation of the Symmetric Kernel $\mathcal{S}(\theta_1,\theta_2)$} \label{app:derivation-S}

In this appendix we derive the symmetric kernel $\mathcal{S}$ appearing in \eqref{eq:kernel-S}. The starting point is the relation between the continuous big $q$-Hermite polynomials and the Al Salam--Chihara polynomials,
\be\label{eq:HQ-relation}
H_{n}(x,\ka|q^{2})=Q_{n}(x;\ka,0|q^{2}),
\ee
where $Q_{n}(x)=Q_{n}(x;a,b|q^{2})$ is defined by the three-term recursion relation
\be \label{eq:recur-Q}
2xQ_{n}(x)=Q_{n+1}(x)+(a+b)q^{2n}Q_{n}+(1-q^{2 n})(1-abq^{2n-2})Q_{n-1}(x),
\ee
with initial conditions $Q_{0}=1$ and $Q_{-1}=0$. Setting $b=0$ reduces \eqref{eq:recur-Q} to the recursion \eqref{eq:recur}, which establishes \eqref{eq:HQ-relation}.
The Al Salam--Chihara polynomials admit a Poisson kernel expressed in terms of a very-well-poised basic hypergeometric series~\cite{Berkooz:2018jqr}:
\be \label{eq:kernel-Q}
\begin{aligned}
\sum_{n=0}^{\infty} \frac{Q_n\left(x ; a, b \mid q^2\right) Q_n\left(y ; \alpha, \beta \mid q^2\right)}{\left(a b, q^2 ; q^2\right)_n} t^n & =\frac{\left(\beta t / a, \alpha t e^{ \pm i \theta}, a t e^{ \pm i \phi} ; q^2\right)_{\infty}}{\left(\alpha a t, t e^{ \pm i \theta \pm i \phi} ; q^2\right)_{\infty}} \times \\
& \times{ }_8 W_7\left(\alpha a t q^{-2} ; \alpha t / b, a e^{ \pm i \theta}, \alpha e^{ \pm i \phi} ; q^2, \beta t / a\right),
\end{aligned}
\ee
valid when $ab=\alpha\beta$, and where we set $x=\cos\theta$ and $y=\cos\phi$. We will specialize to $a=\alpha$ and then take the limit $b=\beta\to0$. To take this limit, we use the series expansion of ${}_8W_7$~\cite{Stokman_2012}:
\be
\begin{split}
    _{8}W_{7}(a^{2}tq^{-1};at/b,ae^{\pm i\theta},& ae^{\pm i\phi};q,bt/a)	= \\
    &\sum_{r=0}^{\infty}\frac{1-a^{2}tq^{2r-1}}{1-a^{2}tq^{-1}}\frac{\left(a^{2}tq^{-1},\frac{at}{b},ae^{\pm i\theta},ae^{\pm i\phi};q\right)_{r}}{\left(ab,ate^{\pm i\theta},ate^{\pm i\phi},q;q\right)_{r}}\left(\frac{bt}{a}\right)^{r}.
\end{split}
\ee
In the limit $b\to0$, the factor $(ab;q)_r$ in the denominator becomes $(0;q)_r=1$. Meanwhile, the divergent factor $(at/b;q)_r$ combines with the vanishing factor $(bt/a)^r$ to produce a finite result:
\be
\lim_{b\to0}\left(\frac{bt}{a}\right)^{r}\left(\frac{at}{b};q\right)_{r}=\lim_{b\to0}\prod_{j=1}^{r}(\frac{bt}{a}-t^{2}q^{j-1})=(-t^{2})^{r}q^{\binom{r}{2}}.
\ee
Therefore,
\be
\begin{aligned}
\lim _{b \rightarrow 0} {}_8 W_7\left(a^2 t q^{-1} ; a t / b, a e^{ \pm i \theta}, a e^{ \pm i \phi} ; q, b t / a\right) & =\sum_{r=0}^{\infty} \frac{1-a^2 t q^{2 r-1}}{1-a^2 t q^{-1}}(-1)^r q^{\binom{r}{2}} \\
& \times \frac{\left(a^2 t q^{-1}, a e^{ \pm i \theta}, a e^{ \pm i \phi} ; q\right)_r}{\left(a t e^{ \pm i \theta}, a t e^{ \pm i \phi}, q ; q\right)_r} t^{2 r}.
\end{aligned}
\ee
Now set $a=\alpha=\kappa$ and take $b=\beta\to0$ in \eqref{eq:kernel-Q}. Using \eqref{eq:HQ-relation} to rewrite $Q_n$ in terms of $H_n$, we obtain
\be
\begin{aligned}\sum_{n=0}^{\infty}\frac{H_{n}(x;\ka|q^{2})H_{n}(y;\ka|q^{2})}{\left(q^{2};q^{2}\right)_{n}}q^{2\Del n} & =\frac{\left(\ka q^{2\Del}e^{\pm i\theta},\ka q^{2\Del}e^{\pm i\phi};q^{2}\right)_{\infty}}{\left(\ka^{2}q^{2\Del},q^{2\Del}e^{\pm i\theta\pm i\phi};q^{2}\right)_{\infty}}\mathcal{S}(\theta,\phi),
\end{aligned}
\ee
where the symmetric kernel is given by
\be
\begin{aligned}
\mathcal{S}(\theta, \phi) & =\lim_{b\rightarrow 0} {}_8 W_7\left(\kappa^2 q^{2 \Delta-2} ; \kappa q^{2 \Delta} / b, \kappa e^{ \pm i \theta}, \kappa e^{ \pm i \phi}, ; q^2, b q^{2 \Delta} / \kappa\right) \\
& =\sum_{n=0}^{\infty} \frac{1-\kappa^2 q^{2 \Delta+4 n-2}}{1-\kappa^2 q^{2 \Delta-2}} \frac{\left(\kappa^2 q^{2 \Delta-2}, \kappa e^{ \pm i \theta}, \kappa e^{ \pm i \phi} ; q^2\right)_n}{\left(\kappa q^{2 \Delta} e^{ \pm i \theta}, \kappa q^{2 \Delta} e^{ \pm i \phi}, q^2 ; q^2\right)_n}(-1)^n q^{n(n-1)} q^{4 \Delta n}.
\end{aligned}
\ee
This reproduces the expression for $\mathcal{S}$ stated in \eqref{eq:kernel-S}.

\section{Calculating the Distance to the End of the World} \label{app:distance}

In this section, we compute the expectation value of the length operator $\hat l$ in the bound states. We define
\be\label{eq:distance-def}
\la\hat{l}\ra_{n}:=\frac{\int_{-\infty}^{\infty}\dd l\thinspace l\thinspace|\psi_{k_{n},\nu}(l)|^{2}}{\int_{-\infty}^{\infty}\dd l\left|\psi_{k_{n},\nu}(l)\right|^{2}},
\qquad 
k_n = -\left(\nu + n +\frac{1}{2}\right).
\ee
It is convenient to rewrite the wavefunction in terms of Laguerre polynomials. Introducing the variable $x=e^{-l}$, one finds
\be
\psi_{k_{n},\nu}(l)=e^{l/2}W_{-\nu,k_{n}}\left(e^{-l}\right)=(-1)^{n}n!e^{-x/2}x^{k_{n}}L_{n}^{\left(2k_{n}\right)}(x).
\ee
With this change of variables, both the numerator and denominator in~\eqref{eq:distance-def} reduce to integrals over $x\in(0,\infty)$. In particular, the normalization integral becomes
\be \label{eq:to-compute-1}
\int_{-\infty}^{\infty}\dd l\left|\psi_{k_{n},\nu}(l)\right|^{2}
=(n!)^{2}\int_{0}^{\infty}\dd x\,e^{-x}x^{2k_{n}-1}\left(L_{n}^{(2k_{n})}(x)\right)^{2},
\ee
while the logarithmic moment needed for $\la \hat{l}\ra_n$ is
\be \label{eq:to-compute-2}
\int_{-\infty}^{\infty}\dd l\, l \left|\psi_{k_{n},\nu}(l)\right|^{2}
= - (n!)^{2}\int_{0}^{\infty}\dd x\,e^{-x}x^{2k_{n}-1}\left(L_{n}^{(2k_{n})}(x)\right)^{2}\ln x.
\ee
We will repeatedly use the standard orthogonality relation for Laguerre polynomials,
\be 
\int_{0}^{\infty}e^{-x}x^{\beta}L_{j}^{(\beta)}(x)L_{k}^{(\beta)}(x)\dd x
=
\delta_{jk}\frac{\Gamma(k+\beta+1)}{k!}.
\ee

A minor mismatch arises because the power of $x$ in~\eqref{eq:to-compute-1} is $2k_n-1$, whereas the upper index of the Laguerre polynomial is $2k_n$. To align the two, we lower the upper index by one using the identity
\be \label{eq:relation1-L}
L_{j}^{(\beta)}(x)=\sum_{r=0}^{j}L_{r}^{(\beta-1)}(x).
\ee
Substituting this into~\eqref{eq:to-compute-1} and applying orthogonality gives
\be \label{eq:result-compute-1}
\int_{0}^{\infty}\dd x\,e^{-x}x^{2k_{n}-1}\left(L_{n}^{(2k_{n})}(x)\right)^{2}
=
\sum_{r=0}^{n}\frac{\Gamma(r+2k_{n})}{r!}
=
\frac{\Gamma(n+2k_{n}+1)}{n!\,2k_{n}}.
\ee
We next compute the corresponding integral with a $\ln x$ insertion. A convenient way to generate $\ln x$ is to differentiate with respect to the exponent of $x$, namely with respect to a parameter $\beta$. The only subtlety is that $L_n^{(\beta)}(x)$ also depends on $\beta$. For this, we use the identity
\be \label{eq:relation2-L}
\frac{\partial}{\partial\beta}L_{n}^{(\beta)}(x)=\sum_{j=0}^{n-1}\frac{L_{j}^{(\beta)}(x)}{n-j}.
\ee
For simplicity, let us write $\beta=2k_n$. Differentiating~\eqref{eq:result-compute-1} with respect to $\beta$, the left-hand side becomes
\be \label{eq:beta-derivative}
\begin{aligned}
\frac{\partial}{\partial\beta}\int_{0}^{\infty}\mathrm{d}x\,e^{-x}x^{\beta-1}\left(L_{n}^{(\beta)}(x)\right)^{2}
=
-\frac{\Gamma(n+\beta+1)}{n!\beta}\la\hat{l}\ra
+2\int_{0}^{\infty}\dd x\, e^{-x}x^{\beta-1}L_{n}^{(\beta)}(x)\partial_{\beta}L_{n}^{(\beta)}(x),
\end{aligned}
\ee
where $\la\hat l\ra$ is the desired expectation value, to be evaluated at $\beta=2k_n$.

Substituting~\eqref{eq:relation1-L} and~\eqref{eq:relation2-L} into the second term gives
\be \label{eq:l-compute-2}
\begin{aligned}
\int_0^{\infty} \dd x\, e^{-x} x^{\beta-1} L_n^{(\beta)}(x)\,\partial_\beta L_n^{(\beta)}(x)
& =\sum_{j=0}^{n-1} \frac{1}{n-j} \int_0^{\infty} \dd x\, e^{-x} x^{\beta-1} L_n^{(\beta)}(x) L_j^{(\beta)}(x) \\
& =\sum_{j=0}^{n-1} \frac{1}{n-j} \sum_{l_1=0}^n \sum_{l_2=0}^j \int_0^{\infty} \dd x\, e^{-x} x^{\beta-1} L_{l_1}^{(\beta-1)}(x) L_{l_2}^{(\beta-1)}(x) \\
& =\sum_{j=0}^{n-1} \frac{1}{n-j} \sum_{l_1=0}^n \sum_{l_2=0}^j \delta_{l_1 l_2} \frac{\Gamma(\beta+l_1)}{l_1!}.
\end{aligned}
\ee
We may decompose the sum over $l_1$ as
\be
\sum_{l_{1}=0}^{n}=\sum_{l_{1}=0}^{j}+\sum_{l_{1}=j+1}^{n}.
\ee
Since $j<n$ throughout the sum, the Kronecker delta $\delta_{l_1l_2}$ has no support when $l_1>j$, for any allowed value of $l_2$. Therefore the last line of~\eqref{eq:l-compute-2} reduces to
\be \label{eq:result-compute-2}
\begin{aligned}
\sum_{j=0}^{n-1} \sum_{l=0}^j \frac{1}{n-j} \frac{\Gamma(\beta+l)}{l!}
& =\frac{1}{\beta} \sum_{j=0}^{n-1} \frac{1}{n-j} \frac{\Gamma(\beta+j+1)}{j!} \\
& =\frac{\Gamma(\beta+n+1)}{n!\beta}\big(\psi(\beta+n+1)-\psi(\beta+1)\big),
\end{aligned}
\ee
where $\psi(x)$ is the digamma function,
\be
\psi(x)= \frac{\Gamma^\prime (x)}{\Gamma (x)}=\frac{\dd}{\dd x}\ln\Gamma(x).
\ee

Substituting~\eqref{eq:result-compute-2} back into~\eqref{eq:beta-derivative}, we obtain
\be \label{eq:result-compute-3}
\left[- \langle\hat{l}\rangle+2\big(\psi(\beta+n+1)-\psi(\beta+1)\big)\right]
=
\frac{n!\beta}{\Gamma(n+\beta+1)}
\frac{\partial}{\partial\beta}
\int_{0}^{\infty}\mathrm{d}x\,e^{-x}x^{\beta-1}\left(L_{n}^{(\beta)}(x)\right)^{2}.
\ee
The $\beta$-derivative on the right-hand side can be evaluated directly from the right-hand side of~\eqref{eq:result-compute-1}:
\be \label{eq:result-compute-4}
\frac{n! \beta}{\Gamma(\beta+n+1)}
\frac{\partial}{\partial\beta}
\left[\frac{\Gamma\left(n+\beta+1\right)}{n!\beta}\right]
=
\psi(n+\beta+1)-\frac{1}{\beta}.
\ee
Substituting~\eqref{eq:result-compute-4} into~\eqref{eq:result-compute-3}, and then setting $\beta=2k_n=-2n-2\nu-1$, we arrive at the expectation value of $\hat l$ in the $n$th bound state:
\be
\la\hat{l}\ra_{n}
= - \psi(-2n-2\nu) -\psi(-2n-2\nu-1) + \psi(-n-2\nu).
\ee

\bibliography{ref}
\bibliographystyle{JHEP}

\end{document}